\documentclass{ar-1col}

\usepackage[comma]{natbib}
\usepackage{url}
\usepackage{amsmath,verbatim,amssymb,bm,graphicx}

\def\bi{\begin{itemize}}
\def\ei{\end{itemize}}
\def\be{\begin{equation}}
\def\ee{\end{equation}}
\newcommand{\bcu}{}
\newcommand{\sectn}[1]{\ref{sec:#1}}
\newcommand{\sect}[1]{Section \ref{sec:#1}}
\newcommand{\eq}[1]{Equation \ref{eq:#1}}
\newcommand{\figg}[1]{\textbf{Figure~\ref{fig:#1}}}

\newcommand{\sigmah}{\sigma_{\rm h}}
\newcommand{\sigmac}{\sigma_{\rm c}}
\newcommand{\gammarad}{\gamma_{\rm rad}}
\newcommand{\gammacool}{\gamma_{\rm cool}}
\newcommand{\gammainj}{\gamma_{\rm inj}}
\newcommand{\gammabr}{\gamma_{\rm br}}
\newcommand{\gammacut}{\gamma_{\rm cut}}
\newcommand{\sigmae}{\sigma_{{\rm c},e}}
\newcommand{\sigmacs}{\sigma_{{\rm c},s}}

\jname{Xxxx. Xxx. Xxx. Xxx.}
\jvol{AA}
\jyear{YYYY}
\doi{10.1146/((please add article doi))}

\begin{document}

\markboth{Sironi et al.}{Relativistic Reconnection in Astrophysics}

\title{Relativistic Magnetic Reconnection in Astrophysical Plasmas: A Powerful Mechanism of Nonthermal Emission}
\author{Lorenzo Sironi,$^{1,2}$ Dmitri A. Uzdensky,$^{3,4}$ and Dimitrios Giannios$^5$
\affil{$^1$Department of Astronomy and Columbia Astrophysics Laboratory, Columbia University,
New York, NY, 10027, USA, email: lsironi@astro.columbia.edu}
\affil{$^2$Center for Computational Astrophysics, Flatiron Institute,
New York, NY, 10010, USA}
\affil{$^3$Center for Integrated Plasma Studies, Univ. of Colorado, Boulder, CO, USA}
\affil{$^4$Rudolf Peierls Centre for Theoretical Physics, Clarendon Laboratory, University of Oxford, Parks Road, Oxford OX1 3PU, United Kingdom}
\affil{$^5$Department of Physics, Purdue University, West Lafayette, IN, 47907, USA}}

\begin{abstract}
Magnetic reconnection---a fundamental plasma physics process, where magnetic field lines of opposite polarity annihilate---is invoked in astrophysical plasmas as a powerful mechanism of nonthermal particle acceleration, able to explain fast-evolving, bright high-energy flares. Near black holes and neutron stars, reconnection occurs in the ``relativistic'' regime, in which the mean magnetic energy per particle exceeds the rest mass energy. This review reports recent advances in our understanding of the kinetic physics of relativistic reconnection:
(1) Kinetic simulations have elucidated the physics of plasma heating and nonthermal particle acceleration in relativistic reconnection;
(2) The physics of radiative relativistic reconnection, with its self-consistent interplay between photons and reconnection-accelerated particles---a peculiarity of luminous, high-energy astrophysical sources---is the new frontier of research; (3) Relativistic reconnection plays a key role in global models of high-energy sources, both in terms of global-scale layers, as well as of reconnection sites generated as a byproduct of local magnetohydrodynamic instabilities. We summarize themes of active investigation and future directions, emphasizing the role of upcoming observational capabilities, laboratory experiments, and new computational tools.
\end{abstract}

\maketitle

\tableofcontents

\section{INTRODUCTION}
\label{sec:intro}
Astrophysical high-energy sources universally display nonthermal photon spectra.
Nonthermal energy spectra also characterize high-energy cosmic rays (CRs) and astrophysical neutrinos.
Thus, the multi-messenger signatures of astrophysical high-energy sources require a mechanism to generate nonthermal particles, typically with power-law energy distributions.

Particle acceleration has historically been attributed to shocks via the Fermi process, in which charged particles are energized by scattering back and forth across the shock \citep{bell_78,blandford_ostriker_78,drury_83,blandford_eichler_87}. The Fermi mechanism provides a compelling explanation for the acceleration of electrons and ions at shocks of supernova remnants (SNRs; e.g., \citealt{caprioli_15,marcowith_16}) and gamma-ray burst (GRB) afterglows (e.g., \citealt{waxman_06,sironi_keshet_15}). Both refer to ``external'' shocks, formed by the interaction of the blast wave with the surrounding medium.
However, shock models face severe difficulties in explaining emission originating from within magnetized relativistic flows, i.e., in the case of ``internal'' shocks. Astrophysical relativistic outflows are commonly launched as Poynting-flux-dominated, driven by a strong magnetic field threading a rotating compact object---either a spinning black hole and/or its accretion disk \citep{blandford_77,blandford_payne_82,meier_01,vlahakis_03} or a magnetized neutron star \citep{rees_gunn_74,kennel_coroniti_84,bogovalov_99,usov_92,metzger_11}. The flow starts as magnetically dominated, and some form of internal dissipation of the dominant magnetic energy is required to mediate the tremendous, rapid release of energy inferred from observations.

The most dramatic examples of such phenomena are the flaring episodes observed from relativistic jets powered by supermassive black holes in active galactic nuclei (AGN), especially blazars, and from the Crab Nebula. Blazar flares at GeV and TeV energies can be as short as tens of minutes \citep{aharonian_07,ackermann_16}, which is shorter than the light-crossing time of the black hole powering the jet. The Crab Nebula, a prototypical high-energy source, has long been considered a standard candle, yet it recently showed tremendous GeV flares on timescales of tens of hours \citep{abdo_11,tavani_11,buehler_12}. Shock models cannot adequately explain the properties of these flares. In magnetically dominated outflows, internal shocks are generally inefficient, as regard to both plasma heating \citep{kennel_coroniti_84b,giannios_08,narayan_11} and particle acceleration \citep[e.g.,][]{sironi_spitkovsky_09, sironi_spitkovsky_11b, sironi_13,sironi_keshet_15}. 
In application to the Crab Nebula flares, shocks cannot account for the extreme photon energies and the hard power-law spectra.
Thus, observations of high-energy astrophysical flares from relativistic outflows have called the shock paradigm into question. 

\begin{marginnote}
\entry{Hard and soft spectrum}{A power-law energy spectrum $dN/d \varepsilon\propto \varepsilon^{-p}$, where $dN$ is the number of particles or photons in the energy interval $[\varepsilon,\varepsilon+d \varepsilon]$, is defined as hard (respectively, soft) if the power-law slope $p$ is small (respectively, large).}
\end{marginnote}

Dramatic,  bright and rapid flares, often with nonthermal spectra, are also observed from magnetically dominated astrophysical environments that are not expected to have supersonic and super-magnetosonic flows. These are the magnetospheres and coronae of relativistic compact objects---neutron stars and accreting black holes.  Prominent examples of flaring phenomena from such sources include giant hard X-ray and gamma-ray flares from Soft Gamma Repeaters (SGRs, a class of magnetars; \citealt{hurley_05}), pulsed high-energy emission from pulsar magnetospheres \citep{abdo_10}, and flaring emission from accretion disk coronae and magnetospheres of stellar-mass and supermassive black holes, including Sgr~A* \citep{genzel_03,baganoff_01,nielsen_13,gravity_18} and~M87 \citep{aharonian_06,abramowski_12}.

This suggests that, instead of shocks, another robust and ubiquitous dissipation mechanism must be operating efficiently in a broad variety of strongly magnetized environments, tapping directly into the free energy of a stressed magnetic field rather than the bulk kinetic energy of a fast outflow. 
This mechanism is {\it magnetic reconnection}---a fundamental plasma-physical process entailing a rapid rearrangement of magnetic field topology, which enables a violent release of the stored magnetic energy \citep[see][for recent reviews]{zweibel_yamada_09, yamada_review-2010, hesse_review,hantao_review}.
Simply put, reconnection
involves sudden breaking and reconnecting of misaligned magnetic field lines in a plasma, which releases the accumulated magnetic stresses and rapidly converts magnetic energy into plasma energy. 
Magnetic reconnection is widely recognized as one of the most important plasma processes in a broad range of space, astrophysical, and laboratory contexts and has therefore been the subject of intensive study over the last several decades using analytical, numerical, experimental, and observational methods. 
It is commonly accepted to be the mechanism powering solar flares; it regulates space weather---the complex dynamical behavior of the Earth's magnetosphere, including its interaction with the solar wind and the triggering of intense geomagnetic storms; and it is responsible for sawtooth crashes in tokamak magnetic fusion devices \citep{yamada_review-2010,hesse_review,hantao_review}. 

This review focuses on relativistic reconnection [RR], where the term ``relativistic'' indicates that some aspects cannot be described with a non-relativistic, Newtonian approach: the plasma is heated to (ultra)-relativistic temperatures, and/or bulk flows can be (ultra)-relativistic \citep[for reviews of RR, see] []{hoshino_lyubarsky_12,kagan_15,guo_review_20,guo_review}. As we will describe, RR has some peculiarities as compared to its non-relativistic counterpart, including fast dissipation rates and rapid motions. Most importantly, it enables efficient nonthermal particle acceleration, which is essential for explaining the extreme emission signatures of high-energy astrophysical sources.

In tackling the physics of collisionless RR, a fully kinetic approach is essential for two main reasons. First, the change in the macroscopic topology of the magnetic field is due to non-ideal plasma effects on microscopic scales of, e.g., the electron Larmor radius. In fact, the resistive magnetohydrodynamic (MHD) approach, which treats the plasma as a single fluid and neglects two-fluid and kinetic physics, yields reconnection rates that are much lower than equivalent two-fluid or kinetic calculations \citep{birn_01, cassak_review}.
Second, the formation of nonthermal particle distributions, potentially with hard slopes, can be captured self-consistently only within a kinetic framework. 
The most popular and versatile computational tool for describing the complex nonlinear interplay between electromagnetic fields and particles at the kinetic level is the Particle-in-Cell (PIC) method \citep[e.g.,][]{birdsall_langdon_91}, which is widely used for studies of reconnection as well as other collective plasma processes (e.g., shocks, turbulence, plasma instabilities).

This review primarily focuses on the insights obtained from recent PIC simulation studies regarding energy dissipation, particle acceleration, and the resulting nonthermal emission, and by extension, the multi-messenger signatures of RR in high-energy astrophysical sources. It is organized as follows: in \sect{basic} we introduce the basic geometry and fundamental physics of magnetic reconnection and the main parameters governing the system; in \sect{mhd} we discuss the widespread evidence for reconnection layers in relativistic astrophysical objects; in \sect{micro} we describe the plasma microphysics of RR, including its complex dynamics driven by the interplay of several instabilities and its role in plasma heating and particle acceleration; \sect{radiation} is devoted to radiative RR and the self-consistent interplay of photons and RR-accelerated particles; in \sect{turb} we summarize recent PIC efforts to model global-scale astrophysical systems harboring RR layers; in \sect{astro} we describe how recent advancements in our understanding of RR have allowed us to gain unprecedented insight into the emission physics of astrophysical high-energy sources; finally, in \sect{summary} we summarize the main points discussed in this review and present current themes of active investigation and future directions, emphasizing the role of novel computational tools and upcoming laboratory experiments.

\section{FUNDAMENTALS OF MAGNETIC RECONNECTION}
\label{sec:basic}
\subsection{Governing Equations and Generalized Ohm's Law}
\label{sec:ohm}
Magnetic reconnection refers to the breaking and reconnecting of oppositely directed magnetic field lines in a plasma. 
As we justify below, this requires the  ``ideal'' condition
\begin{equation}
\label{eq:ideal}
{\bm E}+\frac{\langle{\bm v}_s\rangle}{c}\times {\bm B}=0
\end{equation}
to be violated somewhere in
the domain for each relevant plasma species $s$. Here, ${\bm E}$ and ${\bm B}$ are the electromagnetic fields, while $\langle {\bm v}_s\rangle=\int\! ({\bm u}/\gamma) f_sd^3 u/\int \!f_s d^3u$ is the mean 3-velocity of species $s$. Here, $f_s$ is the phase-space distribution function of species $s$, ${\bm u}$ is the 4-velocity and $\gamma=\sqrt{1+u^2/c^2}$ is the Lorentz factor. 
In {\it collisional} plasmas, the ideal condition can be broken by resistive effects due to binary particle collisions. This is encoded in Ohm's law, which in {\it non-relativistic, resistive, single-fluid} MHD reads
\begin{equation}
{\bm E}+\frac{{\bm v}_f}{c}\times {\bm B}=\eta {\bm J} ~~~[\text{non-relativistic}]
\end{equation}
where $\eta$ is the resistivity due to particle collisions, ${\bm v}_f$ the velocity of the MHD fluid, 
and ${\bm J}$ the electric current. 
Its extension to {\it relativistic, resistive, single-fluid} MHD reads
\begin{eqnarray}
   \Gamma_f\left[{\bm E}+ \frac{{\bm v}_f}{c}\times {\bm B}-\frac{({\bm E}\cdot {\bm v}_f) {\bm v}_f}{c^2}\right]= \eta({\bm J}-\rho_e {\bm v}_f) ~~~[\text{relativistic}]
\end{eqnarray}
where $\rho_e={\bm \nabla} \cdot {\bm E}/4\pi$ is the electric charge density and $\Gamma_f=1/\sqrt{1-(v_f/c)^2}$ the fluid bulk Lorentz factor \citep[e.g.,][]
{komissarov_07b,bucciantini_13}. 
The numerical implementation of resistive effects in relativistic MHD codes is a non-trivial task, requiring an implicit–explicit scheme to treat the stiff resistive source terms and preserve causality \citep[e.g.,][]{delzanna_07,ripperda_19}. 

The combination of Ohm's law and Faraday's equation leads to the magnetic induction equation, which governs the field evolution. For non-relativistic, resistive MHD,
\begin{equation}
\label{eq:induction}
\frac{\partial {\bm B}}{\partial {t}}= \nabla \times ({\bm v}_f \times {\bm B})- \nabla \times \left(\frac{\eta c^2}{4\pi}\, \nabla \times {\bm B}\right)~~~[\text{non-relativistic}]
\end{equation}
In the case of vanishingly small resistivity ($\eta\rightarrow 0$, or equivalently ${\bm E}+ ({\bm v}_f/c)\times {\bm B}\rightarrow0$), the first term on the right-hand side dominates. Then, the fluid can be called a perfect or ideal fluid and the magnetic field is ``frozen'' into the fluid, as prescribed by Alfv\'en's theorem, also known as the flux freezing theorem \citep{alfven_43,pegoraro_12,asenjo_comisso_17,comisso_asenjo_20}. More precisely, \eq{induction} implies that the perfectly-conducting limit (i.e., $\eta\rightarrow 0$) should apply when the typical scale over which the field varies is much larger than resistive scales. This limit is violated in regions of large magnetic shear (equivalently, sheets of intense electric current), where the second term on the right-hand side of \eq{induction} dominates. There, resistive effects become important, invalidating Alfv\'en's theorem and causing the magnetic field to diffuse through the plasma. This is what reconnection fundamentally entails: a {\it macroscopic} change in the magnetic field topology due to {\it microscopic} effects---where ``macroscopic'' refers to the extent of the system along the field direction, while ``microscopic'' to the scale of the field reversal, where resistive effects become important.

\begin{marginnote}
\entry{Alfv\'en theorem}{Alfv\'en's theorem, or the frozen-in flux theorem, states that electrically conducting fluids and embedded magnetic fields are constrained to move together in the limit of vanishing resistivity.}
\end{marginnote}

In dilute astrophysical plasmas, binary collisions are rare, so the collisional resistivity~$\eta$ is often insufficient to break flux freezing on interesting time and length scales. A large body of work has focused on identifying the processes that can supply additional electric field terms, so the ideal condition in \eq{ideal} is violated and reconnection may take place in collisionless or weakly collisional plasmas. The terms that break flux freezing can be derived from the momentum equation of species~$s$ (with particle mass $m_s$ and charge~$q_s$). This gives the generalized Ohm's law, which in  the collisionless limit reads
\begin{equation}
\label{eq:gener}
{\bm E}+\frac{\langle{\bm v}_s\rangle}{c}\times {\bm B}=\frac{{\nabla} \cdot {\bm P}_s}{q_s n_s}+\frac{m_s}{q_s}\left[
\frac{\partial\langle {\bm u}_s \rangle}{\partial t}+ (\langle{\bm v}_s\rangle\cdot \nabla) \langle {\bm u}_s \rangle\right]
~~~[\text{relativistic}]
\end{equation}
where $n_s=\int\!f_s d^3 u$ is the particle number density, $\langle{\bm u}_s\rangle=\int\! {\bm u} f_s d^3 u/n_s$ the mean particle 4-velocity, and ${\bm P}_s=m_s[\int\! {\bm u} ({\bm u}/\gamma)f_sd^3 u-n_s \langle {\bm v}_s \rangle\langle {\bm u}_s\rangle ] $ the pressure tensor, representing thermal momentum transport (\citealt{hesse_07}; see \citealt{most_22b} for a fully  covariant general-relativistic formulation). Following \citet{hesse_07} we note that, while \eq{gener} is relativistically covariant as a whole, the decomposition of momentum transport between bulk ($n_s m_s\langle{\bm u}_s\rangle\langle{\bm v}_s\rangle$) and thermal (${\bm P}_s$) contributions in \eq{gener} depends on the frame in which it is performed. While this may seem undesirable, it is the only choice that allows to smoothly transition \eq{gener} to the corresponding non-relativistic expression. In electron-ion plasmas, the generalized Ohm's law is derived from the electron momentum equation. In the collisionless limit and for non-relativistic conditions, it reads
\begin{equation}
\label{eq:gener2}
{\bm E}+\frac{{\bm v}_f}{c}\times {\bm B}=
\frac{{\bm J} \times {\bm B}}{e n_e}
-\frac{{\nabla} \cdot {\bm P}_e}{en_e}
-\frac{m_e}{e} 
\left[
\frac{\partial\langle {\bm u}_e \rangle}{\partial t}\!+ \!(\langle{\bm v}_e\rangle\!\cdot\!\nabla) \langle {\bm u}_e \rangle\right]
~~[\text{non-rel., electron-ion}]
\end{equation}
where we have assumed charge quasi-neutrality, $n_e\simeq n_i$, and used $\langle {\bm v}_e\rangle\simeq{\bm v}_f-{\bm J}/en_e$. 

The Hall term---the first term on the right-hand side of \eq{gener2}---plays a very important role for fast reconnection in electron-ion non-relativistic plasmas \citep{birn_01,liu_22}. In pair plasmas, the Hall effect is absent, due to the mass symmetry between electrons and positrons. There, fast reconnection is mediated by the off-diagonal terms of the pressure tensor \citep{bessho_05,hesse_07,comisso_asenjo_14,
melzani_14a,goodbred_22}, which are also important for electron-ion plasmas in the small, electron-scale diffusion region right at the X-point \citep{lyons_90,horiuchi_94,cai_97,kuznetsova_98,
egedal_19}. 
By identifying the dominant contributors to the breaking of flux freezing, it may be possible to write the corresponding terms as $\eta_{\rm eff} {\bm J}$---here, $\eta_{\rm eff}$ is an effective
resistivity---which could then be incorporated in resistive, single-fluid MHD approaches as a kinetically motivated subgrid prescription. While not the main focus of this review, we emphasize that this is an area of extremely active research, with encouraging results especially for electron-positron plasmas \citep{bessho_07,bessho_12,selvi_23,moran_25}.

\subsection{Geometry of the Reconnection Layer}

\begin{figure}[t]
\includegraphics[width=0.8\textwidth]{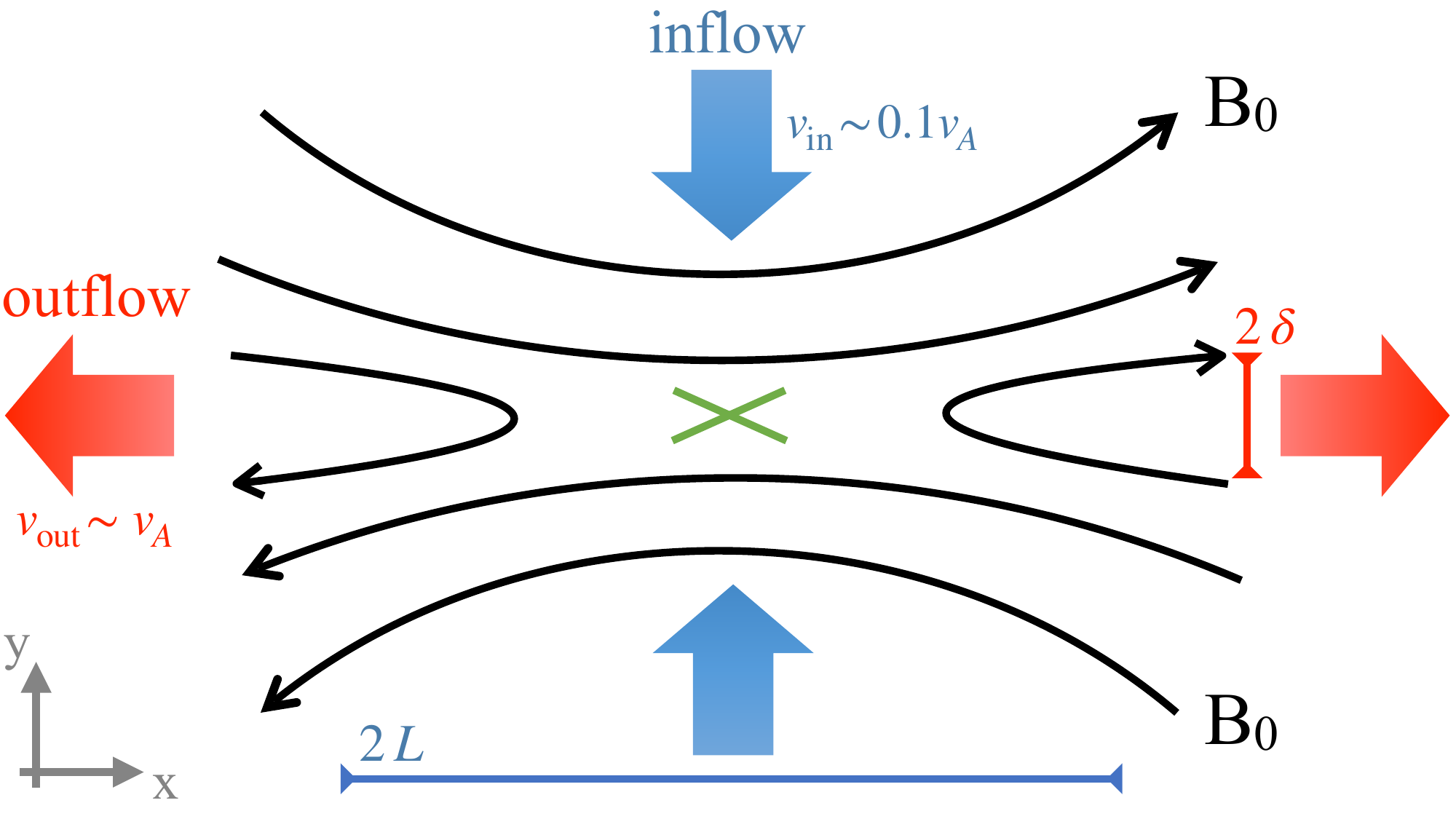}
\caption{Illustration of the reconnection layer geometry, for a reconnecting field of strength $B_0$. The half-length along the field is $L$, while the half-thickness of the field reversal is $\delta\ll L$. The upstream flows into the layer along the $\pm \hat{y}$ direction with speed $v_{\rm in}\simeq 0.1 v_{A}$, while the downstream flows out along $\pm \hat{x}$ with speed $v_{\rm out}\sim v_{A}$. Here, $v_A$ is the Alfv\'en speed, which in the relativistic regime approaches the speed of light. The change in magnetic topology occurs at the central X-point (green cross). The electric current is directed along $-\hat{z}$.}
\label{fig:sketch}
\end{figure}

The basic geometry of the reconnection layer is illustrated in \figg{sketch}, showing magnetic field lines that reverse over a width $\delta\ll L$, where $L$ is the typical size of the reconnection region.
The anti-parallel field lines in the ``upstream'' (pre-reconnection) region are carried toward the current sheet by the inflowing plasma, which moves along $\pm \hat{y}$ with speed~$v_{\rm in}$. By virtue of the flux-freezing theorem, the upstream plasma carries a motional electric field directed along~$z$, of strength $E_{\rm up}=(v_{\rm in}/c) B_0$. At the main X-point (green cross), the flux-freezing constraint breaks down due to non-ideal microscopic effects, allowing the field lines to tear and reconnect. 
Propelled by plasma pressure gradients and magnetic tension, the ``downstream'' (post-reconnection) plasma flows out along $\pm \hat{x}$ with speed~$v_{\rm out}$. The outflow velocity is of the order of the Alfv\'en speed $v_A$ associated with the reconnecting field $B_0$ (defined in \sect{params}). The Alfv\'en speed $v_A$ sets the basic scale for all velocities in reconnection, including the inflow velocity $v_{\rm in}$, which in collisionless plasmas equals $v_{\rm in}\simeq 0.1\, v_{A}$ \citep[e.g.,][]{cassak_review,comisso_16}.

\begin{marginnote}
\entry{Alfv\'en speed}{The group speed of an Alfv\'en wave, a type of plasma wave in which ions oscillate in response to a restoring force provided by magnetic field tension.}
\end{marginnote}

Traditionally, starting with the application of reconnection to solar flares, the reconnection rate has been viewed as the most important characteristic of magnetic reconnection. 
The reconnection rate is commonly defined as the rate of transfer of magnetic flux from the upstream to the downstream through a magnetic X-point [or a neutral X-line in three dimensions~(3D)]. Since most reconnection studies have focused on two-dimensional (2D) systems, or on 3D systems whose initial configuration possesses translational invariance in the out-of-plane ($z$) direction, 
one commonly characterizes the reconnection rate as the rate of reduction of the upstream magnetic flux per unit length in the $z$-direction. From  Faraday's law, this equals $\dot{\Psi}=c E_{\rm rec}$, where $E_{\rm rec}$  is the  absolute value of the $z$-component of the electric field at the main X-point.
The reconnection rate can also be intuitively expressed as $\dot{\Psi} = v_{\rm in} B_0=c E_{\rm up}$.
In a (statistical) steady state, the magnetic flux does not accumulate anywhere, so the two estimates must be the same, implying that an electric field $E_z=E_{\rm up} = E_{\rm rec}=(v_{\rm in}/c) B_0$ fills the whole volume.\footnote{ Mathematically, this follows from Faraday's law, which in steady state reads $c\, \nabla \times \boldsymbol{E} = -\partial{\boldsymbol{B}}/\partial t =0$; in 2D (so, $\partial_z = 0$), this implies $\partial_x E_z = 0 = \partial_y E_z$, i.e., $E_z$ is uniform.}
Given that $\dot{\Psi}=v_{\rm in} B_0$, it is customary to identify $v_{\rm in}$ as {a convenient measure of} the reconnection rate (note that rather than a rate, it is a velocity). We will adopt this widespread choice in the rest of our review.

\begin{marginnote}
    \entry{Reconnection rate}{The speed at which plasma and magnetic flux are carried into the reconnection layer (note that it is a velocity, not a rate).}
\end{marginnote}

One of the main reasons why the reconnection rate has received so much attention in the literature \citep{cassak_review,comisso_16,liu_22,goodbred_22} is that it also characterizes the reconnection power---i.e., the rate at which magnetic energy is brought into the layer, processed there, and ultimately converted to plasma energy. Indeed, the magnetic {enthalpy inflow} per unit area  of the layer (in the $xz$ plane) from both sides is {given by the Poynting flux} $2 c E_{\rm up} B_0/4\pi = 4 v_{\rm in} (B_0^2/8\pi)$.
The reconnection rate also determines the timescale $\tau_{\rm rec} = L/v_{\rm in}$ of a reconnection event---and hence the duration of a reconnection-powered flare. 
The first (Sweet-Parker, {non-relativistic resistive~MHD}) theory of magnetic reconnection to explain solar flares immediately led to great discrepancy between the predicted (months) and observed (minutes) duration of solar flares \citep{parker_57,sweet_58}, and the effort to resolve this reconnection-rate discrepancy---i.e., the search for a mechanism of fast reconnection, see \sect{ohm}---has been the main driver of reconnection research over several decades. This question is also extremely important in high-energy astrophysics, where reconnection is often invoked to explain rapid bright flares, which require fast energy dissipation and efficient particle acceleration. As regard to particle energization, the reconnection rate sets {the basic scale for} the rate of energy gain, $\dot{\varepsilon}\simeq e E_{\rm rec} c\simeq e (v_{\rm in}/c)B_0 c$, 
for a particle with energy $\varepsilon$ and charge~$e$.
Thus, the reconnection rate controls the highest energies that reconnection-accelerated particles can reach and their ability to overcome various (e.g., radiative) losses.

\subsection{Basic Reconnection Parameters}
\label{sec:params}
Reconnection dynamics depends chiefly on the upstream Alfv\'en speed, whose magnitude is governed by two dimensionless parameters: the ``hot magnetization'' $\sigmah$ and the ratio $B_g/B_0$ between the ``guide'' magnetic field $B_g$ (non-reconnecting, oriented along the $z$-direction of the electric current) and the alternating (reconnecting) field $B_0$. The hot magnetization $\sigmah$ quantifies the ratio between the magnetic enthalpy density of the alternating field and the  enthalpy density of the upstream medium:
\be
\sigmah = \frac{B_0^2}{4\pi h}
\ee
where $h=\rho c^2+ [\gamma_{\rm ad}/(\gamma_{\rm ad}-1)] p$ 
 is the plasma enthalpy density, with $\rho$ the rest mass density, $p$ the pressure, and $\gamma_{\rm ad}$ the adiabatic index. In the limit $p\gg \rho c^2$ of relativistically hot upstream plasma, the hot magnetization is simply related to the upstream plasma beta ($\beta_p \equiv 8 \pi p/B_0^2$), namely $\sigmah=1/(2\beta_p)$. In the opposite case of non-relativistic upstream temperatures ($p\ll \rho c^2$), $\sigmah$ reduces to the ``cold magnetization'' 
\be
\sigmac=\frac{B_0^2}{4\pi \rho c^2}~ .
\ee
It is also convenient to define species-specific magnetizations, where the hot magnetization $\sigma_{{\rm h},s}$ is normalized to the enthalpy density of  species~$s$, and the cold magnetization $\sigma_{{\rm c},s}$ to its rest mass energy density~$\rho_s c^2$ (here, $s=i$ for ions and $s=e$ for electrons). In the case of efficient conversion of magnetic energy to plasma {internal} energy, the typical particle Lorentz factor of species $s$ reaches $\sim \sigma_{{\rm c},s}/2$. It follows that particles can attain ultra-relativistic energies as long as the cold magnetization $\sigma_{{\rm c},s}\gg1$. We also note that the cold magnetization is the ratio of two important frequencies: $\sigma_{{\rm c},s}=\omega_{{\rm c},s}^2/\omega_{{\rm p},s}^2$, where $\omega_{{\rm c},s}=q_sB_0/m_sc$ is the {non-relativistic} gyro-frequency and $\omega_{{\rm p},s}=\sqrt{4\pi n_s q_s^2/m_s}$ is the plasma oscillation frequency without relativistic corrections. From the latter, one defines the plasma skin depth (or, inertial length) of species $s$ as~$c/\omega_{{\rm p},s}$. 
\begin{marginnote}
    \entry{Hot and cold magnetization}{Two alternative ways of quantifying the strength of the reconnecting field, by normalizing its enthalpy density to the plasma relativistic enthalpy density (hot magnetization) or its rest mass energy density (cold magnetization).}
    \entry{Guide field}{Uniform (so, non-reversing) component of the magnetic field, oriented along the direction of the electric current.}
\end{marginnote}

The relevant Alfv\'en speed for reconnection is the component along the outflow direction,
\be
\label{eq:alfven1}
v_A = c\, \frac{B_0}{\sqrt{4\pi h+B_0^2+B_g^2}} \, , 
\ee
where the numerator represents the tension force of reconnected field lines (which propel the flow away from the X-point, as a slingshot), while the denominator accounts for the overall inertia of the downstream flow \citep{melzani_14a, werner_17,werner_21}. Equivalently, it can be written as a function of $\sigmah$ and $B_g/B_0$ as
\be
\label{eq:alfven}
v_A = c\, \sqrt{\frac{\sigmah}{1+\sigmah(1+B_g^2/B_0^2)}}
\ee
where the term $\propto B_g^2$ in the denominator accounts for the inertia associated with the guide field. Note that the guide field affects the Alfv\'en speed for $B_g/B_0\gg \max[1,1/\sqrt{\sigma_{\rm h}}]$, with stronger guide fields yielding slower Alfv\'en speeds; {in contrast}, in the limit $B_g/B_0\ll1$ of weak guide fields, \eq{alfven} takes the commonly-used relativistic form $v_A=c\sqrt{\sigmah/(1+\sigmah)}$.
In the non-relativistic case $p\ll \rho c^2$ and $\sigmah\simeq \sigmac\ll1$, \eq{alfven} reduces to the familiar non-relativistic expression $v_A = B_0/\sqrt{4\pi \rho}$, which is independent of the guide field strength. For  $\sigmah\gg1$ and $B_g\ll B_0$ the Alfv\'en speed in \eq{alfven} approaches the speed of light. Then, the reconnection inflow velocity reaches $v_{\rm in}\simeq 0.1\,v_A\sim  0.1 c$, i.e., plasma and magnetic flux flow into the layer at $\sim 10\%$ of the speed of light, and the reconnection electric field approaches $E_{\rm rec} \sim 0.1\, B_0$.

In summary, the ``relativistic'' nature of reconnection \citep{blackman_field_94,lyutikov_uzdensky_03,lyubarsky_05} can manifest in two ways. First, the mean energy per particle can be much larger than the rest-mass energy (so, if the magnetic energy is efficiently converted to plasma energy, then the particles become ultra-relativistic); second, bulk motions at the Alfv\'en speed can be ultra-relativistic. 
This regime---where at least one of these two ``relativistic'' aspects is realized---is the focus of our review.

\section{DISSIPATION BY RR IN ASTROPHYSICAL HIGH-ENERGY SOURCES}
\label{sec:mhd} 
A large body of global fluid-type (MHD or force-free) simulations have revealed the development of intense current layers in compact-object magnetospheres, which become potential reconnection sites. In general terms, current-sheet formation is driven by the opening up of magnetic field-line/flux surfaces (i.e., when the magnetospheric topology transitions from closed/dipolar to open/monopolar) driven by the toroidal-field pressure that builds up due to the accumulation of toroidal magnetic flux by differential rotation (e.g., \citealt{babcock_61}).

Broadly speaking, we can classify reconnection layers according to the typical length scale of the reversing fields when the layer first develops: (A) {\it macroscopic} layers, whose extent is comparable to the global size of the system; and (B) {\it mesoscopic} layers, on scales smaller than the source size, but still much larger than plasma scales (e.g., the Larmor radius). Beyond a characterization of size (macroscale vs mesoscale), this classification also proves useful for visualizing the role and location of reconnection layers in global astrophysical systems. In fact, the presence of case-(A)
layers is determined by the global electromagnetic structure of the source, so their location is often ``predictable,'' and they persist as long as the magnetospheric geometry does not appreciably evolve. Case-(B) layers, on the other hand, may arise in different locations at different times, so their occurrence is more erratic and variable, in both  space and time.  As a result, case-(A) reconnection often leads to a drastic reconfiguration of the overall magnetospheric structure, releasing a significant fraction of the total magnetic energy. In contrast, in case (B), there is usually a broad (possibly, power-law) distribution of magnetic structures and sizes and, hence, of magnetic energies released by reconnection events. Finally, the distinction between case-(A) and (B) layers, while based on their size and location, correlates well with the strength of the guide field in the upstream region: case-(A) examples typically possess weak guide fields ($B_g\ll B_0$), while case-(B) layers have guide fields comparable to (or larger than) the reversing component ($B_g\gtrsim B_0$).  

We note that the distinction between case (A) and (B) might be employed not only in astrophysics, but also in, e.g., space physics. There, reconnection in the Earth's magnetotail would be classified as case (A), while current sheets in the solar corona as case~(B).

\begin{figure}[t]
\begin{minipage}{0.5\textwidth}
\includegraphics[width=\textwidth]{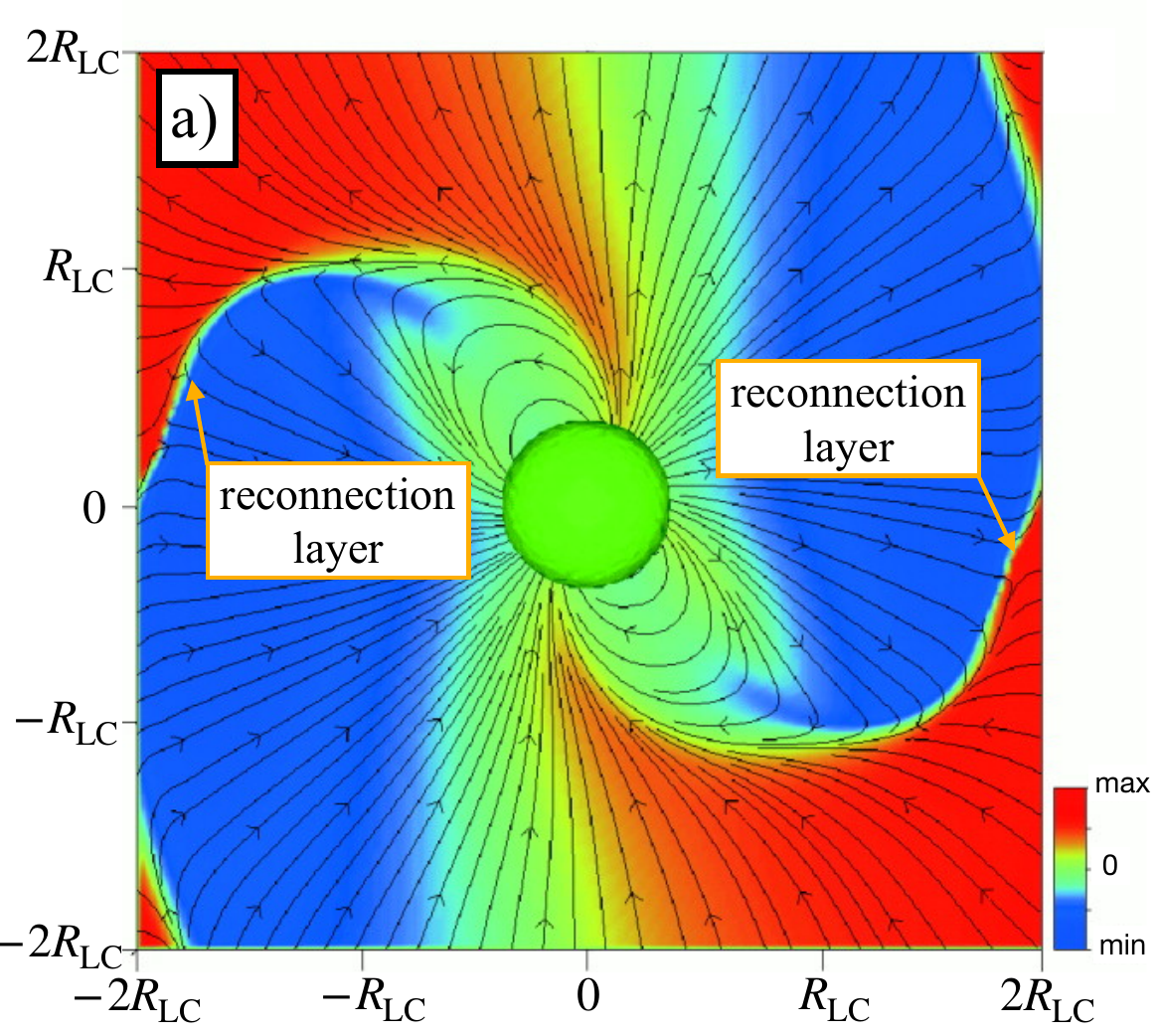}
\end{minipage}
\begin{minipage}{0.5\textwidth}
\includegraphics[width=\textwidth]{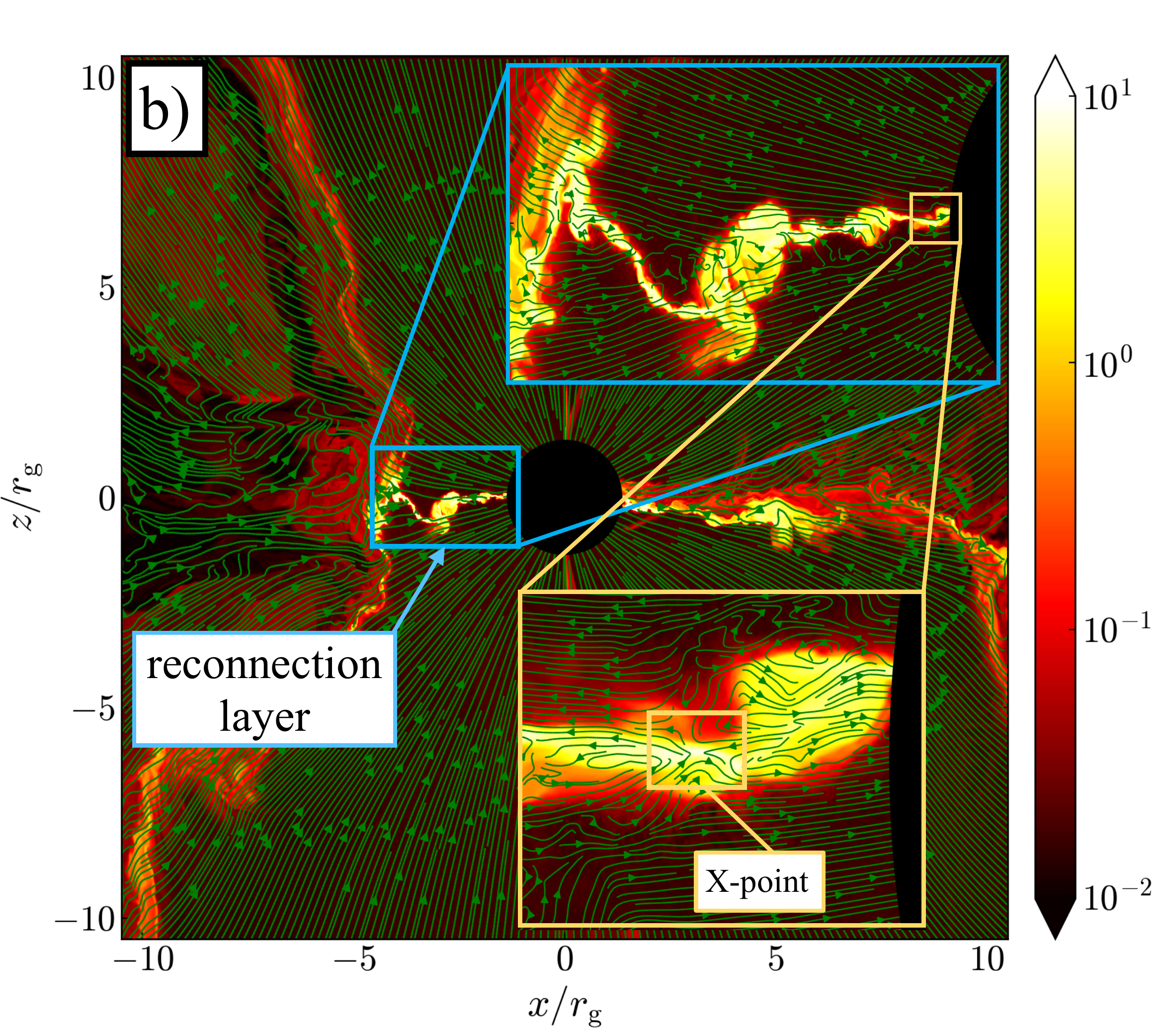}
\end{minipage}
\caption{Representative examples of case-(A) reconnection layers. (a) Magnetospheric structure of an oblique pulsar, from a 3D force-free simulation \citep{spitkovsky_06}; in-plane field lines are overlaid, while color represents the strength of the out-of-plane field, which reverses direction across the current sheet. (b) Black hole magnetosphere during a MAD eruption, from a 3D general relativistic MHD (GRMHD) simulation \citep{ripperda_22}; color represents the plasma temperature, revealing a hot current sheet at the equator, that separates the accretion disk from the event horizon. In both panels, we mark with arrows the location of current sheets / reconnection layers.}
\label{fig:anatoly}
\end{figure}

\subsection{Case-(A) Reconnection Layers}
The most notable example of case-(A) layers is the {current sheet extending beyond the light cylinder of rotation-powered pulsars}. For an aligned rotator with axisymmetric fields, the force-free solution of the pulsar magnetosphere requires an equatorial current sheet at cylindrical radii $R>R_{\rm LC}$, as demonstrated analytically \citep{contopoulos_99} and with fluid-type simulations \citep{spitkovsky_06,komissarov_06,mckinney_06}. {This current sheet} plays an essential role for the pulsar's electrodynamic life: 
the field-aligned currents flowing from the polar caps to infinity along the open magnetic field lines return back toward the light cylinder along the equatorial current sheet. In oblique pulsars (where the magnetic and rotational axes are misaligned), the current sheet wobbles with the pulsar period forming a ``ballerina skirt''-shaped layer [\figg{anatoly}(a)] analogous to the heliospheric current sheet \citep{spitkovsky_06,mckinney_06,tchekhovskoy_spitkovsky_13}. 
The observed pulsed GeV and TeV emission \citep{bai_10,Chen_14,philippov_14,uzdensky_spitkovsky_14, cerutti_15,philippov_15,cerutti_philippov_16,philippov_18,Kalapotharakos_18,hu_beloborodov_22,hakobyan_spitkovsky_23} and the simultaneous radio emission \citep{uzdensky_spitkovsky_14, lyubarsky_19,philippov_19,bransgrove_23} are attributed to the magnetic dissipation in the current sheet, just beyond the light cylinder. In oblique pulsars, the ``ballerina skirt''-wobbling results in a folded sheet that is carried outwards by the pulsar wind; in the equatorial plane, the wind appears as a sequence of magnetically-dominated stripes of alternating toroidal field, separated by current sheets---a striped wind. Magnetic dissipation of the stripes starts near the light cylinder \citep{lyubarsky_kirk_01,kirk_03,petri_05}, nearly exhausting the magnetic energy by $\sim 10^2-10^4\,R_{\rm LC}$ \citep{cerutti_philippov_17,Cerutti_20}, well before the pulsar wind termination shock (at $\sim 10^9\,R_{\rm LC}$, for the Crab pulsar).

\begin{marginnote}
\entry{Light cylinder}{Boundary of the inner pulsar magnetosphere, at the cylindrical radius $R_{\rm LC}=cP/2\pi$ (where $P$ is the pulsar spin period) beyond which plasma frozen to magnetic field lines can no longer co-rotate with the star.}
\end{marginnote}
 
\begin{figure}[t]
\begin{minipage}{0.5\textwidth}
\includegraphics[width=\textwidth]{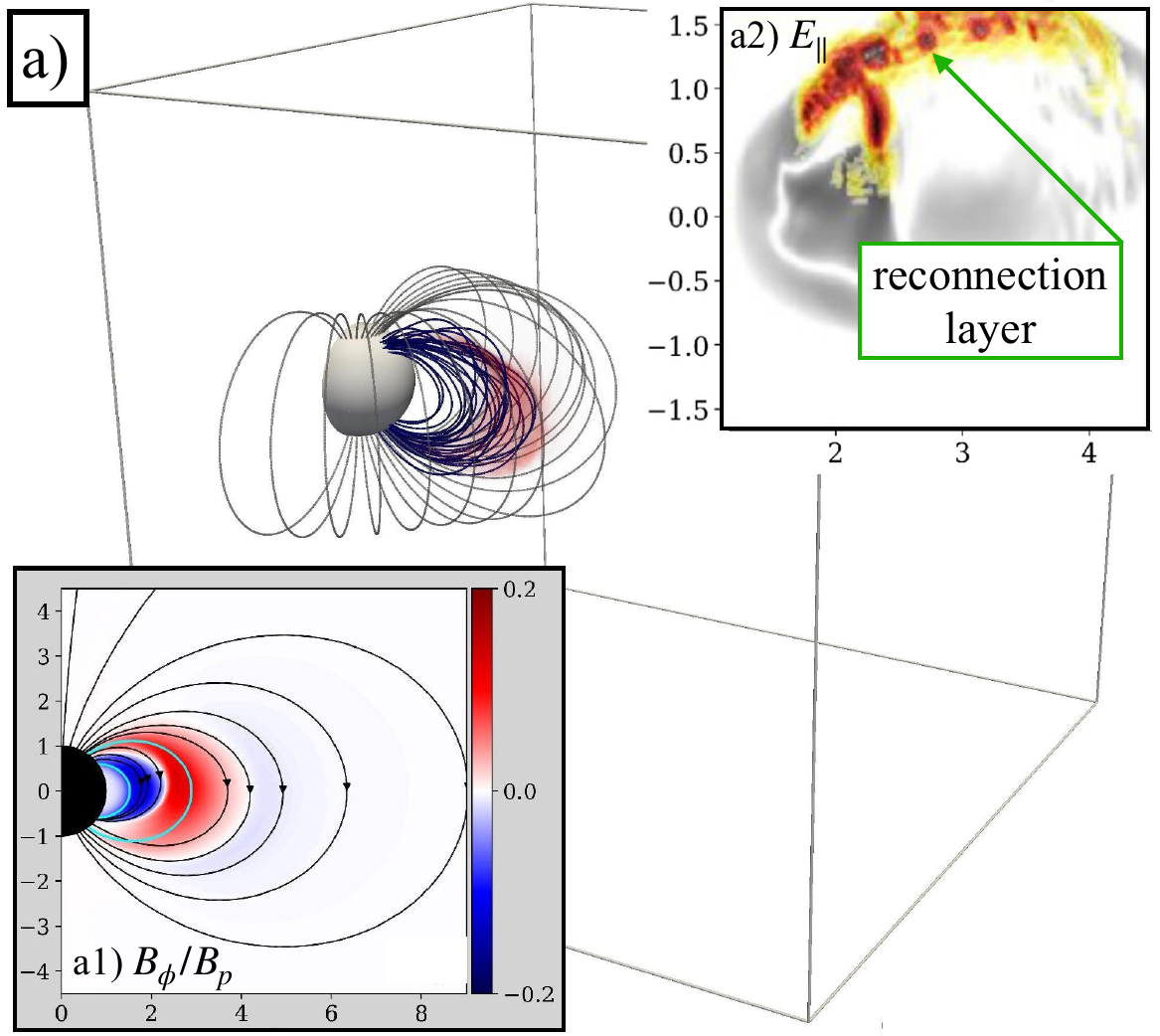}
\end{minipage}
\begin{minipage}{0.5\textwidth}
\includegraphics[width=\textwidth]{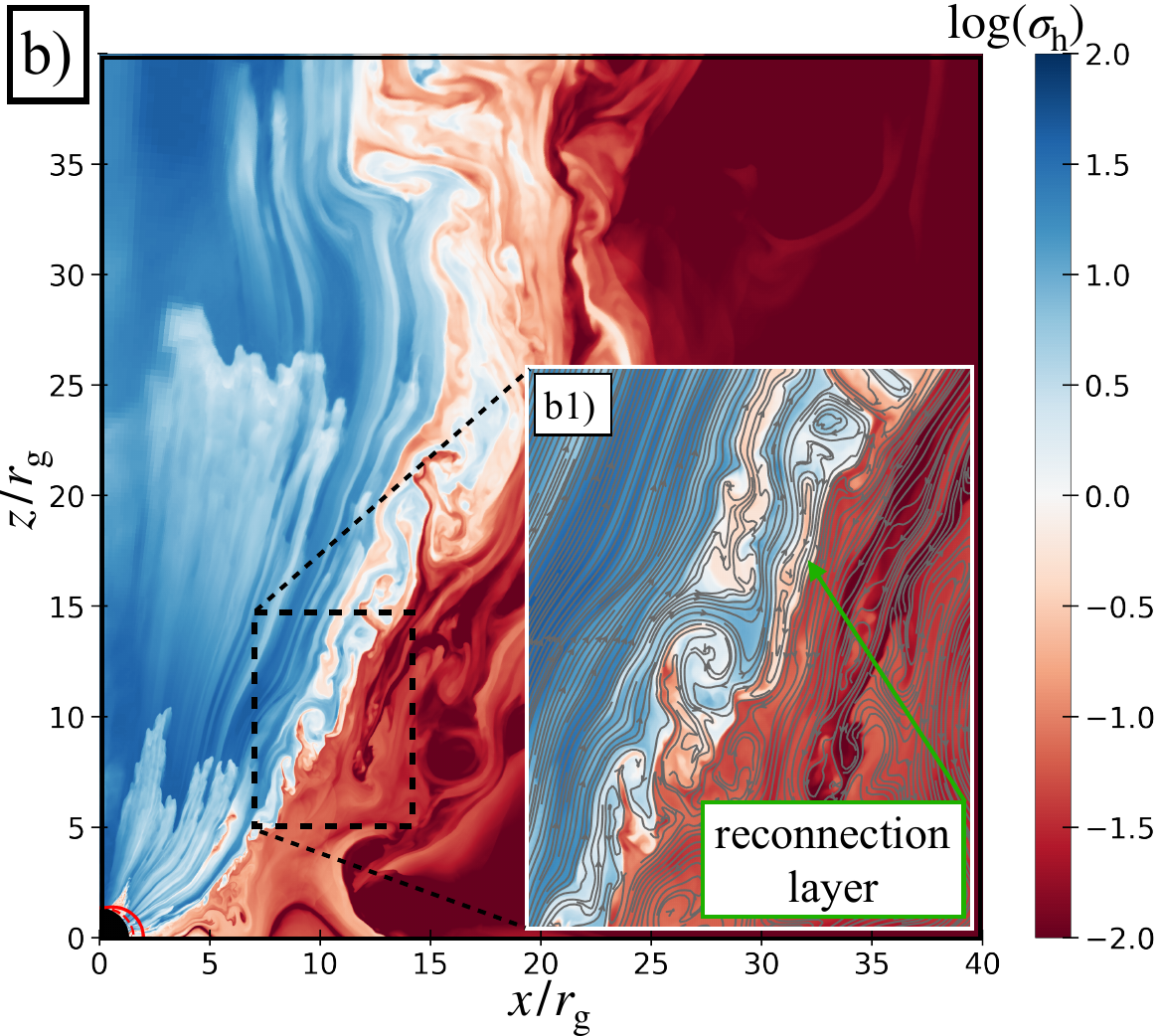}
\end{minipage}
\caption{Representative examples of case-(B) reconnection layers. (a) Development of the kink instability in the twisted flux bundle of a magnetar magnetosphere, from a 3D force-free simulation \citep{mahlmann_23}; inset (a1) shows the meridional-plane strength of  the toroidal $B_{\phi}$ (in units of the field $B_p$ at the pole), with poloidal field lines overlaid; inset (a2) shows in color the strength of the parallel electric field $E_{\parallel}={\bm E}\cdot {\bm B}/B$ in the equatorial plane. All axes are in units of the magnetar radius. (b)  Development of KH-like vortices at the boundary of a relativistic black hole jet, from a 2D GRMHD simulation \citep{sridhar_24}. The nonlinear evolution of KH rolls leads to current sheets on scales comparable to the shear layer width. Color indicates the hot magnetization $\sigmah$, with field lines overlaid in inset (b1).
In both panels, we mark with arrows the location of current sheets / reconnection layers.}
\label{fig:jens}
\end{figure}

A close analogue of the pulsar current sheet---so, of case (A)---is the {equatorial reconnection layer in black hole magnetically-arrested disks (MAD)}. 
MAD flows exhibit long phases of accretion where magnetic flux accumulates on the horizon and shorter eruptive phases where magnetic flux is ejected \citep{igumenschchev_03,narayan_03,igumenschchev_08,tchekhovskoy_11}. Spinning black holes in the MAD state show the formation of strong jets, due to the large accumulation of flux on the black hole horizon \citep{blandford_77}. As the flux piles up, the jet base widens, pushing the accreting gas away and forming a current sheet at the equator \citep{parfrey_19,crinquand_20,ripperda_20,ripperda_22}, as shown in \figg{anatoly}(b).  
Reconnection in this current sheet results in large quasi-periodic flux eruptions where field-line bundles are ejected from the black hole into the  disk \citep{lyutikov_mckinney_11,bransgrove_21}. 
The large reservoir of magnetic energy dissipated by reconnection may power the high-energy flares observed from Sgr A* and M87 \citep[e.g.,][]{hakobyan_ripperda_23,chen_23}.

\begin{marginnote}
    \entry{Gravitational radius}{The gravitational radius for a source of mass $M$ is defined as $r_{\rm g}=GM/c^2$. The event horizon of a Schwarzschild (i.e., zero-spin) black hole is $2 r_{\rm g}$.}
\end{marginnote}

Another case-(A) example is the current sheet forming in the magnetosphere of accreting neutron stars, e.g., in X-ray pulsars. Magnetic interaction between the star and the accretion disk may lead---because of the differential rotation between the two field-line footpoints, one on the star and the other on the disk---to field-line opening, resulting in the formation of a conical current sheet, which subsequently reconnects \citep{lovelace_95,uzdensky_etal_02a,uzdensky_etal_02b,uzdensky_02a,uzdensky_02b, uzdensky_04b, parfrey_17, parfrey_23}. 
A similar situation may occur in accreting black holes: magnetic field lines with one footpoint on the black hole event horizon and the other in the accretion disk get twisted by differential rotation, building up toroidal  field whose outward pressure inflates and opens up the magnetosphere. This creates a conical current sheet a few gravitational radii above the disk, setting the stage for subsequent global-scale case-(A) reconnection \citep{uzdensky_04a,uzdensky_05,parfrey_15,elmellah_22,elmellah_23}. Large magnetic flux bundles formed in this conical layer have been invoked to explain the orbiting hot spots observed by the GRAVITY Collaboration from Sgr~A* \citep{gravity_18}.

Case (A) also includes the current sheets generated by a global, dramatic opening-up of twisted neutron star magnetospheres, in both single and  binary systems. Thus, in isolated magnetars, crustal surface motions shift the frozen-in footpoints of magnetospheric field lines, generating twisted current-carrying flux bundles. If the footpoint displacement is sufficiently fast/large, this results in a global eruption that opens up the magnetosphere, creating extended current sheets and ultra-relativistic ejecta \citep{thompson_02,parfrey_12,parfrey_13,yuan_20,yuan_22,mahlmann_23}. Magnetic reconnection in these current sheets can generate high-frequency radio waves, potentially leading to Fast Radio Burst (FRB) emission \citep{lyubarsky_20,mahlmann_22}. In {interacting} binary neutron-star systems, global-scale reconnection layers result from the build-up of twist in the magnetic flux tube connecting the two merging stars. The flux tube inflates and eventually opens up, forming a reconnection layer that could produce intense ``precursor'' radio waves prior to gravitational wave events \citep{lyutikov_19b,sridhar_21b,most_20,most_22,most_23}. Potentially, a similar scenario might occur in black hole--neutron star mergers \citep{most_23b}.

We also classify as case (A) current sheets in striped jets, i.e., relativistic jets launched by black holes in which the magnetic field threading the ergosphere frequently changes polarity \citep{giannios_uzdensky_19,zhang_giannios_21}. When they first appear, such current sheets extend on global scales (i.e., the black hole horizon), hence their case-(A) classification. While 3D GRMHD simulations of black hole accretion flows have demonstrated the emergence of relativistic jets with polarity reversals \citep{christie_lalakos_19,mahlmann_20,chashkina_21,kaufman_23}, no global numerical 3D study has so far been able to follow these current sheets from the jet base to large distances. One-dimensional models indicate that, in striped jets, dissipation of magnetic energy through reconnection is expected to start close to the compact object, but it is initially slow because the jet expansion rate is much faster than the reconnection rate \citep{drenkhahn_02a,drenkhahn_02b}. Then, dissipation peaks at large distances from the engine (there, the jet expansion rate is significantly smaller), where it may be responsible for powering the GRB ``prompt'' phase and the blazar-zone emission \citep{giannios_06b,giannios_12,mckinney_uzdensky_12, begue_17,giannios_uzdensky_19, gill_20}.

\begin{marginnote}
    \entry{Ergosphere}{The surface inside which a physical observer cannot be stationary with respect to infinity. It extends from the event horizon at the poles up to the Schwarzschild radius at the equator. Any particle inside this region (or ergoregion) is forced into rotation around the black hole spin axis.}
\end{marginnote}

\subsection{Case-(B) Reconnection Layers}
Case-(B) layers can be generated by the nonlinear evolution of MHD instabilities (e.g., kink, Kelvin-Helmholtz, Rayleigh-Taylor) and MHD turbulence. 
Relativistic magnetized jets in AGN (e.g., blazars and M87 jet), GRBs, microquasars (e.g., GRS 1915+105), and Pulsar Wind Nebulae (PWNe, such as the Crab Nebula)---launched by rotating, magnetized compact objects---exhibit a tightly wound-up magnetic field that is prone to current-driven kink instabilities, analogous to those appearing in tokamak plasmas \citep{eichler_93,begelman_98,appl_00,giannios_06,uzdensky_macfadyen_07,moll_08,mckinney_09,mizuno_09,mizuno_11,mizuno_12,mignone_10,nalewajko_begelman_12,porth_15,bromberg_16,barniol_17,das_begelman_19}. 
At nonlinear stages, kink instabilities introduce field reversals on scales comparable to the
jet cross-sectional radius [much shorter than the global system size, i.e.,  the distance to the central engine; hence, case~(B)]. This leads to dissipation of magnetic energy via reconnection, and ensuing nonthermal emission \citep{alves_18,davelaar_20,ortuno-macias_22}.

Case-(B) reconnection layers induced by the kink instability may also occur in twisting flux tubes of magnetars and accretion disk coronae. In magnetars, the twist comes from crustal motions [see \figg{jens}(a)], whereas coronal flux tubes anchored to a Keplerian accretion disk will be sheared and twisted by the disk's differential rotation. The ensuing reconnection-mediated dissipation of field energy has been invoked, respectively, to explain X-ray emission from magnetars \citep{beloborodov_21,mahlmann_23} and the hard state of accreting X-ray binaries (XRBs) and AGN \citep{goodman_uzdensky_08,beloborodov_17,sironi_beloborodov_20,sridhar_21,sridhar_23}. These layers have typical lengths of order the cross-sectional radius of the twisted flux tube, which is substantially smaller than the system size; hence, we classify them as case-(B) layers.
 
 At relativistic jet boundaries, the velocity shear between the relativistic magnetically dominated jet and the ambient plasma can trigger Kelvin-Helmholtz (KH) instabilities (\figg{jens}(b)). In their nonlinear stages, magnetized KH vortices generate reconnection layers on scales comparable to the width of the shear layer \citep{sironi_21,davelaar_23,sridhar_24}. Electrons accelerated by KH-driven reconnection can explain the limb-brightening of the radio jet in~M87 \citep{walker_18,lu_asada_23}. Case-(B) layers also result from the nonlinear development of magnetic Rayleigh-Taylor (RT) modes. Flux bundles ejected into a MAD disk after a flux eruption appear as ``tubes'' of low density, magnetically dominated plasma \citep{porth_21}, which get  shredded by magnetic RT modes due to the inward-pointing gravity. RT-driven reconnection may power the IR and X-ray flares observed from Sgr~A* \citep{zhdankin_23}.

Case-(B) layers are also a natural by-product of MHD turbulence \citep[e.g.,][]{zhdankin_13}. Fluid motions at the driving scale lead to the formation of interfaces of velocity and magnetic shear. Intermittent current sheets, on scales comparable to the driving scale, can give a significant contribution to the overall dissipation \citep{zhdankin_13} and process a large fraction of particles in the turbulent volume \citep{comisso_18,comisso_19}. If turbulence is magnetically dominated, the ensuing reconnection operates in the relativistic regime. Reconnection within turbulence has been invoked to explain the hardest segments of PWNe spectra \citep{uzdensky_etal_11, cerutti_13, cerutti_14a, cerutti_14b, lyutikov_19,comisso_20}.  In accretion disk coronae, turbulence-driven reconnection is expected, in analogy to the solar corona, from interactions between neighboring magnetic flux loops with different/random orientations \citep{galeev_79,dimatteo_98,miller_00,uzdensky_goodman_08}. There, the interplay between turbulence and reconnection may explain the hard state of X-ray binaries and AGN \citep{groselj_24,nattila_24} and the generation of high-energy neutrinos \citep{fiorillo_24,mbarek_24} observed by {\it IceCube} from Seyfert galaxies \citep{icecube_22}.

\section{THE PLASMA PHYSICS OF RR}\label{sec:micro}
While useful to elucidate the global dynamics of astrophysical high-energy sources, fluid-type simulations suffer important shortcomings. First, 
they cannot properly model the  physics of particle acceleration and the development of nonthermal distributions. 
Second, the rate of reconnection-mediated field dissipation in single-fluid MHD is roughly one order of magnitude lower than in more sophisticated and realistic two-fluid or kinetic approaches \citep{birn_01, cassak_review,comisso_16}. Third, both ideal-MHD and force-free approaches fail to properly describe regions with $E>B$ and/or $E_\parallel=\boldsymbol{E}\cdot \boldsymbol{B}/B\neq 0$, which can be important for the early stages of particle acceleration, as we discuss below. 

The plasma in compact object magnetospheres and outflows is nearly collisionless, so fluid approaches are not, strictly speaking, applicable, and a kinetic treatment is required. The most
versatile and convenient computational tool for describing the  kinetic physics of plasmas is the PIC method. In this section, we summarize how PIC simulations have led to important advances in our understanding of the kinetic physics of RR. PIC codes can model collisionless plasmas in the most fundamental way \citep{birdsall_langdon_91}, as a collection of charged macro-particles (each one representing a large number of real particles) moving under the influence of the Lorentz force. The electric charges and currents deposited by the macro-particles on the computational grid are used to solve for the electromagnetic fields via Maxwell's equations. The loop is closed self-consistently by extrapolating the fields to the macro-particle locations, where the Lorentz force is computed. 

\begin{marginnote}
    \entry{PIC method}{Computational method for solving the Vlasov-Maxwell system. Individual macro-particles in a Lagrangian frame are tracked in continuous phase space, whereas moments of the distribution such as charge densities and electric currents are computed on Eulerian (stationary) mesh points.}
\end{marginnote}

The initial equilibrium in most simulations (both fluid and kinetic) of isolated reconnection layers is the Harris current sheet \citep{harris_62,kirk_03}, which is an exact 1D equilibrium of the Vlasov-Maxwell system. It is characterized by the field profile
\begin{equation}
{\bm B}=B_0 \tanh \frac{y}{\delta}\ \hat{{\bm x}} +B_g \hat{{\bm z}},
\label{eq:harris_field}
\end{equation} 
where $\delta$ is the half-thickness of the layer.
The particles within the sheet are initialized in a drifting Maxwell-J\"uttner distribution \citep{juttner_11} in which positive and negative charges move in opposite directions.
The density profile of current sheet particles is
\begin{equation}
\label{eq:density_profile}
n=n_{\rm cs} \ {\rm sech}^2\ \frac{y}{\delta},
\end{equation}
Pressure equilibrium requires that $B_0^2=8 \pi n_{\rm cs} k_B T_{\rm cs}$, where $T_{\rm cs}$ is the temperature of current sheet particles.
An additional uniform background population with density $n_0$ and no drift velocity is  added to the current sheet population. Thus, the total density in the middle of the sheet is $n_{\rm cs} +n_{\rm 0}$, whereas far from the current sheet is $n_{\rm 0}$.

\begin{marginnote}
        \entry{Harris equilibrium}{The most commonly employed equilibrium for a pressure-supported current sheet.}
\end{marginnote}

While the Harris sheet is the most common initial condition, reconnection can also be initialized using a force-free equilibrium \citep[e.g.,][]{bobrova_01}; dynamical scenarios such as X-point collapse have also been explored \citep{yuan_16,lyutikov_17a,lyutikov_17b,lyutikov_17c}. We also note that the standard Harris equilibrium assumes identical plasma conditions on the two sides of the sheet. In a number of cases this is not realistic, e.g., at the interface between a magnetically dominated low-density jet (likely dominated by electron-positron pairs) and a weakly magnetized, dense electron-proton wind. In such a configuration, reconnection proceeds in an ``asymmetric'' configuration. The study of relativistic asymmetric reconnection is still in its infancy \citep{mbarek_22,figueiredo_24}. \citet{mbarek_22} found that the reconnection dynamics and the physics of particle acceleration are mostly determined by the side having lower magnetization. Below, we report results obtained in the symmetric case.

Most choices for initializing a current sheet in PIC simulations are designed to rapidly result in active reconnection. In fact, most simulations start from a thin layer already on the verge of being
unstable, so it is questionable how such an unstable configuration
would be realized in the first place in a realistic system (see \citealt{uzdensky_loureiro_16, comisso_etal_16}, for a non-relativistic treatment of reconnection onset in a compressing sheet).
The goal of this section is to capture the general properties of a well-developed layer in quasi-steady state, rather than the processes that led to reconnection onset (which, most likely, are controlled by the global system dynamics).
We discuss in \sect{turb} a few cases where reconnection layers are not set up as initial conditions, but rather form as a result of MHD-scale dynamics.

\subsection{Fluid-level Description of Collisionless RR}
\label{sec:fluid}
While in PIC simulations one has access to the full phase-space information, i.e., a kinetic-level description, it is still useful to reduce it to the simpler level of two-fluid-MHD or even single-fluid-MHD models, which are easier to interpret. 
One can use this fluid-level picture to describe some of the most important quantitative characteristics of reconnection (e.g., reconnection rate) and make comparisons with fluid-type numerical and analytical models. Furthermore, a fluid-like description becomes useful in large systems where the reconnection layer becomes unstable to a slew of fluid-type instabilities and develops a rich hierarchy of plasma and field structures of different sizes, necessitating a statistical approach. This subsection is devoted to such a fluid-level description. 

Early analytical studies of resistive-MHD reconnection in the relativistic limit \citep{blackman_field_94,lyutikov_uzdensky_03, lyubarsky_05} explored the key dynamical aspects of RR in the laminar, single-X-point regime, thus developing relativistic generalizations of the classical Sweet-Parker \citep{parker_57,sweet_58} and Petschek \citep{petschek_64} models. In astronomical sources, the global (macroscopic) system size---i.e., the current-layer length~$L$ along the outflow direction---is much larger than microphysical (or microscopic) plasma scales, such as the collisionless skin depth or the characteristic Larmor radius of energized particles.  It is at these scales that the ideal-MHD flux-freezing constraint breaks down, enabling magnetic field lines to reconnect. These microscopic scales control the thickness $\delta\ll L$ of the non-ideal, dissipative sub-regions. 
Mass (or number) conservation would then 
imply a very low inflow velocity $v_{\rm in} \sim v_{\rm out} \delta/L\sim v_{A} \delta/L\ll v_A$ \citep{lyubarsky_05}, and hence a very low reconnection rate, inconsistent with observations.

Fortunately, however, this does not happen in reality because thin and long current sheets have been shown to be unstable to several rapidly-growing instabilities, which lead to the development of fine substructure.
This in turn enables fast reconnection, but also casts doubt on the physical realizability of very thin and long current sheets with huge aspect ratios.  This concern is indeed well founded. These rapidly-growing instabilities invalidate the classical picture of stationary, smooth, laminar, steadily reconnecting current sheets and instead result in a much more complex, much more dynamic picture.  

The problem of reconnection thus effectively becomes the problem of determining the dynamic behavior of thin current sheets governed by the nonlinear interplay of several instabilities, occurring at the same time and competing with (or enhancing) each other.
In full 3D there are at least four main instabilities (and in reality even more) at play: tearing (aka plasmoid), plasmoid-coalescence, drift-kink, and the flux-rope kink instability, all of them fed by magnetic energy. Of these instabilities two are primary (tearing and drift-kink), developing in the initial thin current sheet itself, and two are secondary or parasitic (coalescence and flux-rope kink), involving the structures created by the two primary instabilities. We now quickly describe these four instabilities. 

\begin{figure}[t]
\begin{minipage}{0.4\textwidth}
\includegraphics[width=\textwidth]{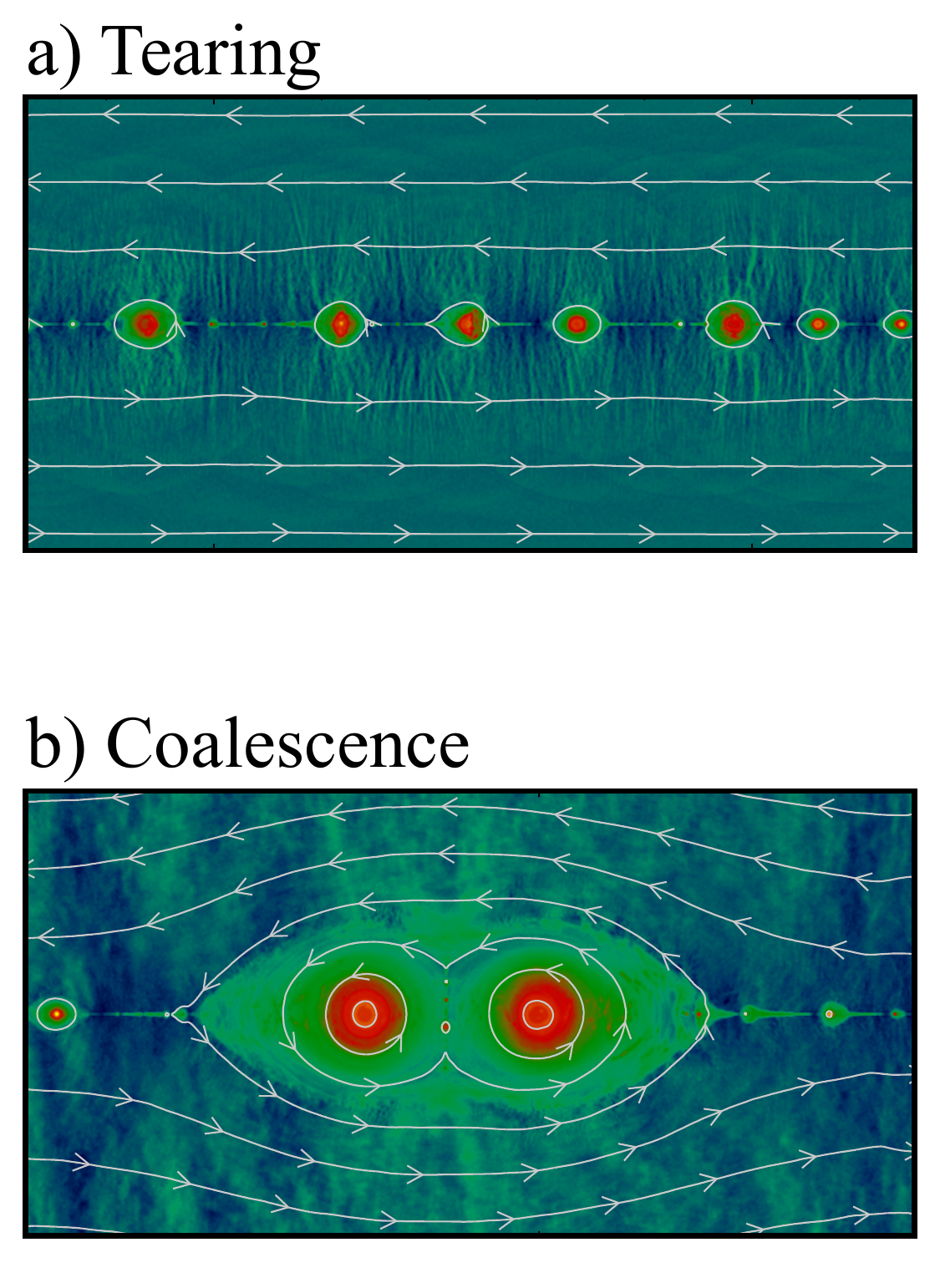}
\end{minipage}
\begin{minipage}{0.5\textwidth}
\includegraphics[width=\textwidth]{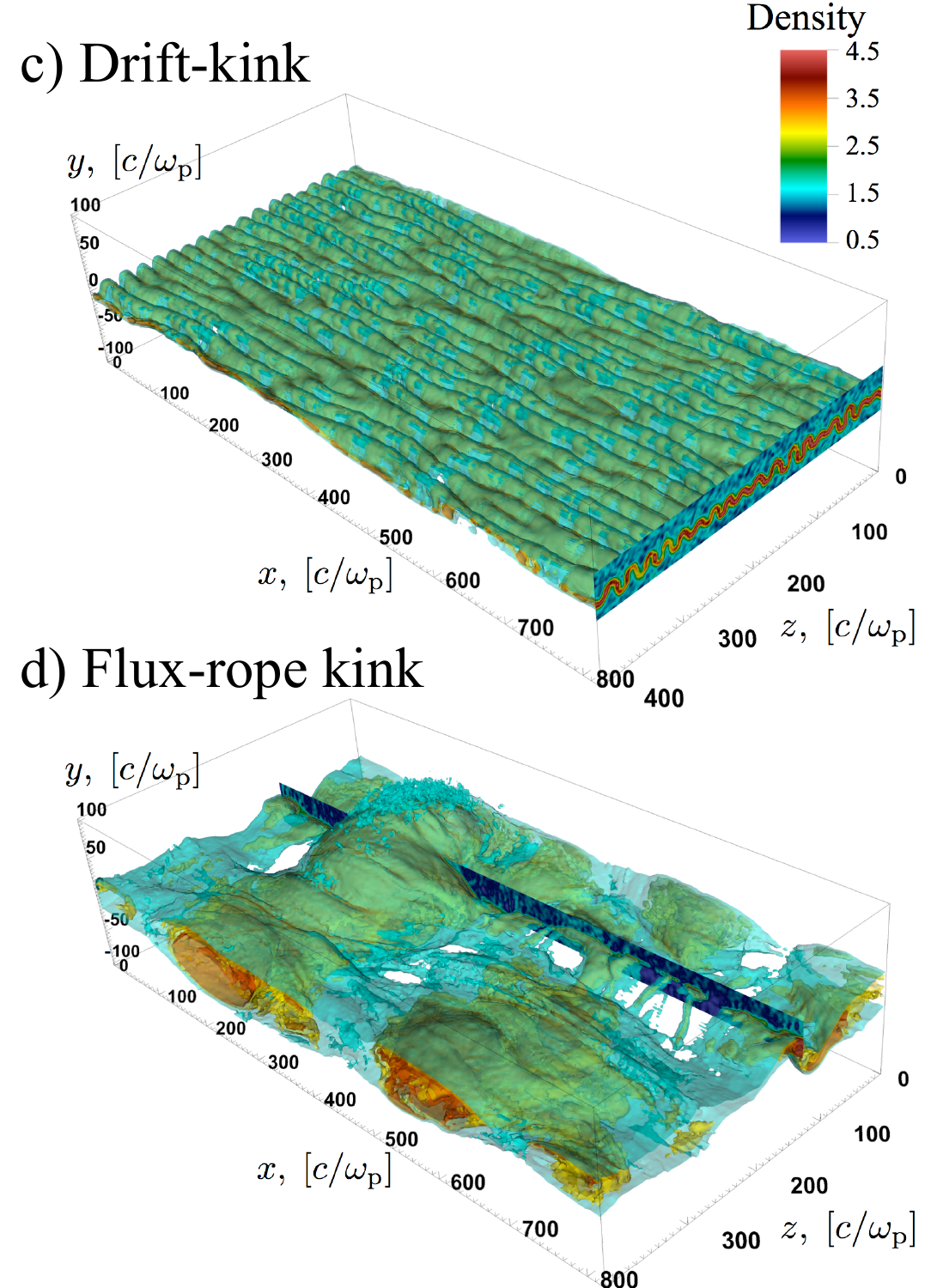}
\end{minipage}
\caption{The four main instabilities governing the evolution of 3D layers: (a) tearing, (b) coalescence, (c) drift-kink, and (d) flux-rope kink. The left column illustrates 2D instabilities (figure adapted from \citet{petropoulou_18}), the right column 3D instabilities (figure adapted from \citet{sironi_spitkovsky_14}); the top row presents primary instabilities, while the bottom row parasitic ones. Color denotes particle density; in the left column, field lines are overlaid.}
\label{fig:instab}
\end{figure}   

\subsubsection{2D Reconnection}
\label{sec:2drec}
The most prominent instability is the tearing (aka plasmoid) instability, which is synonymous with reconnection itself \citep{loureiro_07}. 
This instability is non-ideal, i.e., it is forbidden in ideal MHD and requires at least resistive MHD (or other non-ideal effects). It breaks the translational symmetry of a thin current sheet in the $x$ direction and leads to the formation of a necklace of magnetic islands, lowering the overall magnetic energy. A smooth and long current sheet is then replaced by a 1D chain of discrete parallel filaments of electric current (flowing in the out-of-plane $z$ direction) corresponding to these islands \citep{bhattacharjee_09, ULS-2010}.
Each island consists of closed nested magnetic flux surfaces centered around a magnetic O-point, and the islands are separated from each other by X-points, see \figg{instab}(a). Small, elementary current sheets extend around each X-point (i.e., in between each pair of islands); each is an active reconnection site, feeding reconnected magnetic flux and plasma into the two islands flanking it on each side, thus sourcing their growth. Since they get filled with plasma, the magnetic islands are often called ``plasmoids,'' and the two terms are often used interchangeably (strictly speaking, the term ``magnetic island'' refers to the magnetic structure, whereas ``plasmoid'' to the plasma content). 

\begin{marginnote}
    \entry{Tearing instability}{Non-ideal instability that breaks a thin current sheet into a chain of magnetic islands.}    
    \entry{Plasmoids}{Magnetic islands, filled with plasma, generated by the tearing mode.}
\end{marginnote}

Since the out-of-plane electric currents in the islands are parallel, they attract each other. This---in combination with the stochasticity of plasmoid motions---drives the so-called coalescence instability [\figg{instab}(b)], which pulls the islands left and right along the 1D chain in the $x$-direction, with speeds comparable to the Alfv\'en speed~$v_A$. In the case of RR, where $v_A \simeq c$, the fastest bulk motions can be ultra-relativistic.
Since the equilibrium on which this instability grows is not the original, smooth current sheet but the nonlinear plasmoid chain produced by the tearing mode, the coalescence instability can be considered secondary, or parasitic, with respect to the primary tearing mode. 

\begin{marginnote}
    \entry{Coalescence instability}{Attraction between the parallel currents of neighboring plasmoids, leading to their merger.}
\end{marginnote}

The interplay between these two instabilities leads to a very complex, chaotic dynamics in 2D reconnection. 
Two adjacent plasmoids moving towards each other may collide and eventually merge,
creating another current sheet at the merger interface. 
These plasmoid-merger events involve an ``anti"-reconnection current sheet perpendicular to the original one (so, in the $y$ direction) and with an electric field that is opposite to the main reconnection electric field. The time-varying electric currents sourced at the interface of merging plasmoids generate fast-mode waves that propagate back into the upstream region. Upon conversion to electromagnetic waves, they have been invoked to explain the pulsar radio ``nanoshots'' coming in
phase coincidence with higher-energy emission \citep{lyubarsky_19,philippov_19}, and proposed as a candidate mechanism for powering FRBs \citep{lyubarsky_20,mahlmann_22}.

When two plasmoids move away from each other, attracted by their other neighbors, the inter-plasmoid current sheet linking them stretches. At some point, it becomes unstable to the tearing instability, giving birth to one or more new plasmoids. An interesting chaotic process ensues, generating a highly dynamic hierarchical self-similar chain of plasmoids of different sizes. The plasmoid size ($w$) distribution resembles a broken power law $dN/dw\propto w^{-s}$ with index $s\simeq 1$ below the break and $s\simeq 2$ at the largest sizes \citep{sironi_16}. The size distribution spans from the smallest plasmoids---just a bit larger than the thickness of the smallest elementary current layers---up to large, macroscopic plasmoids with $w\sim 0.1\,L$.
The rare largest, so-called ``monster,'' plasmoids, born near the central X-point and therefore having a long time to grow, can reach up to a substantial fraction of the system size~$L$.

The overall, time-averaged reconnection rate in the plasmoid-dominated, large-system-size regime is governed by the local reconnection rate in the small-scale elementary layers at the bottom of the hierarchy. For collisionless 2D reconnection, this is $\dot{\Psi} = c E_{\rm rec} = v_{\rm in} B_0 \simeq 0.1 v_A B_0$.
This fundamental result seems to be quite universal, seen in both non-relativistic and relativistic two-fluid and fully-kinetic (PIC) simulations; it is also consistent with observations of solar flares and geomagnetic storms in the Earth magnetosphere, believed to be powered by collisionless reconnection. It is worth noting that this universal value of 0.1 for the normalized collisionless reconnection rate $\dot{\Psi}/v_A B_0$ is about one order of magnitude higher than its collisional counterpart, where $\dot{\Psi}/v_A B_0 \simeq 0.01$ in the plasmoid-dominated resistive-MHD regime \citep[e.g.,][]{birn_01,cassak_review,galishnikova_23}. 

In the relativistic case where $v_A \simeq c$, 
the reconnection inflow velocity attains a sizeable fraction (about 10\%) of the speed of light, $v_{\rm in} \simeq 0.1 c$. Consequently, the reconnection electric field reaches $\sim 10\%$
of the upstream magnetic field, $E_{\rm rec} \simeq 0.1 B_0$. A peculiarity of the relativistic regime ($\sigmah\gg1$) is that a strong guide field $B_g \gtrsim B_0$ slows down reconnection. This is because the guide field is not destroyed, but is largely (outside of the immediate vicinity of X-points) frozen into the plasma and hence moves with it. Then, the effective inertia of the guide field needs to be taken into account when evaluating the Alfv\'en velocity $v_A$, see \eq{alfven}. At fixed $\sigmah$, the Alfv\'en speed decreases for increasing $B_g/B_0$, which reduces the typical reconnection outflow and plasmoid velocities, the reconnection electric field, and the reconnection rate. In contrast, in the non-relativistic limit ($p\ll \rho c^2$ and $\sigmah\simeq \sigmac\ll1$), the Alfv\'en speed is $v_A/c=\sqrt{\sigmac}$ regardless of $B_g/B_0$.

\subsubsection{3D Reconnection}
\noindent
In three dimensions---i.e., allowing for variations in the out-of-plane $z$-direction---
the layer dynamics is even richer and more complex. Two new instabilities with non-zero $z$-wavenumber ($k_z\neq 0$) come into play:
the (relativistic) drift-kink and the flux-rope kink.
First, the initial thin current sheet is prone to the relativistic drift-kink instability (RDKI), see \figg{instab}(c). This is a non-ideal MHD instability. It involves kinking, or rippling, of the current sheet in the $z$-direction, with fastest-growing modes having wavelength in the $z$-direction comparable to the sheet thickness. This instability was first investigated in the context of two-fluid non-relativistic reconnection in traditional 
electron-ion plasmas, for space-physics applications \citep{pritchett_96}. However, it was shown that its growth rate becomes small for large ion-to-electron scale separations (hence, large mass ratios), and as a result it has not received much attention from the space plasma community. Since there is no inter-species scale separation in pair plasmas---which have ample relevance in relativistic astrophysical settings---the RDKI growth rate can be very large. In fact, the RDKI grows 
faster than the tearing mode for current sheets with weak guide fields \citep{zenitani_07}, and it may dominate the 3D dynamics at early times \citep{sironi_spitkovsky_14}.

\begin{marginnote}
    \entry{Drift-kink instability}{Non-ideal instability that leads to rippling of the current sheet in the direction of the electic current.}
\end{marginnote}

In addition to the primary RDKI, which rapidly destabilizes the initial thin current sheet, another powerful instability becomes active at later times and governs the development of structures with finite extent in the $z$-direction and the potential transition to 3D turbulence. 
This is the flux-rope kink instability---essentially, the ideal-MHD kink instability of the magnetic flux ropes formed by the primary tearing mode (see \figg{instab}(d)). These flux ropes are the 3D counterparts of 2D magnetic islands/plasmoids.
Their internal structure is that of a Z-pinch (in the absence of a guide field) or a screw-pinch (if a finite guide field is present). In 3D, once tearing-generated flux ropes become sufficiently large, they naturally get kink-unstable. Thus, the flux-rope kink instability, just as the coalescence instability described above, is a parasitic, secondary instability feeding on the outcome of the primary tearing mode. Its nonlinear development limits the coherence scale of magnetic and plasma structures in the $z$-direction. This results---at least in PIC simulations employing periodic boundary conditions along the outflow direction \citep{werner_21}---in a chaotic, turbulent 3D dynamics. The degree at which the nonlinear, long-term dynamics of a reconnecting layer may be faithfully described by the well-known statistical properties of homogeneous turbulence---i.e., in terms of Fourier spectra of density and magnetic fluctuations---is a frontier of modern research. While the reconnection layer is not homogeneous in the inflow direction, it should be fairly homogeneous along the $z$-direction of the electric current, and possibly along the outflow $x$-direction. We therefore argue that power spectra measured within the reconnection midplane as a function of $k_z$ (or $k_z$ and~$k_x$) may help assess the (potentially) turbulent nature of well-evolved reconnection layers.  

\begin{marginnote}
    \entry{Flux-rope kink instability}{Akin to the MHD kink instability, it causes transverse non-axisymmetric
displacements of a flux rope.}
    \entry{Z-pinch and screw-pinch}{The Z-pinch is a plasma column threaded by purely toroidal fields. The screw-pinch has both toroidal and poloidal fields.}
\end{marginnote}

Both the RDKI and the flux-rope kink instability are strongly affected by the guide field. Even a moderate guide field essentially inhibits the RDKI, while a strong $B_g \gtrsim B_0$ suppresses flux-rope kink modes with out-of-plane wavenumbers $k_z$ greater than the limit imposed by the Grad-Shafranov condition, $k_z w \gtrsim B_g/B_0$ \citep{bateman_78}, with $w$ the diameter of the given flux rope (this also corresponds to the  critical balance condition by \citealt{goldreich_sridhar_95}).

\subsection{Kinetic-level Description of Collisionless RR}
\label{sec:accel}
The rich dynamics of the reconnection layer, as described in \sect{fluid}, influences how the upstream magnetic energy gets partitioned within the downstream flow. Some of this energy remains in electromagnetic form, while the dissipated magnetic energy is divided among various species (protons, electrons, positrons, and potentially heavier ions), and within each species, it is further partitioned into bulk, thermal, and nonthermal components.

This is fundamentally a kinetic question, and the main goal of this subsection is to summarize recent insights on this topic, as revealed by PIC simulations. First, we discuss {(in \sect{heat})} how reconnection divides the available energy between electromagnetic fields and different plasma species, independent of the shape of the distribution function (i.e., focusing solely on the mean energy). Next, we present our current understanding of nonthermal particle acceleration in~RR and the resulting energy spectrum {(\sect{plaw})}. Finally, we comment on the momentum-space anisotropies in the distribution of accelerated particles {(\sect{pitch})}.

\subsubsection{Energy partition and plasma heating}
\label{sec:heat}
We first discuss the fraction of upstream magnetic energy that is converted into plasma energy in the downstream region, followed by how the plasma energy is divided between bulk (ordered) and internal (random) components. Finally, we address how the dissipated heat is partitioned between electrons and ions.

For weak guide fields, 2D reconnection dissipates roughly half of the upstream magnetic energy, so the downstream flow is approximately in equipartition between plasma energy (including all species) and the energy of the reconnected magnetic field \citep{sironi_15}. The dissipation efficiency decreases with stronger guide fields \citep{sironi_15, werner_17, werner_21, werner_24}. 
In 3D, the nonlinear development of the flux-rope kink instability opens an additional channel for field dissipation, destroying the flux ropes and allowing more magnetic energy to be released than in~2D \citep{werner_21}.

For each species, the dissipated energy is divided between bulk kinetic and internal components (the latter includes both thermal and nonthermal contributions). As discussed in \sect{fluid}, the fastest motions in the reconnection outflow can approach the Alfvén limit of \eq{alfven}. However, most of the plasma flows out at trans-relativistic speeds, as  derived by \citet{lyubarsky_05}: pressure balance across the layer implies that the downstream plasma pressure equals the upstream magnetic pressure, $p\sim B_0^2/8\pi$ (for simplicity, we assume weak guide fields and a cold upstream); the momentum equation implies that $h \,(\Gamma_{\rm out} v_{\rm out}/c)^2\sim B_0^2/4\pi$, and since the enthalpy density $h\simeq 4\,p$ for a relativistically hot plasma, the typical motions in the outflow direction are only marginally relativistic, $\Gamma_{\rm out} v_{\rm out}/c\sim 1$. This result---confirmed by PIC simulations \citep{melzani_14a,sironi_16,kagan_16,sironi_beloborodov_20,gupta_24}---implies that most of the dissipated magnetic energy is converted into internal motions, i.e., plasma heating.

The partitioning of the dissipated energy between electrons and ions---beyond being a fundamental plasma physics question---has significant implications for hot, collisionless accretion flows around supermassive black holes, such as Sgr~A* and M87*. Kinetically motivated models for the electron heating rate in RR can be integrated into GRMHD simulations, allowing to produce synthetic images and spectra that can be compared with observations from the Event Horizon Telescope (EHT) \citep[e.g.,][]{chael_18}.

In the ultra-relativistic limit $\sigmah\gg1$, particle rest-mass does not matter and electrons and ions receive comparable amounts of energy \citep{melzani_14a,guo_16,werner_18}, so the dynamics of electron-ion reconnection in the $\sigmah\gg1$ limit  resembles the $e^+e^-$ pair case. This ``democratic'' energy partition is also observed in magnetically dominated (i.e., $\sigmah\gg1$) electron-positron-ion plasmas, as long as the upstream enthalpy is dominated by the ion rest mass energy, i.e., $\sigmah\sim \sigma_{{\rm c},i}$. There, the post-reconnection energy is shared roughly equally between magnetic fields, pairs, and ions \citep{petropoulou_19}.  
At fixed magnetization, the mean energy per electron (or positron) then scales inversely with the multiplicity, i.e., with the number of pairs per ion 
 ---interestingly, the same scaling is observed in relativistic pair-loaded shocks in the regime when ions dominate the pre-shock mass budget \citep{groselj_22}.

The situation is more complex in semi-relativistic electron-ion plasmas, where the ``semi-relativistic'' {reconnection} regime is characterized by $\sigma_{{\rm h},i} \ll 1 \ll \sigma_{{\rm h},e}$ \citep{melzani_14a,melzani_14b,werner_18}.
Here, the electron vs. ion heating efficiency is a function of plasma beta and guide field strength. For weak guide fields, {electron and ions} 
heating at high beta is dominated by adiabatic compression (``adiabatic heating''), while at low beta it is accompanied by a genuine increase in entropy (``irreversible heating'') \citep{rowan_17}. Ions are heated more efficiently than electrons at low and moderate beta, whereas the electron and ion heating efficiencies become comparable at $\beta_p\sim 1/(2\sigmah)$, when both species start already relativistically hot in the upstream. The dependence on the guide field strength has been mostly explored in the low-beta limit \citep{rowan_19,comisso_24,werner_24}. 2D PIC simulations of semi-relativistic reconnection show that the fraction of initial magnetic energy converted into electron heat is nearly independent of~$B_g/B_0$, whereas ions are heated less as the guide field increases. Consequently, for large guide fields ($B_g\gtrsim B_0$), electrons receive most of the dissipated energy, while for weak or moderate guide fields ($B_g\lesssim B_0$) ions are heated more than electrons. 3D simulations with weak guide fields lead to a higher degree of electron heating---thus, the heating ratio approaches unity---than corresponding 2D counterparts \citep{werner_24}.

\subsubsection{Particle acceleration}
\label{sec:plaw}
Since the very first PIC simulations, the question of whether RR  is capable of generating nonthermal particles has been a crucial focus.  
The history of PIC investigations of particle acceleration in RR starts in the early 2000s, thanks to a few seminal works, mostly by Zenitani \& Hoshino \citep{zenitani_01,zenitani_05, zenitani_07,zenitani_08,jaroschek_04,lyubarsky_liverts_08,cerutti_12b,kagan_13}. They observed the generation of nonthermal spectra in 2D simulations, but the dynamic range of those early-day simulations was insufficient to draw firm general conclusions, especially in 3D,---in fact, \citet{zenitani_08} incorrectly concluded that 3D reconnection, which is dominated by the RDKI at early times, would lead only to thermal heating rather than nonthermal energization, since the RDKI would disrupt the trajectories of particles accelerating along the reconnection electric field. A watershed moment came in $\sim 2014-2015$, when three independent groups convincingly established RR as an efficient particle accelerator, in both 2D and 3D \citep{sironi_spitkovsky_14,guo_14,guo_15,werner_16}. Still, the nonthermal energy spectra did not extend more than a factor of a few to ten beyond $\sim\sigma_{{\rm c},s}m_s c^2$. It is only after $\sim2020$ that the physics of particle acceleration to Lorentz factors $\gg \sigma_{{\rm c},s}$---which is crucially different between 2D and 3D---has been elucidated \citep{zhang_21,zhang_23}. 

Electric fields in reconnection can be conveniently divided between ideal fields $\boldsymbol{E}_{\rm id}=-(\boldsymbol{v}_f/c) \times \boldsymbol{B}$, where $\boldsymbol{v}_f$ is the fluid velocity, and non-ideal fields $\boldsymbol{E}_{\rm nid}=\boldsymbol{E}-\boldsymbol{E}_{\rm id}$, which cannot be captured in ideal MHD. Non-ideal regions prominently appear at/near X-points. In fact, X-points have long been  considered prime candidates for generating nonthermal spectra in~RR {\cite[e.g.,][]{larrabee_03, zenitani_01, lyubarsky_liverts_08}---
\citet{zenitani_01} proposed that nonthermal distributions result as particles are accelerated by the X-point non-ideal electric field for a time proportional to their gyration period in the reconnected magnetic field.
Non-ideal fields necessarily occur in regions of electric dominance ($E>B$) or in the presence of a field-aligned electric field ($E_\parallel=\boldsymbol{E}\cdot \boldsymbol{B}/B\neq 0$). While such conditions are sufficient, regions where $\boldsymbol{E}\neq \boldsymbol{E}_{\rm id}$ also exist without electric dominance or $E_\parallel$ \citep{totorica_23}. The importance of electric dominance becomes greater as $v_A$ increases:
near X-points, the reconnected magnetic field scales as $B_{\rm rec}\propto  B_0x$, whereas the reconnection electric field is $E_{\rm rec}\sim 0.1 v_A B_0$, so the extent of the region of electric dominance scales as $\propto v_A$. 

While intense non-ideal fields are confined to small regions, ideal electric fields driven by fluid motions pervade the reconnection volume and are thus prime candidates for governing particle acceleration to high energies \citep{guo_15,guo_19, uzdensky_22}. Fast outflows in the reconnection downstream carry strong ideal fields, with $E \sim B_0$. Particles can be accelerated by scattering back and forth between coalescing plasmoids via a Fermi-like process \citep{guo_14,guo_15}, or between plasmoids and fast outflows (see \citealt{nalewajko_15} for a comprehensive review of 2D acceleration mechanisms). The converging upstream inflows carry an ideal  field $E \sim \eta_{\rm rec} B_0$, where $\eta_{\rm rec} = v_{\rm in}/c$. As we discuss below, the latter plays an important role for particle acceleration to the highest energies in~3D.

\begin{figure}[t]
\includegraphics[width=\textwidth]{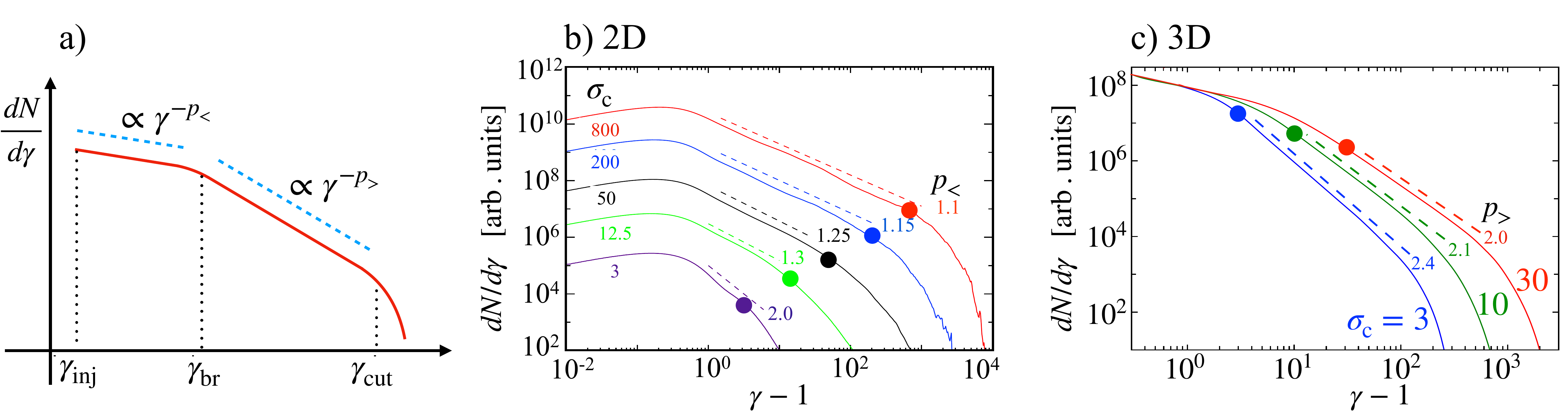}
\caption{Particle spectra in relativistic electron-positron reconnection. (a) Sketch of the particle energy spectrum (adapted from \citealt{comisso_24}), modeled as a broken power law $dN/d\gamma \propto \gamma^{-p_<}$ below the break Lorentz factor $\gammabr$ and $dN/d\gamma \propto \gamma^{-p_>}$ above the break. The spectrum extends from the injection Lorentz factor $\gammainj$ up to the cutoff $\gammacut$. (b) Particle energy spectra from 2D PIC simulations \citep{guo_15}. Dashed lines show the best-fit power-law slopes $p_<$ for the low-energy range $\gammainj<\gamma<\gammabr$. (c) Particle energy spectra from 3D PIC simulations \citep{zhang_23}. Dashed lines show the best-fit power-law slopes $p_>$ for the high-energy range $\gammabr<\gamma<\gammacut$. Both panels (b) and (c) refer to cases with zero guide field; the cold magnetization $\sigmac$ is indicated in the legend, with filled circles marking $\gamma=\sigmac$.
}
\label{fig:spec}
\end{figure}

\paragraph{Electron-positron plasmas} Most advances in our understanding of particle acceleration in RR come from electron-positron PIC simulations, which are computationally cheaper than electron-ion runs. The particle energy spectrum in the downstream region is typically modeled as a broken power law (sketch in \figg{spec}(a)), with a low-energy range $dN/d\gamma \propto \gamma^{-p_<}$  extending from the injection Lorentz factor $\gammainj$ to the break~$\gammabr$, and a high-energy range $dN/d\gamma \propto \gamma^{-p_>}$ from $\gammabr$ to the cutoff~$\gammacut$. The injection Lorentz factor is $\gammainj\sim 0.1-0.3\,\sigmac$ \citep{french_23}, while the break is roughly one order of magnitude higher, $\gammabr\sim1-3\,\sigmac$ \citep{werner_16,comisso_23}. We now discuss the physics of particle acceleration and the formation of power-law spectra, first in the low-energy range and then for Lorentz factors above~$\gammabr$. Unless otherwise noted, the following discussion applies to cases with weak or vanishing guide fields.

The physics of power-law formation in the low-energy range ($\gammainj<\gamma<\gammabr$) is described by an analytical model proposed by \citet{uzdensky_22}, building on \citet{zenitani_01}: acceleration by the X-point electric field proceeds until the particles become magnetized by the reconnected magnetic field and get advected away. 
For weak guide fields, the resulting power-law index is $p_< \simeq \sqrt{(1+\sigmah)/\sigmah}$ \citep{uzdensky_22}, which asymptotes to unity for $\sigmah\gg1$ and is in very good agreement with PIC simulation results, see \figg{spec}(b) \citep{sironi_spitkovsky_14,guo_14,guo_15,werner_16}. 

For particles in this low-energy range, acceleration is then a rapid, one-shot, ``injection'' process, which boosts them from the low, e.g., non-relativistic, upstream energies up to ultra-relativistic energies ($1\ll\gammainj<\gamma<\gammabr$). 
For moderate/strong guide fields ($B_g\gtrsim \eta_{\rm rec} B_0$), $E_\parallel$ plays a dominant role for particle injection \citep{sironi_22,french_23}. 
In weak guide fields ($B_g\lesssim \eta_{\rm rec} B_0$),  most of the
particles ending up with high energies have passed
through regions of electric dominance, $E>B$ \citep{zenitani_01,larrabee_03,lyubarsky_liverts_08,sironi_22,chernoglazov_23}. However, it remains debated whether such regions provide most of the energization during the injection stage---\citet{french_23} found that Fermi reflection and pick-up acceleration in the snapping reconnected field lines dominate over direct $E>B$ acceleration in the injection phase of $B_g=0$ reconnection. Still, broadly-defined non-ideal fields ($\boldsymbol{E}\neq\boldsymbol{E}_{\rm id}$, but not necessarily $E>B$) appear to be essential for particle injection even for vanishing guide fields \citep{totorica_23}.  

While the injection physics is similar in 2D and 3D \citep[e.g.,][]{sironi_22}, the process of particle acceleration to energies beyond the break differs dramatically between 2D and~3D. In 2D, the highest-energy particles are trapped in plasmoids \citep{werner_16,uzdensky_22}, and their energy increases as $\propto \sqrt{t}$ due to magnetic moment conservation, coupled with a linear increase in the local field strength as plasmoid cores compress during subsequent plasmoid growth \citep{petropoulou_18,hakobyan_21}. The spectrum is predicted to have a slope $p_>\simeq 3$ for weak guide fields \citep{hakobyan_21}. However, since this acceleration proceeds slowly, compelling evidence for such a power-law range has yet to emerge in 2D PIC simulations.

\begin{marginnote}
    \entry{Magnetic moment}{$\mu=m_s u_{s,\perp}^2/2 B$, for a particle with mass $m_s$ and 4-velocity $ u_{s,\perp}$ orthogonal to a field of strength $B$.}
\end{marginnote}

In 3D, the flux-rope kink instability breaks the $z$-invariance of magnetic flux ropes, allowing some particles to escape back upstream by moving along~$z$. 
Escaping particles with Lorentz factors $\gamma\gtrsim \,\sigmac$ have Larmor radii large enough that they can sample both upstream sides of the reconnection layer (left in \figg{accel}) and get efficiently accelerated \citep{zhang_21}. The acceleration process can be described in two equivalent ways: an energetic particle crosses the current sheet from one upstream region to the other, completing an arch-shaped segment of its cyclotron orbit in each region. Since the field reverses, these arches always return the particle back to the layer, and its $z$-displacement is always in the same direction, which allows the particle to move nearly along the upstream motional electric field $E\sim\eta_{\rm rec} B_0$, accelerating continuously at nearly the maximum rate \citep[][]{speiser_65,uzdensky_11,cerutti_12a}. Equivalently, the particle is confined in between the two converging upstream flows, and gets energized via a Fermi-like process \citep{giannios_10}. 
This results in fast acceleration, with $\dot{\gamma}\sim \eta_{\rm rec}\omega_{\rm c}$ (right in \figg{accel}). 
The maximum particle energy---in the absence of cooling losses---is obtained by balancing the acceleration time $t_{\rm acc} =\gamma/\dot{\gamma}\sim \gamma/ \eta_{\rm rec}\omega_{\rm c}$ with the advection time out of the system $t_{\rm adv}\sim L/c$, yielding a maximum Lorentz factor $\gamma_{\rm max}\sim \eta_{\rm rec} (\omega_{\rm c}L/c)$. Equivalently, the Larmor radius of the highest energy particles is $\sim \eta_{\rm rec} L\sim 0.1\,L$, which is comparable to the size of the largest plasmoids (see \sect{fluid}). In the absence of cooling losses, the maximum Lorentz factor sets the high-energy cutoff of the spectrum, $\gamma_{\rm cut}\sim \gamma_{\rm max}$.

The 3D physics of power-law formation in the high-energy range ($\gammabr<\gamma<\gammacut$) has been described by \citet{zhang_23} for weak guide fields (see \figg{spec}(c)): high-energy particles gain most of their energy in the upstream region, during a short-lived ``free phase'' where they meander between the two sides of the layer, as described above; 
they leave the region of active acceleration after a time $t_{\rm esc}$, when they get captured/trapped by the downstream flux ropes; during the subsequent ``trapped phase,'' no significant energization occurs. 3D PIC simulations show that $t_{\rm esc}\sim t_{\rm acc}$ for $\sigmah\gtrsim {\rm few}$, which leads to a universal (i.e., nearly $\sigmah$-independent) power-law spectrum $dN_{\rm free}/d\gamma\propto \gamma^{-1}$ for the free particles undergoing active acceleration. The spectrum of trapped particles---which dominate the overall particle count, since the free phase is extremely short-lived---can be obtained by assuming a quasi-steady state: at each energy, the rate of free particles getting trapped (after $t_{\rm esc}\sim t_{\rm acc}\propto \gamma$) equals the rate of trapped particles leaving the system (on $t_{\rm adv}\sim L/c$), yielding $dN/d\gamma\propto (t_{\rm adv}/t_{\rm esc})\; dN_{\rm free}/d\gamma \propto \gamma^{-2}$, i.e., a universal power-law slope $p_>\simeq 2$. 

We note that most of the findings discussed in this subsection apply to cases with vanishing or weak guide fields. At a fixed $\sigmah$, stronger guide fields tend to steepen the power-law slopes, both in the low-energy range and above the break \citep{werner_17,werner_21,comisso_23}. However, a comprehensive understanding of the physics of power-law formation in the presence of strong guide fields is still lacking.

\begin{figure}[t]
\begin{minipage}{0.55\textwidth}
\includegraphics[width=\textwidth]{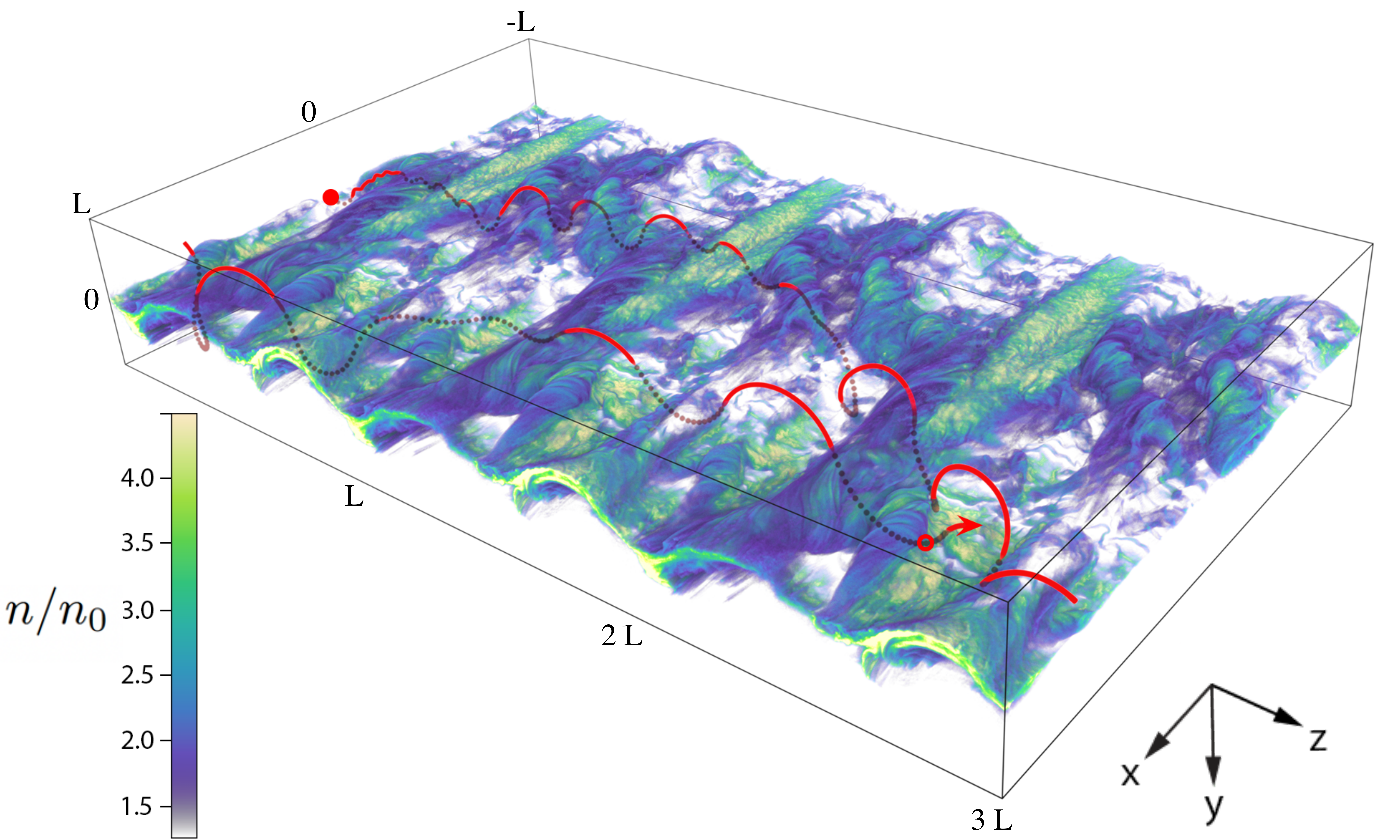}
\end{minipage}
\begin{minipage}{0.45\textwidth}
\includegraphics[width=\textwidth]{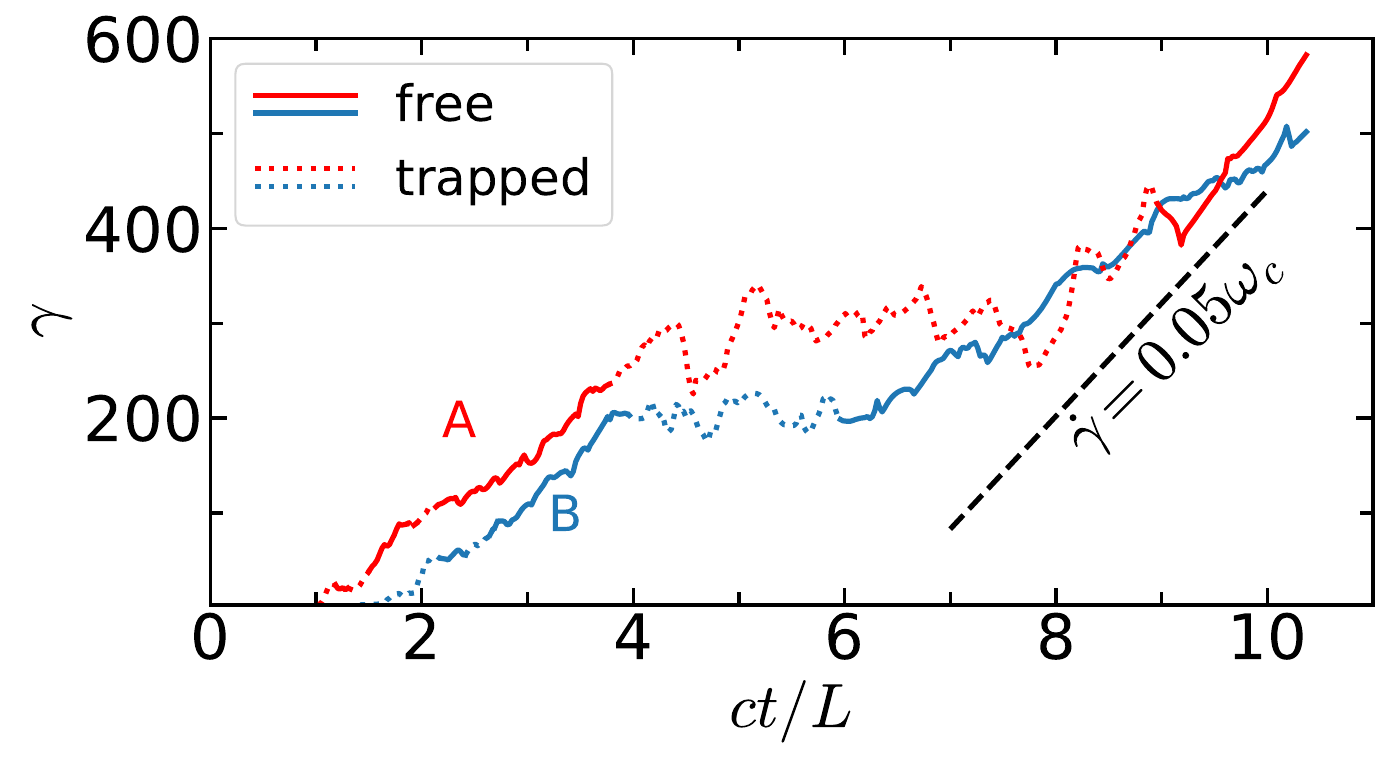}
\end{minipage}
\caption{{\bf Left:} Isosurfaces of plasma density from a 3D PIC simulation \citep{zhang_23}. 
    The trajectory of a representative high-energy positron is overlaid, starting at the filled red circle and ending at the tip of the red arrow, after looping
once through the periodic $z$ boundary. It is colored in solid red if the positron is in the upstream (inflow) region above the midplane (i.e., on the same side as the observer), otherwise it is dotted black. The red open circle indicates the particle position at the time of the density isosurfaces. {\bf Right:} Time evolution of the Lorentz factor of two representative positrons, from the same study. Solid lines indicate when the positron is free, dotted when trapped (see text). The spatial track of positron A is the one shown in the left panel.}
\label{fig:accel}
\end{figure}

\paragraph{Electron-ion and electron-positron-ion plasmas} 
In the ultra-relativistic limit $\sigmah\gg1$, the properties of electron-ion reconnection are nearly the same as in pair plasmas, and the two species have similar spectra \citep{melzani_14a,melzani_14b,guo_16}. However, electron and ion spectra differ in the semi-relativistic regime $\sigma_{{\rm h},i} \ll 1 \ll \sigma_{{\rm h},e}$ \citep{werner_18,ball_18,kilian_20,werner_24,comisso_24}. There, the break in the ion spectrum lies at $\gamma_{{\rm br},i}\sim \sigma_{{\rm c},i}$, while for electrons it is at $\gamma_{{\rm br},e}\sim 0.1\sigma_{{\rm c},e}$ \citep{comisso_24}; electron slopes are generally harder than ion slopes \citep{werner_18,ball_18,comisso_24}. As in pair plasmas, the energy spectra of both species steepen with increasing guide field strength \citep{ball_sironi_19,comisso_24}. 

The physics of particle acceleration in 2D simulations of electron-positron-ion reconnection with $\sigmah\gtrsim 1$  was investigated by \citet{petropoulou_19} for vanishing guide fields. They found that electrons and positrons generally display similar spectra, and that the power-law slope is mainly controlled by $\sigma_{{\rm h},e}$ (with harder power laws for higher~$\sigma_{{\rm h},e}$). The 3D physics of ion acceleration in a pair-dominated plasma (with $B_g=0)$ was studied by \citet{chernoglazov_23}, who considered synchrotron-cooled pairs and a small mass fraction of non-radiating ions. They showed that the ion acceleration physics depends sensitively on the strength of the pair cooling losses, as these losses alter the downstream pressure balance, thereby changing the overall properties of the plasmoid chain---stronger cooling leads to greater plasma compression and smaller flux ropes \citep{schoeffler_19}. The high-energy ion spectrum is $dN/d\gamma \propto \gamma^{-2}$ for weak pair cooling (as in the uncooled cases studied by \citealt{zhang_21,zhang_23}), and becomes as hard as $dN/d\gamma \propto \gamma^{-1}$ for strong pair cooling (more on radiative RR in \sect{radiation}). Since a power-law spectrum as hard as $dN/d\gamma \propto \gamma^{-1}$ extending to arbitrarily high energies would violate energy conservation, we argue that the ion spectrum in $\sigmah\gg1$ electron-positron-ion reconnection with strongly cooled pairs and non-radiating ions should follow a broken power law, with $dN/d\gamma \propto \gamma^{-1}$ below the break and $dN/d\gamma \propto \gamma^{-p_>}$ above the break. Here, $p_>$ would be the same as in corresponding uncooled cases. Further work is needed to validate this conjecture.

\subsubsection{Particle anisotropy and kinetic beaming}
\label{sec:pitch}
While most kinetic studies of RR focus on the particle {\it energy distribution}, the {\it angular distribution} of accelerated particles is also crucial. Synchrotron and inverse Compton (IC) radiation from ultra-relativistic particles is strongly beamed along their motion, meaning that the angular distribution of the radiation closely mirrors that of the particles themselves. The observed signature of RR then depends not only on the particle energy spectrum but also on the orientation of the observer's line of sight relative to the angular distribution of the emitting particles.

The first dedicated PIC study of the angular distribution of RR-accelerated particles was conducted by \citet{cerutti_12b}, building on earlier analytical work \citep{uzdensky_etal_11, cerutti_12a}. Since then, most studies---summarized below---have focused on the low-energy spectral range ($\gammainj<\gamma<\gammabr$), typically using 2D simulations \citep[][but see \citealt{cerutti_14a} for 3D cases]{cerutti_12b, cerutti_13, kagan_16, mehlhaff_20}. In 3D, particles accelerated beyond $\gammabr$ during the ``free'' phase of active acceleration are also highly anisotropic, primarily moving along the electric field direction. However, they become nearly isotropic once captured by plasmoids \citep{zhang_21, zhang_23}.

In the low-energy nonthermal range $\gammainj<\gamma<\gammabr$, \citet{cerutti_12b} showed that the angular anisotropy becomes more pronounced at higher particle energies. These particles form narrow, discrete beams whose solid angle shrinks with energy, often following a power law---an effect dubbed ``kinetic beaming.'' In moderate-size systems with no guide field, energetic beams tend to remain in the $xz$ plane of the main layer, first accelerated along $z$ by the reconnection electric field and then deflected by $B_y$ towards $\pm \hat{x}$ \citep{cerutti_12b, cerutti_13,cerutti_14a,kagan_16}. A moderate guide field redirects the beams out of this plane \citep{cerutti_13}. In larger systems, where  major plasmoid mergers frequently occur, the most intense beams lie in the $yz$ plane of secondary anti-reconnection layers and are predominantly directed towards $\pm \hat{y}$ \citep{mehlhaff_20}.

\begin{marginnote}
    \entry{Kinetic beaming}{Energy-dependent focusing of the momentum direction of RR-accelerated particles.}
\end{marginnote}

The observational manifestations of kinetic beaming are strongly influenced by radiative cooling \citep{kagan_16, mehlhaff_20}. Strong angular anisotropy for $\gammainj<\gamma<\gammabr$ persists as long as the particles are actively accelerated within the inter-plasmoid current sheet. However, once they exit the sheet and become magnetized by plasmoids, their anisotropy rapidly diminishes. Lower-energy, slower-cooling particles thus become nearly isotropic before emitting most of their energy. In contrast, higher-energy particles can emit while still being accelerated or immediately afterwards, i.e., before they can fully isotropize. Therefore, the kinetic beaming of radiation is enhanced by radiative cooling and may exhibit an even stronger energy dependence than the particle distribution itself.

Kinetic beaming of RR-accelerated particles and their radiation has significant observational implications, e.g., for gamma-ray flares in the Crab Nebula \citep{uzdensky_etal_11, cerutti_12a, cerutti_13, cerutti_14a, cerutti_14b} and blazar jets \citep{nalewajko_etal-2011, nalewajko_2012, kagan_16, mehlhaff_20}. First, energetic particles with $\gamma\sim \gamma_{\rm br}$ get focused deep inside the current layer, where the magnetic field (more precisely, its component perpendicular to the particle momentum) is weaker, suppressing synchrotron emission and allowing particles to exceed the synchrotron radiation-reaction limit (defined in \sect{radiation}). Second, the strong anisotropy of high-energy emission means that a bright flare is detected only when a narrow beam happens to sweep across our line of sight. As a result, many flares may go undetected, significantly affecting the inferred flare statistics. Third, since these beams are not stationary and can rapidly change direction, the flare light curve may vary faster than the underlying reconnection event---governed instead by the angular size and angular velocity of the sweeping beam---potentially explaining ultra-rapid gamma-ray variability.

So far, we have discussed the angular distribution of RR-accelerated particles relative to the {\it global} geometry of the system (i.e., the $xyz$ axes). It is also important to consider their energy-dependent pitch-angle distribution, defined by the angle $\alpha$ between the particle momentum and the {\it local} magnetic field direction. This is particularly relevant for electrons since synchrotron emission scales as $\propto \sin^2 \alpha$. The pitch-angle anisotropy, quantified by the average $\langle \sin^2 \alpha \rangle$, shows distinct power-law behavior with negative and positive slopes respectively below and above a break Lorentz factor $\gamma_{\rm min,\alpha}$ \citep{comisso_23,comisso_24}. The drop in $\langle {\sin }^{2}\alpha \rangle$ below the expectation of an isotropic particle distribution gets more pronounced for stronger guide fields. For $B_g \gg B_0$, the Lorentz factor $\gamma_{\rm min,\alpha}$ of strongest anisotropy approaches $\sim 2 \sigma_{{\rm c},s}$, and the pitch-angle anisotropy reaches a minimum value of $\langle \sin^2 \alpha \rangle \sim 1/\gamma_{\rm min,\alpha}^2\sim 1/(2\sigma_{{\rm c},s})^2$ \citep{comisso_23,comisso_24}. This pronounced anisotropy at $\gamma \sim \sigma_{{\rm c},s}$ in the presence of strong guide fields is driven by the dominant role of $E_\parallel$ for particle acceleration at these energies \citep{sironi_22}.
The energy-dependent pitch-angle anisotropy has important implications for various astrophysical phenomena, including the hard radio spectrum in PWNe \citep{comisso_20}, orphan gamma-ray flares in blazars \citep{sobacchi_21}, and emission beyond the synchrotron radiation-reaction limit in Crab Nebula flares and GRBs \citep{comisso_21}.

\begin{marginnote}
    \entry{Pitch angle}{Angle between the velocity of a particle and the local magnetic field.}
\end{marginnote}

\section{RADIATIVE RR}
\label{sec:radiation}

Radiative magnetic reconnection is an exciting new frontier of plasma astrophysics, witnessing rapid progress in both analytical \citep{kulsrud_dorman_95, uzdensky_11, uzdensky_mckinney_11, uzdensky_etal_11, cerutti_12a, nalewajko_etal-2011, uzdensky_spitkovsky_14, beloborodov_17, beloborodov_21,mehlhaff_21,hakobyan_ripperda_23,chen_23}, and, especially, numerical realms \citep{jaroschek_hoshino-2009, cerutti_13, cerutti_14a, cerutti_14b, nalewajko_18, werner_19, hosking_20,schoeffler_19,schoeffler_23, hakobyan_21, hakobyan_spitkovsky_23, sironi_beloborodov_20, mehlhaff_20, mehlhaff_24, sridhar_21, sridhar_23, chernoglazov_23}; see \citet{uzdensky_16} for a review. Recently, it has also begun to be explored in laboratory experiments using pulsed-power platforms \citep{datta_24a}.

The physics of radiative reconnection, with its intricate interplay between photons and charged particles, is crucial for understanding extreme plasmas near neutron stars and black holes.
First, radiative cooling alters the energetics and pressure balance in the reconnection layer, potentially leading to strong compression or even a catastrophic radiative collapse \citep{uzdensky_mckinney_11, schoeffler_19, chernoglazov_23, datta_24a}. Second, radiation reaction can influence the downstream plasma dynamics, causing effects like enhanced radiative (Compton) resistivity \citep{goodman_uzdensky_08, schoeffler_23} and Compton drag on fast-moving plasmoids \citep{beloborodov_17, sironi_beloborodov_20, sridhar_21, sridhar_23}. Third, as radiation reaction increases with particle energy, it has a stronger impact on higher-energy particles, altering nonthermal particle acceleration \citep[e.g.,][]{cerutti_12a, cerutti_13,werner_19}.
Finally, in even more extreme environments, gamma-ray ($\gtrsim\,$MeV) photons emitted by reconnection-accelerated particles can create electron-positron pairs, enriching the upstream medium and reducing the inflow magnetization \citep{Lyubarskii_96, uzdensky_11, mckinney_uzdensky_12, hakobyan_19, hakobyan_spitkovsky_23, mehlhaff_21, mehlhaff_24, chen_23}.

This section describes radiative aspects of RR, with a focus on leptonic (electron and positron) radiation losses, as the physics of RR with strong hadronic losses remains largely unexplored. We begin by outlining key electron energy scales (Section~\ref{subsec-rad-parameters}), followed by a summary of recent findings from PIC simulations of radiative RR (Section~\ref{subsec-radiative-sims}).

\subsection{Parameter Space of Radiative RR}
\label{subsec-rad-parameters}

To assess the significance of radiative effects, it is useful to consider a hierarchy of electron energy scales. In addition to the two previously introduced—$\sigma_{{\rm c},e}$ and $\gamma_{\rm max}= \eta_{\rm rec}(\omega_{{\rm c},e}L/c)\gg\sigma_{{\rm c},e}$—four additional Lorentz factors are commonly defined \citep[see, e.g.,][]{uzdensky_16, mehlhaff_21}: $\gammarad$, $\gammacool$, $\gamma_{\rm Q}$, and~$\gamma_{\rm pp}$. The first two  are classical in nature, while the latter two are quantum/QED.
\begin{itemize}
\item The Lorentz factor $\gamma_{\rm rad}$, also known as the classical radiation-reaction or ``burnoff'' limit, at which the radiation-reaction drag force balances the accelerating force from the reconnection electric field, $E_{\rm rec} = \eta_{\rm rec}B_0$. It is given by:
\begin{equation}
    \label{eq:gammarad}
    \gamma_{\rm rad}=\sqrt{\frac{e E_{\rm rec}}{(4/3)\sigma_{\rm T}U}},
\end{equation}
 where $\sigma_{\rm T}$ is the Thomson cross section, and $U$ is the energy density of the magnetic field (for synchrotron cooling) or of the ambient soft radiation field (for IC cooling). Particles generally cannot be accelerated beyond $\gamma_{\rm rad}$,
 since they would be losing more energy than they can gain (exceptions are discussed below). We define cooling as strong when $\gamma_{\rm rad} < \sigma_{{\rm c},e}$, and weak when $\gamma_{\rm rad} > \sigma_{{\rm c},e}$.

For synchrotron cooling, in \eq{gammarad} we have assumed an isotropic particle distribution (i.e., $\langle\sin^2\alpha \rangle=2/3$). If particles cool in a magnetic field comparable to the reconnecting field~$B_0$, i.e., $U\simeq B_0^2/8\pi$, then $\gammarad\propto \sqrt{\eta_{\rm rec}/B_0}$ and the corresponding ``burnoff'' synchrotron photon energy is \begin{equation} 
\epsilon_{\rm rad}\simeq \gammarad^2 \hbar\omega_{{\rm c},e}=\frac{9\,\eta_{\rm rec} m_e c^2}{4\,\alpha_{\rm F}}\simeq 16\,\eta_{\rm rec,-1} {\rm MeV} \, , 
\label{eq:eerad}
\end{equation} 
which is independent of the magnetic field strength. Here, $\alpha_{\rm F}\simeq 1/137$ is the fine structure constant and $\eta_{\rm rec} = 0.1\,\eta_{\rm rec,-1}$. At face value, the synchrotron spectrum cannot extend beyond~$\epsilon_{\rm rad}$. However, this limit can be violated if, in the acceleration region, the magnetic field component orthogonal to the particle motion is $\ll B_0$, either because the overall field is weak (e.g., in electric zones with $E>B$) or because the accelerating particle moves nearly along the local field (so synchrotron losses are suppressed by a factor of $\sin^2\alpha\ll1$). Then, particles can be accelerated beyond $\gammarad$ and emit synchrotron photons that exceed the formal $\epsilon_{\rm rad}$ limit 
\citep{kirk_2004, uzdensky_etal_11,cerutti_12a}.
 
\item The Lorentz factor $\gamma_{\rm cool}$, also known as the ``cooling break,''  at which the radiative cooling time is comparable to the global advection time across the system $t_{\rm adv}=L/c$. By defining the (radiative or magnetic) compactness parameter as $\ell \equiv \sigma_{\rm T} U L/m_e c^2$, one finds that $\gamma_{\rm cool}=\ell^{-1}$ (as long as $\gamma_{\rm cool}\gg1$). We define the fast-cooling regime as $\gamma_{\rm cool}<\sigma_{{\rm c},e}$ and the slow-cooling regime as $\gamma_{\rm cool}>\sigma_{{\rm c},e}$. The classical radiation-reaction limit~$\gamma_{\rm rad}$, the cooling Lorentz factor~$\gamma_{\rm cool}$, and the maximum Lorentz factor $\gamma_{\rm max}$ are linked by a simple relation \citep[e.g.,][]{mehlhaff_21}: 
\begin{equation}
    \gamma_{\rm rad}^2=\gamma_{\rm cool}\gamma_{\rm max}~~.
    \label{eq:ggg}
\end{equation}

\item The Lorentz factor $\gamma_{\rm Q}$ 
above which the energy of the emitted (or up-scattered) photon becomes comparable to the energy of the emitting particle, and hence the discrete, quantum nature of the radiation process comes into play.  In the case of IC scattering, this marks the transition to the Klein-Nishina regime, and we identify $\gamma_{\rm Q}$  with the Klein-Nishina Lorentz factor $\gamma_{\rm KN} \simeq m_e c^2/\epsilon_s$, where $\epsilon_s$ is the energy of soft seed photons \citep{mehlhaff_21}. Its analog for the case of synchrotron emission in a strong magnetic field (nonlinear Compton scattering) is given by a similar expression, $\gamma_{\rm Q,{\rm syn}} \simeq m_e c^2 /(\hbar \omega_{{\rm c},e})$
\citep[see, e.g.,][]{schoeffler_19}.
In both cases, the relevant cross section decreases  above~$\gamma_{\rm Q}$,  and the scaling of the emitted photon energy with the electron Lorentz factor $\gamma$ shifts from quadratic to linear. Importantly, $\gamma_{\rm Q}$ also represents, up to a factor of order unity, the Lorentz factor of reconnection-accelerated particles emitting high-energy gamma-ray photons capable of pair creation by interacting with the same agent that caused their emission: one-photon pair production in a strong magnetic field in the synchrotron case and two-photon pair production on ambient seed photons in the IC case \citep{mehlhaff_24}.

\item The Lorentz factor $\gamma_{\rm pp}$ of electrons capable of generating MeV ($\sim m_ec^2$) photons, which can lead to efficient $\gamma\gamma$ pair production. In the synchrotron case, this is $\gamma_{\rm pp}=\sqrt{m_e c^2/\hbar \omega_{\rm c}}$, while for IC $\gamma_{\rm pp}=\sqrt{m_e c^2/\epsilon_s}$.
\end{itemize}

\noindent
In addition to the individual particle energy scales, a radiative treatment of reconnection should specify the system's Thomson optical depth $\tau_{\rm T} \equiv \sigma_{\rm T} n_e L$ and the optical depth to photon-photon pair production  $\tau_{\gamma\gamma}\simeq \sigma_{\gamma\gamma} n_{\epsilon>\rm MeV} L$, where
 $n_{\epsilon>\rm MeV}$ is the density of photons above $2m_e c^2$ (so, capable of producing pairs), and 
 $ \sigma_{\gamma\gamma}\simeq 0.1 \sigma_{\rm T}$ is the cross section for photon-photon annihilation. The two optical depths can be expressed in terms of parameters we have already introduced, e.g., $\tau_{\rm T}=2 \ell_{\rm B}/\sigmae$, 
 where $\ell_{\rm B_0}=\sigma_{\rm T} (B_0^2/8 \pi) L/m_e c^2$ is the magnetic compactness computed with the reconnecting field~$B_0$.

\begin{figure}[t]
\includegraphics[width=\textwidth]{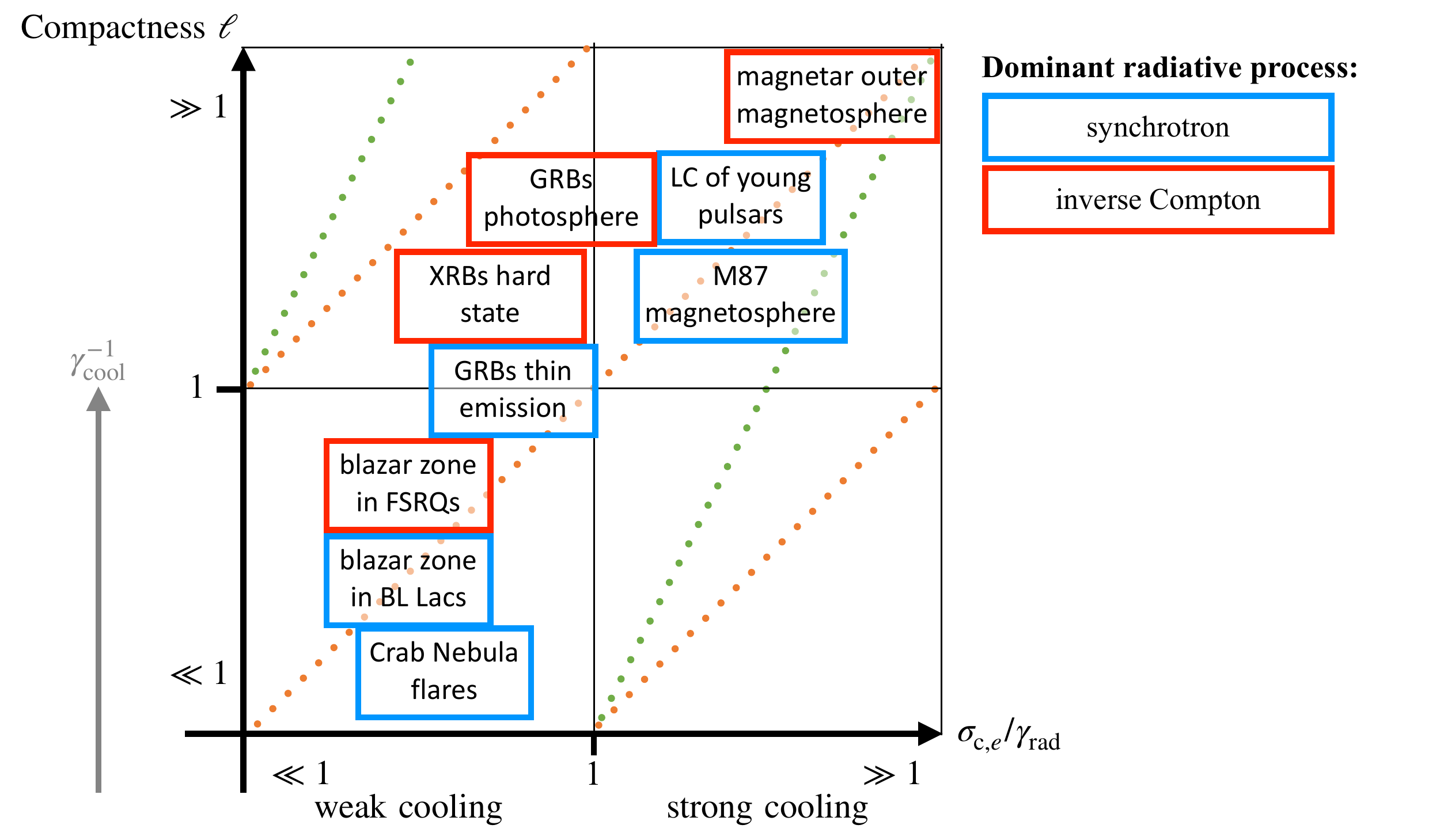}
\caption{Parameter space of radiative RR in the $\sigmae/\gammarad - \ell$ plane, where we place a few boxes of representative X-ray and gamma-ray sources. Here, both $\gammarad$ and $\ell$ are defined with the energy density (either magnetic or soft-photon) that dominates the electron radiative losses. The rightmost grey vertical axis shows $\gammacool^{-1}=\ell$. The borders of the boxes are color-coded based on the dominant radiative process, as shown in the legend: blue for synchrotron and red for inverse Compton.
The dotted green lines are loci of $\ell\propto (\sigmae/\gammarad)^2$ for constant $\gamma_{\rm max}/{\sigmae}^2$, see \eq{ggg}. The dotted orange lines are loci of $\ell_{\rm B_0}\propto \sigmae/\gammarad$ for constant $\tau_{\rm T}/\gammarad$.
}
\label{fig:radiative}
\end{figure}

\begin{table}[h]
\label{tab1}
\tabcolsep7.5pt
\caption{Plasma conditions at the reconnection site for the sources in \figg{radiative}}
\label{tab1}
\begin{center}
\begin{tabular}{@{}l|c|c|c|c@{}}
\hline
Source &$\log U$$^{\rm [a]}$&$\log B_0$&$\log L$&$\log \sigma_{{\rm c},e}$\\
 &[erg$\cdot$cm$^{-3}$] &[G] &[cm] &\\
\hline
blazar zone in BL Lacs&$-2.5\pm$2 &$-0.5\pm$1 &$16.5\pm$1  &$3.5\pm$0.5\\
blazar zone in FSRQs &$-1\pm$2 &$0\pm$1&$1.6\pm$1 &$3\pm$0.5\\
GRBs photosphere &$11.5\pm$2 &$6.5\pm$1 & $9.5\pm$2 &$3.5\pm$1\\
GRBs thin emission &$6.5\pm$2 &$4\pm$1&$12\pm$1 &$3.5\pm$1\\
Crab Nebula flares &$-6.5\pm$1 &$-2.5\pm$0.5 &$16\pm$0.5 &$6\pm$1\\
XRBs hard-state emission &$13.5\pm$1 &$8\pm$0.5 & $6.5\pm$0.5 &$2\pm$1\\
M87 magnetosphere & $2.5\pm$1& $2\pm$0.5 & $16\pm$0.5 &$6.5\pm$1\\
Pulsar light cylinder (LC) & $8.5\pm$2 & $5\pm$1 & $8.5\pm$0.5 & $4.5\pm$1\\
Magnetar outer magnetosphere & $15.5\pm$2 & $9\pm$1 & $7\pm$0.5 & $4\pm$1\\
\hline
\end{tabular}
\end{center}
\begin{tabnote}
$^{\rm [a]}$ $U$ is the energy density that dominates the electron radiative losses. The default is $U=B_0^2/8\pi$, with the following exceptions: for the blazar zone in FSRQs, $U$ is the energy density of external soft photons from the accretion disk, broad line region or dusty torus; for the hard-state emission from XRBs and the emission from the outer regions of magnetar magnetospheres, seed photons for IC scattering are sourced by magnetic dissipation, so $U\sim \eta_{\rm rec} B_0^2/8\pi\sim 0.1 B_0^2/8\pi$ \citep{beloborodov_17,beloborodov_21}. 
\end{tabnote}
\end{table}

The relative ordering of the above particle energy scales defines the different regimes of radiative RR and thus shapes the physical landscape of this process. This allows us to map various astrophysical systems where reconnection plays a major role onto a multi-dimensional parameter space and organize them according to the relative importance of radiative effects. This is illustrated in \figg{radiative} (supported by {\bf Table~\ref{tab1}}), which presents a two-dimensional plane of $\sigmae/\gammarad$ and compactness $\ell$, where both $\gammarad$ and $\ell$ are defined with the energy density (either magnetic or soft-photon) that dominates the electron radiative losses. For example, reconnection in the equatorial current sheet in the inner part of the black hole magnetosphere of M87, as well as in young pulsars, occurs in the regime of strong cooling and large compactness, whereas in the pc-scale dissipation region of blazar jets one typically finds low compactness and weak cooling. The diagram is not meant to be exhaustive but is put forward as a proposal to the high-energy astrophysics community to locate where their favorite nonthermal  source lies in the landscape of radiative~RR.

\subsection{Radiative PIC Simulations}
\label{subsec-radiative-sims}
The last decade has seen an explosion of first-principles numerical studies using radiative-PIC simulations to explore different parts of the multi-dimensional parameter space of radiative~RR. The present subsection describes the main findings of these studies. We first discuss how radiative effects are implemented in modern PIC codes (\sect{numerics}), and then summarize the main  results of PIC studies of radiative~RR (\sect{results}).

\subsubsection{Numerical implementation}
\label{sec:numerics}

While interactions of charged particles with long-wavelength (compared to the grid-cell size) electromagnetic perturbations caused by collective plasma processes are automatically and self-consistently calculated by the basic PIC algorithm, the high-energy emission by individual particles, with wavelengths too short to be resolved by the grid, needs to be treated separately. 

The majority of PIC explorations of radiative RR have been conducted in the classical regime, where radiation effects are treated as a continuous radiation-reaction drag force in the particle equation of motion \citep[e.g.,][]{vranic_16}. This applies to synchrotron emission and IC scattering in soft ambient radiation fields typical of most astrophysical sources. However, as outlined above in Section~\ref{subsec-rad-parameters}, under extreme conditions---very strong magnetic fields and hard seed photons---the classical treatment breaks down and 
the quantum nature of radiation comes into play. 
The classical synchrotron or Thomson treatment is then replaced by the QED nonlinear-Compton or Klein-Nishina framework, respectively.

Numerical implementations of the interaction between charged particles and radiation partly reflect this physical dichotomy, differing based on whether radiation reaction is treated as a smooth, continuous, deterministic drag force  or as a collection of discrete events requiring a probabilistic approach.
The former strategy is simpler and historically was the first to be implemented, e.g., by \citet[][]{jaroschek_hoshino-2009,tamburini_10}  for synchrotron and by \citet{werner_19} for~IC.
Modern radiative PIC codes, however, often adopt a combination of the two approaches and treat the highest-energy photon emission as a sequence of discrete probabilistic events in a Monte Carlo (MC) fashion. This method is essential for describing emission or scattering in the QED regime.
Similarly to how the PIC method represents particle distributions by sampling discrete macroparticles, these simulations represent the emitted (and sometimes ambient/incoming) radiation intensity field via discrete macrophotons. 
One can further distinguish, from a computational/algorithmic perspective, single-macroparticle/macrophoton processes (e.g., synchrotron; external IC scattering in a prescribed ambient photon field; single-photon pair creation in a strong magnetic field; $\gamma\gamma$ pair creation by an energetic gamma-ray photon in an ambient photon field) or two-macroparticle/macrophoton (i.e., binary) processes (e.g., IC scattering including Klein-Nishina effects, $\gamma\gamma$ pair creation from two high-energy photons, pair annihilation).

\subsubsection{PIC results}
\label{sec:results}
We now describe major results of recent PIC studies of radiative~RR, first in the classical regime and then in the QED regime.
\paragraph{Classical regime}
A classical treatment of synchrotron cooling was first incorporated into PIC simulations of RR in the pioneering work by \citet{jaroschek_hoshino-2009} on pulsar magnetospheres. This research direction advanced rapidly a few years later with the studies by \citet{cerutti_13, cerutti_14a}, aimed at explaining the bright gamma-ray flares from the Crab Nebula in the 100\,MeV$-$1\,GeV range, exceeding the standard synchrotron burnoff limit of $\epsilon_{\rm rad}\simeq 16\,\eta_{\rm rec,-1}$~MeV. This requires efficient particle acceleration beyond the radiation-reaction Lorentz factor~$\gamma_{\rm rad}$. As discussed in \sect{pitch}, kinetic beaming can help overcome this limit by focusing energetic particles deep into the current layer, where the perpendicular magnetic field is weak and hence synchrotron losses are suppressed, resulting at the same time in rapid temporal variability \citep{cerutti_13,cerutti_14b, cerutti_14a}.
In a recent 3D study, \citet{chernoglazov_23} revisited this issue and found that 
in the strong cooling regime ($\gammarad < \sigmae$) electrons can indeed accelerate beyond $\gammarad$ (up to $\sim \sigmae$) by moving nearly along the magnetic field. The resulting synchrotron spectral cutoff scales as $\sim \epsilon_{\rm rad} (\sigmae/\gammarad)$, with important implications for Crab Nebula flares and  gamma-ray emission from young pulsars \citep{chernoglazov_23}.

PIC simulations of RR with fast synchrotron cooling ($\gammacool\ll\sigma_{{\rm c},e}$) have been used to explain the rapid, large-amplitude polarization swings observed in blazars \citep{zhang_20,zhang_22,hosking_20}. Plasmoid mergers drive both bright flares and polarization swings. Particles accelerated at the merger interface stream through the post-merger plasmoid, progressively illuminating regions with varying plane-of-sky field direction. This results in the observed rotation of the polarization vector, explaining why high-energy flares in blazar jets often coincide with large optical polarization sweeps.

The first radiative PIC simulations of RR with classical IC losses were conducted by \citet{werner_19} (see also \citealt{nalewajko_18}, who included both synchrotron and IC losses for blazar applications). They found that, while the reconnection rate and overall dynamics remain unchanged, IC cooling strongly affects nonthermal particle acceleration. For moderate cooling ($\gammarad\gtrsim \sigma_{{\rm c},e}$), the electron power-law spectrum steepens at a break energy $\sim \sigma_{{\rm c},e}m_e c^2$, ultimately cutting off at~$\gammarad$.
A follow-up by \citet{mehlhaff_20} revealed strong kinetic beaming at high energies, which could explain rapid variability of gamma-ray blazar flares. 

With a similar approach, \citet{sironi_beloborodov_20} performed large-scale 2D simulations in the regime $\gammacool\ll\sigma_{{\rm c},e}\lesssim\gammarad$. They found that most of the pair plasma in the layer is kept cold by IC losses, trapped in magnetically dominated plasmoids with low thermal pressure. The magnetic energy released by reconnection mostly drives bulk plasmoid motions, with a quasi-Maxwellian distribution with an effective bulk temperature of $\sim100\,\rm keV.$ This suggests that plasmoid-chain Comptonization could power the hard X-ray emission in black hole binaries and~AGN \citep{sridhar_21,sridhar_23}.

\begin{figure}[t]
\centering
\includegraphics[width=0.85\textwidth]{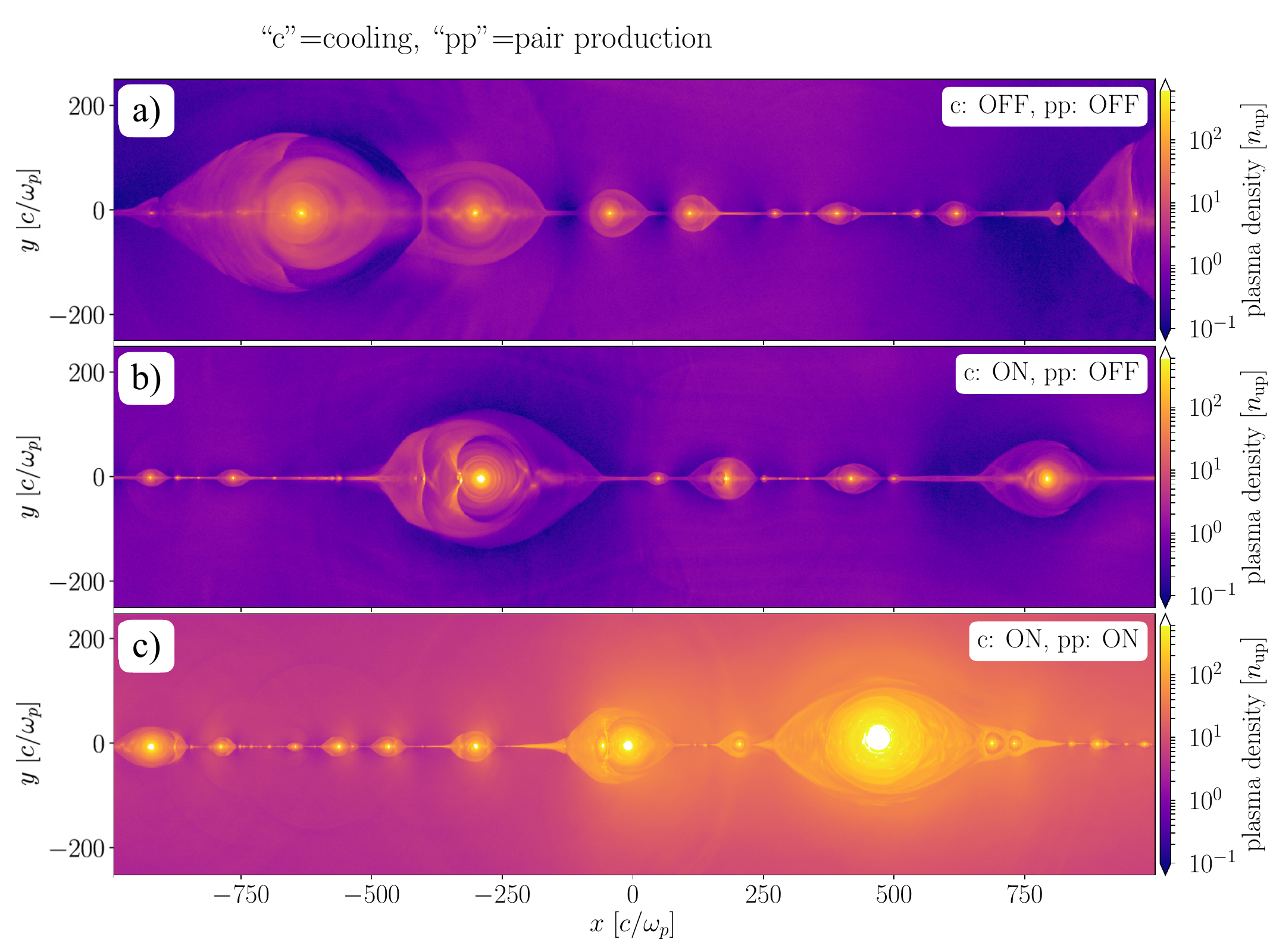}
\caption{PIC simulation of radiative RR including synchrotron cooling and $\gamma-\gamma$ pair production \citep{hakobyan_19}. The plot shows the difference in particle density between scenarios with (a) no cooling or pair production, (b) just cooling, and (c) both cooling and pair production.}
\label{fig:hayk}
\end{figure}

\paragraph{QED regime}
The QED regime of radiative RR represents the cutting-edge frontier of modern theoretical and numerical (PIC) exploration.  
The pioneering radiative-QED PIC studies by \citet{schoeffler_19} and \citet{hakobyan_19} investigated QED effects in the synchrotron case, tracking both the synchrotron photons from reconnection-accelerated particles and the electron-positron pairs created self-consistently via single-photon ($\gamma-B$) and two-photon ($\gamma-\gamma$) pair production, respectively. 
In \citet{hakobyan_19}, synchrotron gamma-ray photons collide with each other, producing hot secondary pairs in the upstream region (\figg{hayk}). 
They found this process to be self-regulating: the increased density of upstream pairs flowing into the sheet reduces the magnetization~$\sigmae$ and hence the mean energy per particle, thus quenching  particle acceleration and lowering the photon spectral cutoff, which in turn weakens subsequent upstream pair production. This offers an explanation for the weak dependence of the observed gamma-ray spectral cutoff in pulsars on the magnetic field at the light cylinder \citep{abdo_10}, as initially proposed by \citet{Lyubarskii_96}.

This self-regulation process also occurs at the magnetospheric current sheet of black-hole MAD accretion flows, such as in~M87 \citep{chen_23, hakobyan_ripperda_23}. These studies demonstrated abundant synchrotron-powered pair production around the reconnection layer, with the produced secondary pairs dominating the reconnection dynamics. The corresponding synchrotron-self-Compton emission by the RR-accelerated particles in this self-regulated regime may power the bright, rapid TeV gamma-ray flares observed from~M87 \citep{hakobyan_ripperda_23}.
 
Radiative-QED reconnection in strong magnetic fields, 
approaching the critical quantum Schwinger field $B_{\rm Q} \equiv m_e^2 c^3/ e \hbar \simeq 4 \times 10^{13}\,{\rm G}$, was investigated by \citet{schoeffler_19,schoeffler_23}. This regime is relevant to the magnetospheres of strongly magnetized neutron stars, e.g., magnetars. These studies included a full quantum treatment of synchrotron emission (nonlinear Compton scattering), incorporating radiation reaction as a continuous drag force at low energies and as discrete photon emission at high energies. \citet{schoeffler_19} also included single-photon pair production in a strong magnetic field.
For weak guide fields, they showed that radiative cooling greatly enhances the compression of plasma and magnetic field inside plasmoids in~2D, leading to an enhancement and strong spatial localization of the resulting emission. In 3D this effect is weakened by the kink instability that disrupts the compressing flux ropes, allowing the radiating particles to escape. 

Reconnection with IC scattering extending into the quantum Klein-Nishina regime was investigated analytically and numerically by \citet{mehlhaff_21, mehlhaff_24}. Photons Comptonized by RR-accelerated particles to very high energies may exceed the threshold for pair production on the soft radiation background and thus return their energy to the plasma as secondary pairs. This process may lead to the same self-regulating feedback loop as described above for the synchrotron pair-production case \citep{mehlhaff_21}.
QED PIC simulations by \citet{mehlhaff_24} showed substantial differences between the observable signatures of reconnection in the Klein-Nishina and Thomson regimes. 
The latter exhibits pronounced harder-when-brighter trends; the former displays a stable spectral shape independent of the overall brightness. This spectral stability is reminiscent of GeV high states in flat-spectrum radio quasars~(FSRQs).

While the recent progress has been impressive, several lines of research are still largely unexplored: 
(\textit{i}) the regime of moderate or large optical depths;
(\textit{ii}) pair annihilation
\citep[][]{nattila_24};
(\textit{iii}) synchrotron self-absorption, which is typically important at low frequencies and may introduce some effective collisionality for low-energy pairs \citep[][]{nattila_24}; 
(\textit{iv}) multi-body reactions, which are neglected because rarer, since their cross section is a higher power of the fine structure constant;
(\textit{v}) hadronic processes, which have important implications for neutrino production; pairs produced by the electromagnetic cascade following photo-meson or Bethe-Heitler interactions might self-regulate the reconnection dynamics and energetics.

\section{KINETIC RR WITHIN MHD-SCALE MODELS}\label{sec:turb}
The results described in \sect{micro} and \sectn{radiation} have been obtained from local studies of~RR, starting from the Harris sheet equilibrium. This section covers cases where reconnection layers are not set up as initial conditions, but rather emerge self-consistently as a result of MHD-scale dynamics. In \sect{local}, we describe the formation of thin  layers as a generic by-product of the relaxation of magnetically dominated systems, ranging from the loss of force-free magnetostatic equilibria, to fundamental MHD instabilities (kink, Kelvin-Helmholtz, Rayleigh-Taylor), to fully developed MHD turbulence. In \sect{global}, we present evidence of RR from global fully kinetic numerical models of astrophysical systems such as pulsar magnetospheres, pulsar winds, and black-hole magnetospheres. In the classification proposed in \sect{mhd}, the former are case-(B) layers (\sect{local}), while the latter are case-(A) layers (\sect{global}).

\subsection{Kinetic RR Driven by MHD Dynamics}
\label{sec:local}
The development of small-scale current sheets in MHD turbulence is a natural by-product of dynamic alignment---turbulent fluctuations can be viewed as current sheets that get progressively more anisotropic at smaller scales, where they can become tearing-unstable \citep{loureiro_17, mallet_17,comisso_18b}. 
While such small-scale layers may dominate overall plasma (thermal) heating, a disproportionally large fraction of energy dissipation occurs in intense intermittent large-scale current sheets, comparable in length to the turbulence driving scale \citep{zhdankin_13}. 
Recent PIC simulations of magnetically dominated turbulence have emphasized the key role of these large current sheets in nonthermal particle acceleration \citep{comisso_18,comisso_19,zhdankin_20}.\footnote{Similar arguments apply to the relaxation of  force-free magnetostatic equilibria \citep{yuan_16,lyutikov_17a,lyutikov_17b,lyutikov_17c}.} Particles are first energized in such layers, and then further accelerated by stochastic interactions with the turbulent fluctuations. The filling fraction and lifetime of the longest current sheets is such that they can process $\sim 10\%$ of the turbulent volume in one large-eddy turnover time.
For fluctuating fields weaker than the mean field at the driving scale, these large-scale layers possess a significant guide-field component.
The work done by parallel electric fields---naturally expected in reconnection layers with strong guide fields---is responsible for most of the initial energy increase, while the subsequent energy gain is governed by the perpendicular electric fields of turbulent fluctuations. The two-stage acceleration process leaves an imprint in the particle pitch-angle distribution: lower-energy particles ($\gamma\lesssim  \sigma_{{\rm c},s}$) are mostly aligned with the field, while higher-energy particles move preferentially orthogonal to it. The energy dependence of the resulting pitch-angle anisotropy has important astrophysical implications, as discussed in \sect{pitch}.

\begin{figure}
\centering
\begin{minipage}{.49\textwidth}
  \centering
        \includegraphics[width=\textwidth]{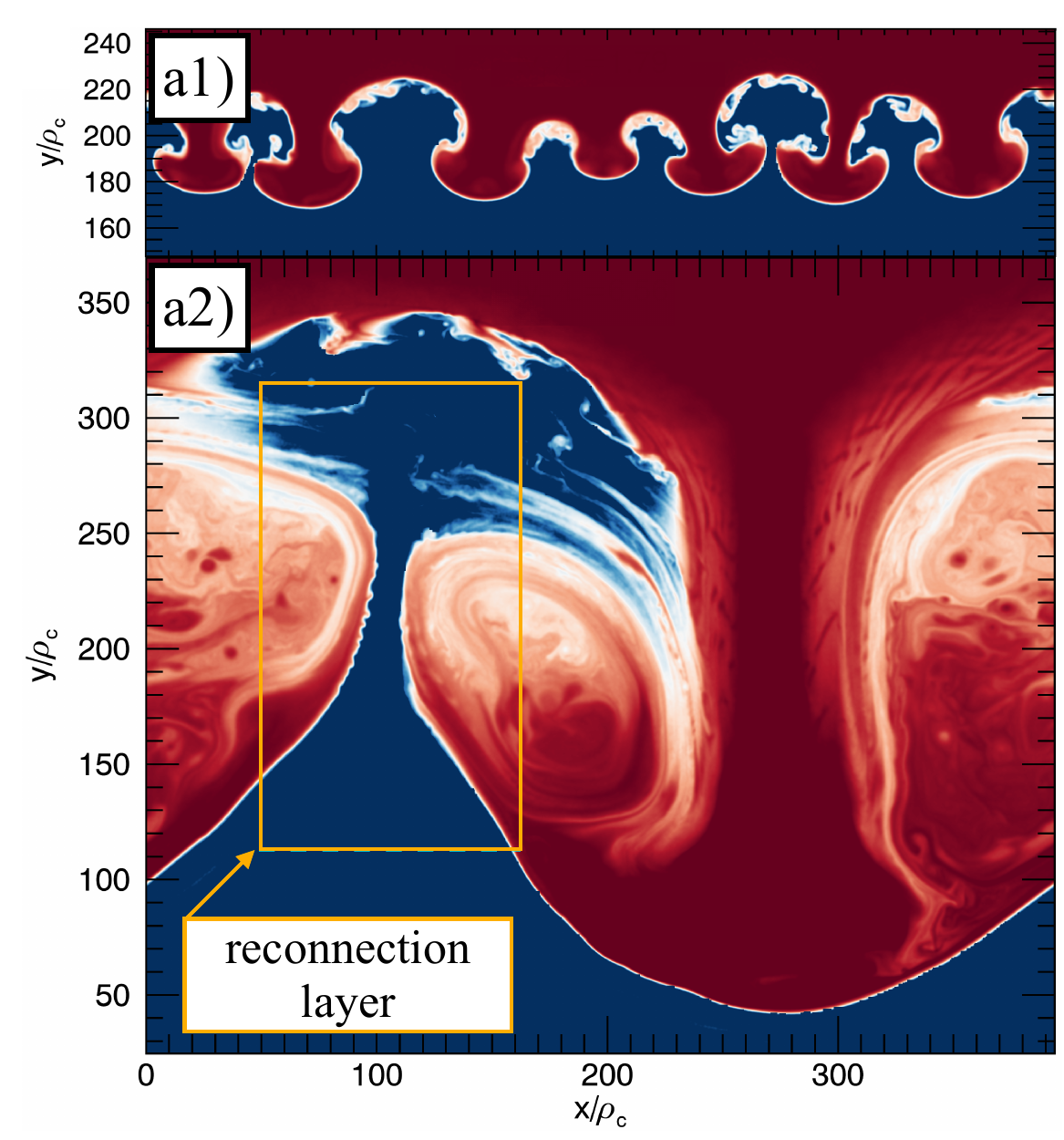}
\end{minipage}
  \begin{minipage}{.49\textwidth}
  \centering
      \includegraphics[width=\textwidth]{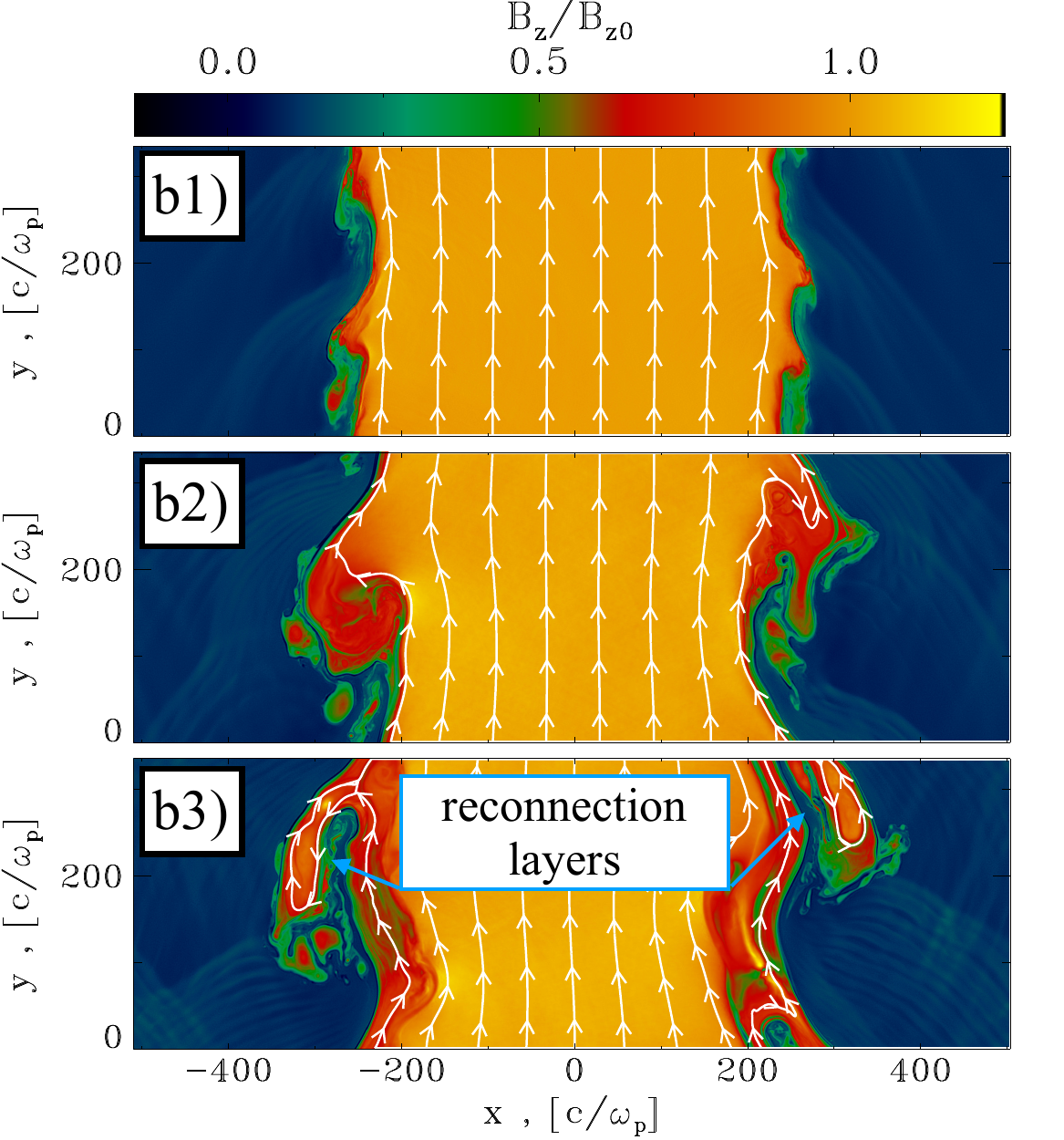}
 \end{minipage}%
    \caption{(a) RT evolution in a 2D PIC simulation \citep{zhdankin_23} having gravitational force in the $-\hat{\boldsymbol{y}}$ direction. The top panel shows small-scale density plumes at early times, while the bottom panel shows large-scale plumes at late times. The location of the self-generated reconnection layer is indicated by the orange box. The unit length is the non-relativistic Larmor radius in the cold, magnetically-dominated region (blue). (b) Temporal evolution (from top to bottom) of the KH instability at the boundary of a relativistic jet, from a 2D PIC simulation with MHD-based initial conditions \citep{sironi_21}. The plot shows the out-of-plane field strength, with in-plane field lines overlaid. The system is initialized with a strongly magnetized jet (yellow central region) moving with respect to the denser, weakly magnetized ambient plasma (blue). The KH waves develop into nonlinear vortices that produce reconnection layers (cyan arrows).}
   \label{fig:rt_kh}
\end{figure}

Similar conclusions hold when large-scale reconnection layers---and potentially MHD turbulence---arise from the nonlinear stages of  MHD instabilities.
In magnetized plasmas, basic MHD instabilities---e.g., kink \citep{alves_18,davelaar_20,ortuno-macias_22}, 
Kelvin-Helmholtz \citep[KH;][]{sironi_21}, magnetorotational instability\citep[MRI;][]{bacchini_22}, Rayleigh-Taylor \citep[RT;][]{zhdankin_23}---inevitably generate reconnecting current sheets, as illustrated in \figg{rt_kh}. This phenomenon is well-established in non-relativistic space and laboratory plasmas and has recently been studied under the relativistic, magnetically dominated conditions relevant to high-energy astrophysics.
Several conclusions hold regardless of the specific MHD instability: (\textit{i}) reconnection occurs in the regime of moderate to strong guide fields, which then implies a preponderance of $E_\parallel$ in the first stages of energization and a pronounced energy-dependent pitch-angle anisotropy; (\textit{ii}) similar to the case of MHD turbulence discussed above, reconnection governs most of the initial energy increase, while the subsequent acceleration mechanism depends on the specific MHD instability:
for kink and~RT, stochastic acceleration in the resulting MHD turbulence; for~KH, systematic shear-driven acceleration in the underlying velocity gradient \citep{rieger_19,sironi_21}.

We conclude this subsection with an important remark. Current PIC studies of reconnection layers resulting from nonlinear MHD dynamics do not possess a dynamic range as large as local Harris-type simulations. In most cases, particle acceleration via reconnection does not seem to achieve Lorentz factors much greater than $\sim \sigmacs$ before another energization mechanism---stochastic or shear-driven acceleration---takes over. 
However, we argue that this limitation is primarily due to the restricted dynamic range of current MHD-scale PIC simulations. We hypothesize that larger boxes will demonstrate that reconnection itself can push particles well beyond~$\sigmacs$---this conjecture may also be explored in RR studies that incorporate some degree of upstream turbulence, as done by \citet{guo_21}. Future work will also clarify the physics that governs the transition between reconnection-driven acceleration and the mechanisms dominating at higher energies.

\subsection{RR in Global Kinetic Models}
\label{sec:global}
In this subsection, we describe global kinetic models of two classes of compact astrophysical sources---pulsars and black holes---and emphasize the role of RR in shaping the electromagnetic structure of their magnetospheres and in producing the observed emission. While local kinetic studies benefit from covering a relatively large dynamic range between the plasma microscopic scales and the length of the reconnection region, they also face significant limitations, e.g., they typically employ an idealized setup that is somewhat disconnected from the system's global properties.
Global PIC simulations overcome these shortcomings by capturing the geometry, orientation, and properties of RR layers from first principles. Unlike fluid approaches (MHD or force-free), they accurately model the reconnection dynamics, including the reconnection rate, and the physics of particle acceleration. However, global PIC simulations often need to sacrifice on the dynamic range, as global scales, such as the neutron star radius or the black hole gravitational radius, are chosen to be only a few orders of magnitude larger than plasma microscales due to computational limitations. Progress in both local and global approaches---and their synergistic interplay---is clearly needed.

\subsubsection{Pulsar Magnetospheres}
\label{sec:pulsars}
Neutron star magnetospheres have gained renewed interest over the last two decades, primarily due to the discovery of pulsed gamma-ray emission, first with the Compton Gamma-Ray Observatory \citep{CGRO_pulsars_99} and later with the Fermi telescope \citep{FERMI2_13}. Remarkably, two of the most famous gamma-ray pulsars---the Crab and Vela---also display pulsed emission at few-to-ten TeV energies \citep{Vela20TeV_23}. The gamma-ray luminosity amounts to a significant fraction ($\sim 0.1-10\%$) of the spindown power, thus requiring a mechanism of efficient energy dissipation that operates in the vicinity of the neutron star. 

A promising candidate is RR in the thin current layer where open magnetic field lines from the northern and southern hemispheres meet beyond the light cylinder. This layer is a persistent feature of the magnetospheric structure [case (A) in \sect{mhd}]. Its properties cannot be studied in isolation from the rest of the magnetosphere (i.e., in a localized Harris-type simulation) because its plasma supply---and consequently, the local magnetization---is tightly coupled to the physics of pair production near the pulsar surface. The first self-consistent kinetic PIC simulations of global pulsar magnetospheres were conducted a decade ago \citep{philippov_14,Chen_14}, with subsequent studies focusing on dissipation and high-energy emission \citep{cerutti_15, cerutti_philippov_16, Kalapotharakos_18, hakobyan_spitkovsky_23}. 
These works showed that the current sheet beyond the light cylinder is plasmoid-unstable, a clear sign of ongoing magnetic dissipation via reconnection. This likely powers the pulsed high-energy emission, as originally proposed by \citet{Lyubarskii_96} and \citet{uzdensky_spitkovsky_14}.

\begin{figure}[t]
\includegraphics[clip,trim=0 50 0 230,width=0.65\textwidth]{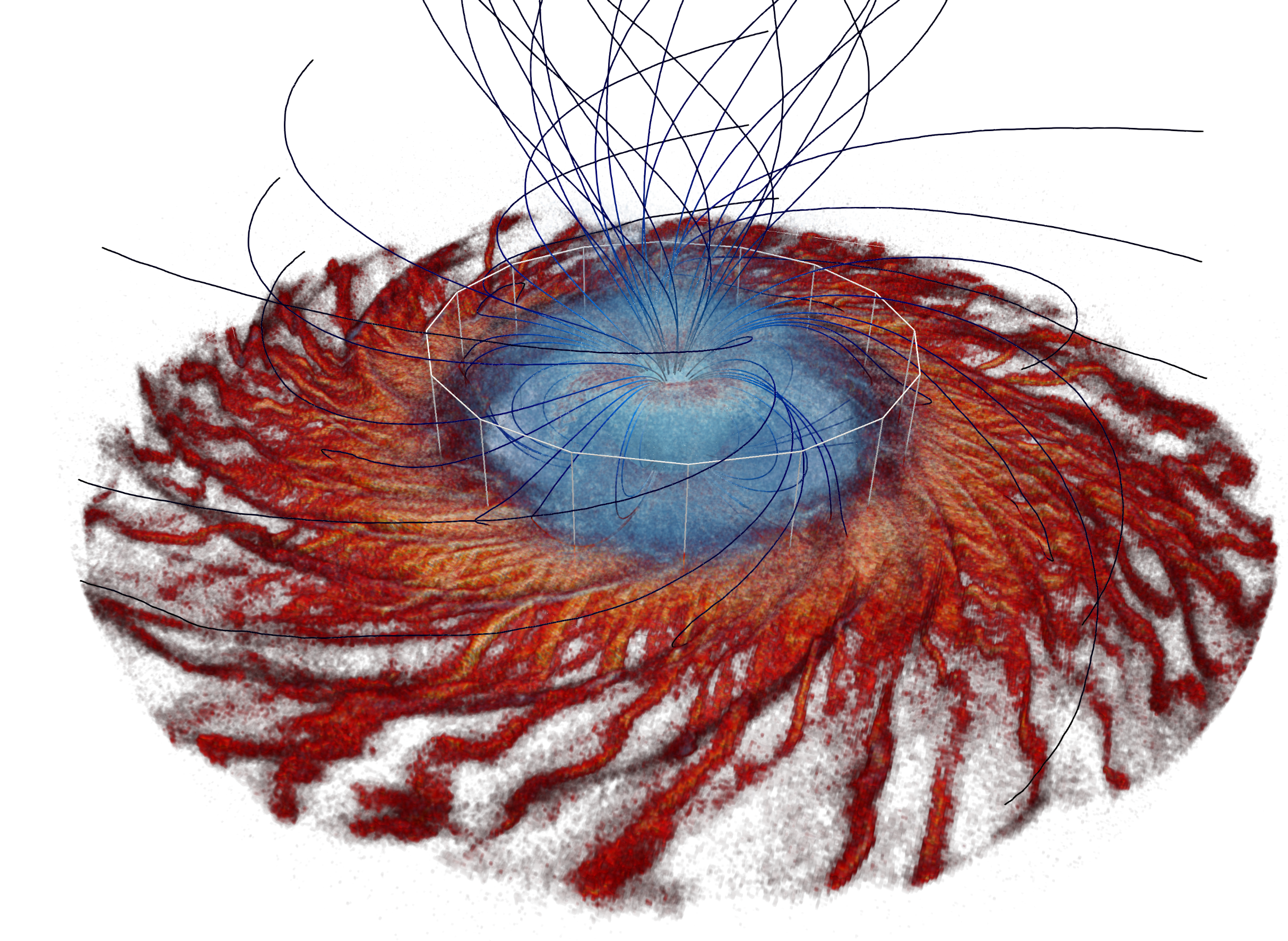}
\caption{Magnetospheric structure of a pulsar with aligned rotational and magnetic axes, from a high-resolution PIC simulation \citep{hakobyan_spitkovsky_23}. The neutron star is in the middle of the box. Volume rendering shows two quantities: the electric current density (in blue) and the pair-plasma density in the reconnection layer (in red), both compensated by a factor of $r^2$. The cylinder in the center indicates the surface of the light cylinder. Blue lines starting from the neutron star surface illustrate magnetic field lines. 
}
\label{fig:3dpsr}
\end{figure}

\figg{3dpsr} shows the 3D structure of a pulsar magnetosphere with aligned rotational and magnetic axes, from a  PIC simulation by~\cite{hakobyan_spitkovsky_23}. 
Reconnection of anti-parallel field lines originating from the two magnetic hemispheres is triggered just beyond the light cylinder. 
There, the magnetic field forms a $45^\circ$ angle with respect to the radial direction, transitioning to a predominantly toroidal geometry further out. The figure reveals prominent overdense tubes (3D flux ropes) that extend radially and propagate outward. The radial undulations in the equatorial current sheet and in the flux ropes are driven by the kinetic drift-kink instability and the flux-rope kink instability, respectively.

PIC simulations have shown that the current layer can dissipate  $\sim1-10\%$ of the spindown power close to the light cylinder \citep{Cerutti_20}. This fraction is insensitive to specific pulsar parameters (e.g., the field strength at the light cylinder), as the dissipation rate only depends on the dimensionless reconnection rate \citep{uzdensky_spitkovsky_14}.\footnote{The fraction of spindown power resulting in gamma rays is, however, sensitive to the obliquity angle between rotational and magnetic axes, with orthogonal pulsars having weaker currents (compensated by the displacement current) and thus lower radiative efficiency.} 
The dissipated magnetic energy is deposited into accelerated pairs, which rapidly radiate it away in the form of synchrotron emission, powering the observed pulsed gamma-ray signal. The emission is beamed along the reconnecting field lines, which allows to reproduce the variety of lightcurves observed in gamma-ray pulsars \citep{bai_10,cerutti_philippov_16}. Particle acceleration and synchrotron emission take place in spatially distinct regions (acceleration happens primarily near X-points, where the magnetic field vanishes, while cooling occurs in plasmoids/flux ropes), which allows to extend the synchrotron spectrum beyond the nominal burnoff limit $\epsilon_{\rm rad}\simeq 16 \,\eta_{\rm rec,-1}\,$MeV \citep{cerutti_12a,cerutti_13}, potentially reaching up to $\sim \epsilon_{\rm rad} (\sigmae/\gammarad)$ \citep{chernoglazov_23}. Since the spectra of most gamma-ray pulsars reach GeV energies, this yields $\sigmae\sim 10-100\;\gamma_{\rm rad}$ at the light cylinder (relativistic parallel bulk flow may also contribute to boosting the emission). IC scattering by energetic pairs with Lorentz factor $\sim \sigmae$ may power the distinct spectral component of pulsed $\sim 20~\textrm{TeV}$ gamma rays recently observed from the Vela pulsar \citep{Vela20TeV_23}.

\subsubsection{Pulsar Winds} Beyond the light cylinder, the pulsar wind propagates as a relativistic, magnetized flow.
Near the equatorial plane of oblique pulsars, the wind starts as a sequence of magnetically-dominated stripes of alternating toroidal field, separated by current sheets---a striped wind. Magnetic dissipation of the stripes starts near the light cylinder, causing the wind to accelerate and thus reducing the dissipation rate \citep{lyubarsky_kirk_01,kirk_03,petri_05}. 
In the case of the Crab pulsar, early analytical works \citep{lyubarsky_kirk_01,kirk_03,petri_05} concluded that the wind should still be Poynting-dominated at the radius at which a standing ``termination'' shock is inferred from observations. 
This question was recently revisited with 2D \citep{cerutti_philippov_17} and 3D \citep{Cerutti_20} global PIC simulations extending up to several tens of light-cylinder radii. 
They found that plasmoid-dominated reconnection consumes the field efficiently at all radii, even past the fast magnetosonic point. Contrary to earlier findings, their results suggest there is a universal dissipation radius solely determined by the reconnection rate, $R_{\rm diss}\sim \exp(\eta_{\rm rec}^{-1})\,R_{\rm LC}\sim 10^2-10^4\,R_{\rm LC}$. In isolated pair-producing pulsars like the Crab, this lies well within the termination shock radius ($R_{\rm sh}\sim 10^{9} R_{\rm LC}$), i.e., the striped component of the wind dissipates well before reaching the termination shock.

The situation is different in the case of pulsars in binary systems, including spider pulsars,
compact binary systems harboring a millisecond pulsar and a low-mass companion. The relativistic magnetically dominated  pulsar wind impacts onto the companion, ablating it and slowly ``devouring'' its atmosphere---hence, the evocative name of these systems. The interaction between the pulsar wind and the companion's stellar wind (or its magnetosphere) forms an intrabinary shock. The shock is at $R_{\rm sh}\sim 10^{2}-10^{4} R_{\rm LC}$, so the pulsar wind arrives at the shock still retaining its striped structure. 
Spider pulsars offer a unique opportunity for global PIC simulations. The shock curvature radius is just three orders of magnitude larger than the characteristic Larmor radius of post-shock particles---a range within reach of modern PIC simulations. With global PIC simulations of the intrabinary shock, \citet{cortes_sironi_22,cortes_sironi_24} found that the magnetic stripes compress at the shock and annihilate via shock-driven reconnection---a mechanism studied previously in local PIC simulations by, e.g., \citet{sironi_spitkovsky_11,lu_21}. Particles accelerated via shock-driven reconnection produce synchrotron spectra and light curves that are in good agreement with X-ray observations \citep{cortes_sironi_22,cortes_sironi_24}.

\subsubsection{Black Hole Magnetospheres}\label{sect:bhmag}
Relativistic magnetospheres can form around spinning black holes in the presence of a large-scale magnetic field. They differ from pulsar magnetospheres in two ways, both due to the existence of an event horizon: (\textit{i}) the magnetic field cannot be maintained indefinitely on an isolated black hole, by virtue of the no-hair theorem \citep{nohair}; and (\textit{ii}) the black hole cannot directly supply plasma. Instead, both the magnetic field and the plasma must come from external sources---e.g., an accretion disk or a companion star---meaning that an isolated black hole magnetosphere cannot exist. In the presence of an external force-free plasma, \citet{blandford_77} showed that the rotation of the black hole event horizon along with the surrounding space-time (through the Lense-Thirring effect) acts like a unipolar inductor, generating strong electric currents in the magnetosphere.
In turn, these currents lead to a global reorganization of the magnetic fields, similar to pulsar magnetospheres. Unlike in pulsars, this rearrangement primarily impacts the innermost regions: field lines threading the ergosphere wind up around the spin axis and are pulled into the event horizon---in a sense, this process mirrors the pulsar current layer, but ``inside-out.'' The resulting field structure resembles a split-monopole configuration, with an equatorial current sheet separating the two hemispheres and carrying the return current. This layer is macroscopic [case (A) in \sect{mhd}], extending from the horizon out to a few gravitational radii, roughly bounded by the ergosphere.

\begin{figure}[t]
\centering
\begin{minipage}{.49\textwidth}
  \centering
        \includegraphics[width=0.85\textwidth]{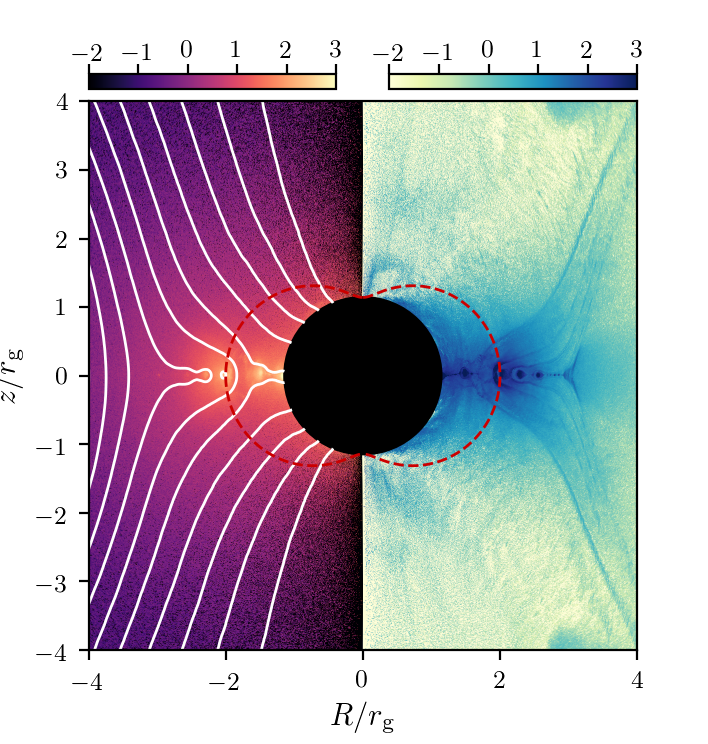}
\end{minipage}
  \begin{minipage}{.49\textwidth}
  \centering
      \includegraphics[width=0.85\textwidth]{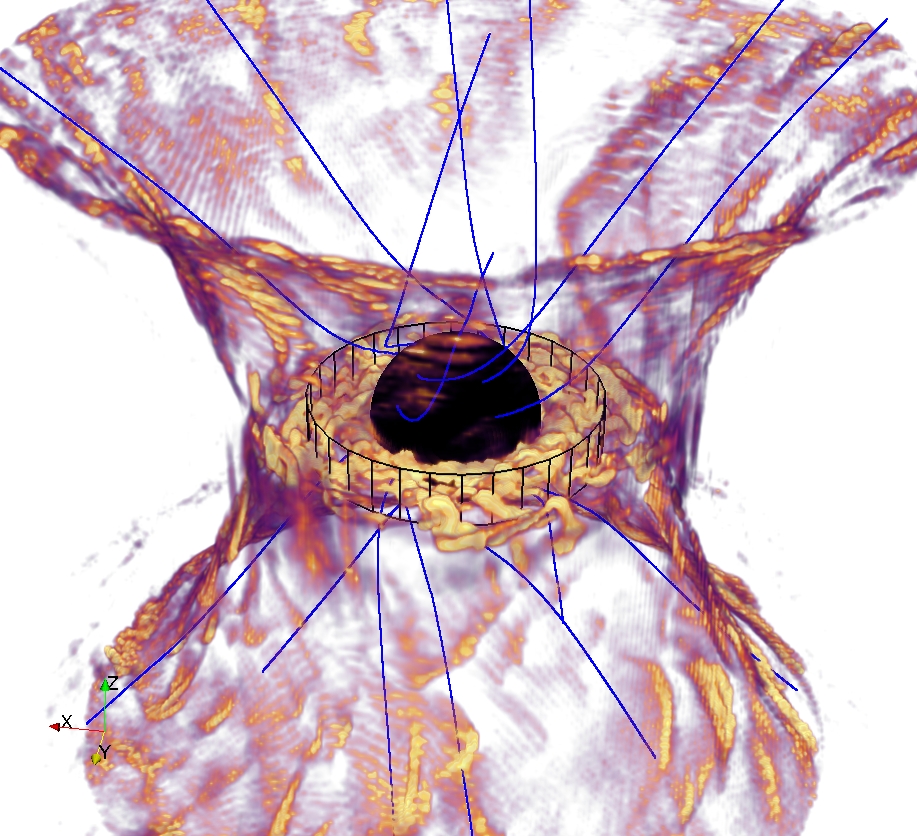}
 \end{minipage}%
\caption{Structure of an aligned black hole magnetosphere for a nearly maximally spinning black hole, from global general relativistic radiative PIC simulations. {\bf Left:} 2D axisymmetric simulation showing the logarithm of the density of gamma-ray photons (in color) and the poloidal field lines (white lines) on the left side and the logarithm of the pair density on the right side (adapted from \citealt{crinquand_21}). Both densities are normalized to the Goldreich-Julian density. {\bf Right:} Volume rendering of the plasma density with magnetic field lines (blue lines) from a 3D simulation (adapted from \citealt{crinquand_22}). The wire-frame cylinder represents the outer edge of the ergosphere.}
\label{fig::bhmag}
\end{figure}

The existence of the ergospheric current layer was first reported in GRMHD studies \citep{komissarov_04, komissarov_07a}. While the importance of reconnection there was expected
\citep{lyutikov_mckinney_11}, it was firmly established with the advent of global general-relativistic PIC (GRPIC) simulations \citep{parfrey_19, crinquand_20, crinquand_21, bransgrove_21, crinquand_22, elmellah_22, elmellah_23, galishnikova_23, niv_23}.  
Figure~\ref{fig::bhmag} shows the overall structure of a black hole magnetosphere in 2D (left) and 3D (right), when the external magnetic field is aligned with the black hole spin axis. The simulations reveal a thin equatorial layer hosting a chain of magnetic islands / flux ropes, along with a powerful Poynting-flux dominated polar outflow (the base of the Blandford-Znajek jet). PIC simulations have emphasized the critical role of RR at the ergospheric current sheet for particle acceleration, gamma-ray emission, and copious pair creation. \citet{crinquand_21, crinquand_22} demonstrated that reconnection could convert a sizeable fraction ($\sim \eta_{\rm rec}$) of the Blandford-Znajek power into high-energy radiation, potentially explaining the rapid TeV gamma-ray flares observed from M87 \citep{hakobyan_ripperda_23} (see, however, 
\citealt{chen_23}). Sub-horizon scale dynamics associated with the largest flux ropes may leave detectable signatures in the image plane, such as moving rings or hotspots \citep{crinquand_22}. \citet{parfrey_19} and \citet{comisso_21b} showed that reconnection within the ergosphere facilitates black-hole spin energy extraction via the Penrose process. The reconnection electric field produces energetic particles and fast plasmoids with negative energy at infinity, which later fall into the black hole, reducing its spin.

As described in \sect{mhd}, the formation of an equatorial current sheet is also a key feature of accretion flows in the MAD state. MAD phases correspond to a saturation of the magnetic flux brought to the black hole by accretion, and they are generally followed by rapid outward diffusion of the excess flux, mediated by reconnection in the equatorial layer. Simulations show that the flux decays exponentially at a rate determined by the reconnection speed \citep{crinquand_21, bransgrove_21}. The much faster rates observed in PIC simulations than in MHD models \citep{bransgrove_21, galishnikova_23} may qualitatively change our  understanding of MAD accretion, which is currently based on MHD approaches. However, GRPIC simulations of black hole accretion are still in their infancy, as they face severe numerical challenges \citep{galishnikova_23}. 

Black hole magnetospheres can also host reconnection layers at higher latitudes. Differential rotation between the black hole event horizon and the inner parts of the accretion flow leads to inflation and opening of the magnetic field lines connecting them \citep{uzdensky_05, parfrey_15, yuan_19}, creating a conical current sheet that separates the jet from the surrounding disk corona or wind. Such a configuration was reproduced in 2D and 3D PIC simulations by \citet{elmellah_22, elmellah_23} in a rather idealized setup, treating the disk as a perfectly conducting rotating boundary. The simulations showed that a large-scale conical current sheet forms and reconnects, leading to a bright sheath-like layer at the interface between the jet and the disk.  Large flux ropes formed in the layer may explain the orbiting hot spots observed by the GRAVITY Collaboration from Sgr~A* \citep{gravity_18}.
More broadly, this process---where reconnection layers form as a result of field-line twisting and opening due to differential rotation---could be operating pervasively above the disk, powering coronal-like activity \citep{uzdensky_goodman_08}.

\section{ASTROPHYSICAL IMPLICATIONS OF RR}
\label{sec:astro}
The radiative signatures of relativistic astrophysical sources exhibit  extreme properties, such as high radiative efficiency, rapid variability, hard photon spectra, and emission exceeding the synchrotron burnoff limit. As outlined in \sect{intro}, these observations pose significant challenges to most particle acceleration mechanisms. In this section, we illustrate how RR has provided solutions to a number of observational puzzles. We frame this discussion in terms of the ``unfair advantages'' RR holds over other dissipation mechanisms and we provide a few key examples of RR-based models applied to astrophysical sources.

\bi
\smallskip
\item {\bf High efficiency.} The radiative efficiency (i.e., the fraction of the system's power converted to radiation) can exceed 10\% in AGN jets, gamma-ray bursts and gamma-ray emitting pulsars---implying a very efficient mechanism for energy dissipation and particle acceleration. In magnetically dominated systems, the dissipative efficiency of case-(A) reconnection layers---namely, the fraction of source luminosity injected into relativistic particles---can approach unity. The portion of the dissipated power that is converted to radiation depends on the model details and the properties of the source. In striped-jet models of GRBs, \cite{drenkhahn_02b} estimated that roughly half of the dissipated magnetic energy is converted into high-energy particles. For typical model parameters, the jet radiative efficiency is quite high, in the $\sim3-50\%$ range. Similarly, RR operates efficiently in equatorial current sheets outside the light cylinder in pulsar magnetospheres, where it dissipates $1-10\%$ of the spindown luminosity \citep[e.g.,][]{Cerutti_20}, powering the observed pulsed gamma-ray emission. 

\smallskip
\item {\bf Broken power-law spectra with  hard low-energy slopes.} In several cases, such as the quiescent emission from high-synchrotron-peak blazars or the flaring activity of the Crab Nebula, the inferred particle spectra are very hard, with spectral indices $p<2$ \citep[e.g.,][]{celotti_08,buehler_12}. While shocks can, under certain conditions, serve as nonthermal particle accelerators \citep[see][for a review of relativistic shocks]{sironi_keshet_15}, they typically produce steeper particle spectra, with $p>2$.
As discussed in \sect{plaw}, RR naturally produces broken power-law energy spectra, with hard ($p$ approaching~1) slopes in the low-energy range. Broadband modeling of blazar emission often requires hard ($p < 2$) electron spectra below GeV energies, followed by a break and a steeper power law with $p \gtrsim 2$ at higher energies \citep{celotti_08,sikora_09}. Such spectra can thus be convincingly explained by the findings of PIC studies of~RR.

\smallskip
\item {\bf Fast time variability.} Many relativistic sources display extremely fast variability, with timescales much shorter than the light-crossing time of the central engine, e.g., $\sim 10$ minute-long blazar flares in the GeV-TeV band \citep{aharonian_07,albert_07,ackermann_16} or infra-day gamma-ray flaring from the Crab Nebula \citep{abdo_11,tavani_11}. Internal shocks in jets exhibit temporal variability that closely follows the one imprinted by the central engine \citep[e.g.,][]{nakar_piran_02}, which fails to explain the ultra-fast blazar flares. In contrast, RR naturally results in rapid flares, for two main reasons. First, reconnection layers are easily fragmented into chains of fast-moving plasmoids, with sizes up to 10\% of the system length. The flare duration is set by the plasmoid size (so, much shorter than the system's light-crossing time), further compressed by the Doppler effect due to the plasmoid fast bulk motion. Synchrotron and IC emission from fast-moving plasmoids can power the bright, fast-evolving flares observed in blazars and GRBs \citep{lyutikov_03,giannios_09, giannios_13, petropoulou_16,christie_19}.\footnote{Trans-relativistic plasmoid bulk motions may also power the hard X-ray emission of black hole binaries and AGN \citep{beloborodov_17,sironi_beloborodov_20,sridhar_21,sridhar_23}.}  
Second, fast variability is enhanced by kinetic beaming---strong energy-dependent anisotropy of reconnection-accelerated particles and their emission, see \sect{pitch}. RR produces tightly focused, swaying beams of energetic particles, which give rise to sharp, intense spikes of high-energy radiation when they cross the observer's line of sight \citep{uzdensky_11,cerutti_12b,mehlhaff_20,sobacchi_23}.

\smallskip
\item {\bf Rapid polarization swings.} Rapid polarization swings observed from blazar jets indicate polarity reversals at the dissipation region, constraining the magnetic field geometry \citep[e.g.,][]{blinov_15}. Large-scale polarization swings can be reproduced in PIC simulations of RR with fast synchrotron cooling \citep{zhang_20,zhang_22,hosking_20}, demonstrating that plasmoid mergers govern both the emission of bright flares and rapid polarization swings. This naturally explains why high-energy flares in blazar jets are often coincident with large polarization sweeps in the optical band \citep[e.g.,][]{blinov_15}.  

\smallskip
\item {\bf Emission above the synchrotron burnoff limit.} Gamma-ray flares from the Crab Nebula are believed to be of synchrotron origin, with emission extending to hundreds of~MeV, thus exceeding the nominal synchrotron burnoff limit $\epsilon_{\rm rad}\simeq 16\,\eta_{\rm rec,-1}\,$MeV \citep{abdo_11,tavani_11}. PIC simulations have demonstrated that in the strong cooling regime electrons can accelerate beyond the standard burnoff Lorentz factor by moving nearly along the magnetic field  (or where the field vanishes), which suppresses synchrotron losses \citep{cerutti_13,cerutti_14a,chernoglazov_23}. The resulting synchrotron spectrum then extends beyond the burnoff limit.

\smallskip
\item {\bf Bright coherent radio emission.} 
RR can produce coherent (typically, radio) emission \citep{uzdensky_spitkovsky_14}.
The time-varying electric currents sourced at the interface of merging plasmoids  generate fast-mode waves that propagate back into the upstream region. Upon converting to electromagnetic waves, they have been invoked to explain the pulsar radio ``nanoshots,'' coming in
phase with higher-energy radiation \citep{lyubarsky_19,philippov_19}, and they have been proposed as a candidate emission mechanism for FRBs \citep{lyubarsky_20,mahlmann_22}. 

\smallskip
\item {\bf Fast cosmic ray acceleration and associated neutrino production.} High-energy particles in RR are accelerated by meandering between the two sides of the current layer, confined in between the converging upstream flows (see \sect{plaw}). This results in fast acceleration, with $\dot{\varepsilon}\simeq e\, \eta_{\rm rec} B_0 c$. Hadrons entering the acceleration process suffer modest radiative losses, so their maximum energy is limited by the length $L$ of the reconnection layer: $\varepsilon_{\rm max}\sim e\,\eta_{\rm rec} B_0 L$. Large-scale RR layers can thus accelerate protons up to $10^{20}$~eV in GRB and powerful AGN jets, while iron nuclei can reach similar energies in AGN jets of more moderate luminosity \citep{giannios_10,zhang_21}. Protons with $\sim$$100$~TeV energies accelerated by RR in AGN coronae may produce the TeV neutrinos \citep[][]{fiorillo_24} observed by {\it IceCube} from Seyfert galaxies \citep{icecube_22}. RR is, therefore, a potent hadronic accelerator, making it a promising mechanism for cosmic-ray acceleration and associated neutrino emission.

\ei
In summary, analytical and numerical insights into the physics of~RR, especially in the radiative regime, have helped solve a number of outstanding observational puzzles, allowing us to deepen our understanding  of the engines powering the most dramatic multi-messenger manifestations of our high-energy Universe.

\section{SUMMARY AND FUTURE PROSPECTS}
\label{sec:summary}
This review reports recent advances in our understanding of the kinetic physics of relativistic magnetic reconnection, with focus on astrophysical motivations and implications. After discussing foundational equations and basic parameters of RR (\sect{basic}), we have presented the widespread occurrence of RR layers in astrophysical high-energy sources (\sect{mhd}). 
After reviewing the modern fluid-level picture of reconnection, we have described how recent kinetic simulations have allowed to assess how RR partitions the dissipated energy among different species, and, within each species, between bulk, thermal and nonthermal components (\sect{micro}). In high-energy astrophysical sources, the interplay between photons and reconnection-accelerated particles plays a major role in controlling the resulting nonthermal signatures (\sect{radiation})---this regime, of radiative RR, represents a new and exciting frontier of plasma astrophysics. 
While most reconnection studies focus on local, isolated layers, recent works have highlighted the role of RR as a by-product of {large-scale MHD dynamics, including MHD instabilities,} and as a key component of global kinetic models of high-energy sources (\sect{turb}). Finally, we have emphasized why RR is uniquely positioned to address a number of observational puzzles (\sect{astro}). Although this review mainly highlights the implications of RR for astrophysical high-energy sources, Sections \sectn{micro} and \sectn{turb}, which focus on the microphysics of RR and its interplay with other fluid instabilities, will also interest the {space-physics and} laboratory communities. In the remainder of this section, we outline a few outstanding theoretical questions and present the landscape of future prospects in RR research, emphasizing the role of observational facilities, laboratory experiments, and novel computational tools.

\subsection{Outstanding Theoretical Questions}

The impressive progress achieved in recent years
opens new tantalizing questions that will guide research efforts in the next decade. In this subsection we discuss some of these critical questions, as well as possible avenues for addressing them.

(1) {\bf Reconnection onset} in global astrophysical contexts. 
Reconnection-powered flares often have a sharp, sudden onset after a prolonged period of quiescence.
This implies that reconnection onset requires some special conditions, e.g., the formation of thin and long current sheets. How such current sheets form, and how and where reconnection is triggered, are crucial questions in plasma astrophysics. While theoretical progress has recently been made in the non-relativistic case, with applications to the solar corona and the Earth's magnetosphere \citep[e.g.,][]{uzdensky_loureiro_16,comisso_etal_16}, how reconnection is triggered in the relativistic environments of black holes and neutron stars---key to predicting the recurrence of reconnection-powered flares---remains poorly understood.

(2) {\bf Interplay of RR with other collective plasma processes} (e.g., instabilities).
The same MHD-scale processes that lead to the formation of current sheets in realistic environments should also create other types of quasi-2D singular structures/discontinuities, e.g., velocity shear layers and shock fronts. These structures may be susceptible to their own instabilities, which then interact with reconnection. The interplay of reconnection with other instabilities has a long history in space and laboratory plasma physics. In RR, this question is even richer, in regard to the dynamics of accelerated particles: e.g., in magnetically dominated environments, the first stages of particle acceleration often occur within reconnection layers; at higher energies, RR-accelerated particles can sample the surrounding flow and electromagnetic-field structure and get energized by other mechanisms. 
By understanding
 where {and how} the transition occurs, i.e., at which point energetic particles leave their ``birth'' layer and get accelerated by larger-scale fluid motions in a Fermi-type process, we will deliver solid predictions on the resulting spectra of particles and photons.

(3) \textbf{RR in extreme astrophysical environments} with intense radiation, pair production, and hadronic losses. Radiative RR with synchrotron losses has been studied in the regime of moderately strong cooling, $0.1\sigmae\lesssim \gamma_{\rm rad} \lesssim \sigmae$, but the case of extremely strong cooling \citep[e.g., $\gamma_{\rm rad}/\sigmae\sim 10^{-6}$ in magnetar winds during an outburst; e.g.,][]{mahlmann_22} is still unexplored. Reconnection with IC drag has been investigated only in the case of an uniform isotropic soft-photon background; if seed photons come from a specific direction, {e.g., in the corona of accretion disks,} they can exert a net force on the layer, driving additional instabilities (e.g., the Kruskal–Schwarzschild instability; e.g., \citealt{lyubarsky_10,groger_24}) which may help triggering reconnection onset. {In addition, the case of synchrotron self-Compton radiation, particularly important for, e.g., blazar jets, remains largely unexplored.}

QED effects such as pair production have been investigated only in the small or moderate optical-depth regime, where the secondary pairs deposited in the upstream inflow region may self-regulate the upstream magnetization \citep[e.g.,][]{schoeffler_19, hakobyan_19, mehlhaff_21, chen_23, hakobyan_ripperda_23, mehlhaff_24}. At large optical depths, however, the produced pairs are produced inside the layer, preventing them from modifying the far-upstream plasma conditions. This regime may be relevant to magnetar flares, GRBs, and neutron-star mergers \citep{uzdensky_11, beloborodov_21}. If the optical depth to Compton scattering is sufficiently large, photons emitted from the layer may exert an outward-directed radiation pressure force on the inflowing plasma, affecting reconnection. Also, strong-field QED processes (e.g., resonant Compton scattering, photon-splitting, etc.), that are important in magnetar magnetospheres \citep{2018ApJ...869...44K}, have yet to be incorporated in PIC studies. All of these effects remain largely unexplored, and yet they are essential for building realistic models of high-energy sources.

The regime of intense hadronic losses is also uncharted territory. High-energy ions cool via synchrotron emission or through interactions with the local radiation field---either sourced externally or resulting from losses of RR-accelerated leptons. Photo-meson interactions can produce high-energy neutrinos. Electron-positron pairs produced by electromagnetic cascades following photo-meson or Bethe-Heitler interactions might self-regulate the upstream magnetization. Progress in modeling RR in the presence of strong hadronic losses is clearly needed to build physically motivated models of multi-messenger sources.

\subsection{Future Prospects: Observational Facilities}
As discussed in \sect{astro}, several astrophysical high-energy sources exhibit telltale signatures of RR, but the theoretical interpretation is often not unique. Observations are frequently spotty, and the instruments lack the sensitivity needed to convincingly distinguish between competing theoretical models. In this section, we briefly outline advances in observational capabilities that will enable progress in the field.

Energization mechanisms are strongly constrained by the maximum rate at which they can accelerate particles. While the synchrotron spectrum of the highest-energy leptons accelerated by most collective plasma processes usually does not exceed the standard burnoff limit of tens of MeV \citep{dejager_harding_92}, it can reach significantly higher energies in~RR \citep[e.g.,][]{uzdensky_etal_11, cerutti_12a}. This makes this spectral range particularly important; however, the  observational capabilities in this band have seen little advancement over the past two to three decades. More sensitive gamma-ray instruments, particularly in the $1-100$~MeV range, are needed to probe gamma-ray spectra, variability, and high-energy cutoffs in sufficient detail and across a large sample of sources. This would allow for the identification of systematic trends and provide more stringent constraints on existing theoretical models.

PIC simulations of RR have shown that particles of different energies may be accelerated by distinct processes and cool in different regions of the reconnection layer, making their radiative signatures highly energy-dependent.
Observations in a single spectral band, while useful, lack the dynamic range necessary to effectively constrain the accelerator. More concerted efforts on simultaneous multi-wavelength (and polarimetric) observations could help bridge this gap. Coordinated radio-to-gamma-ray monitoring of flares and polarization swings could reveal how the accelerator operates  across several decades in energy.

By constraining the field geometry in the dissipation region, polarimetric observations are a key test of RR-powered emission. For example, 
the uncertainty on the dissipation distance in GRB jets (e.g., optically thick versus optically thin conditions) limits progress. The two leading models, photospheric Comptonization and synchrotron emission, predict a markedly different degree of polarization \citep[see, e.g.,][]{kumar_15}. Although RR can power GRB emission in both synchrotron and photospheric models \citep{lyutikov_03,giannios_08},  theoretical/numerical studies of the predicted polarization signatures can be used to distinguish between the two leading models. Thus, a sensitive gamma-ray polarimeter could help pinpoint the prompt GRB emission mechanism and the physical conditions in the dissipation region \citep[for upcoming missions, see][]{bozzo_24}.

After more than a century, the sources of high-energy cosmic rays remain elusive. The main challenge is that the arrival direction of charged particles does not point back to their origin (except possibly for the highest-energy~CRs).
At lower energies, {\it Icecube} has tentatively identified Seyfert galaxies as sources of TeV neutrinos \citep{icecube_22}.
Improved neutrino and ultra-high-energy cosmic rays (UHECRs) statistics will be crucial in identifying candidate CR accelerators. 
Furthermore, correlated observations of neutrino and electromagnetic flares can further solidify these identifications, impose stringent constraints on the underlying particle accelerators, and enable a firm establishment (or refutation) of RR as a potential or even dominant source of these energetic particles.

\subsection{Future Prospects: Laboratory Experiments}
\label{sec:lab}
Laboratory investigation of RR has been lagging behind both its non-relativistic counterpart and recent theoretical advances. A key obstacle is the lack of an experimental platform capable of producing relativistic plasma or sufficiently strong magnetic fields to drive relativistic flows. The main challenge is creating relativistic conditions on macroscopic scales---much larger than the Debye length or the skin-depth---which is needed for collective plasma processes like reconnection. Developing macroscopic relativistic plasma is thus the crucial first step for meaningful experimental studies of~RR. Though difficult, several promising approaches have recently emerged.

Since accelerating ions to relativistic speeds is far more challenging than accelerating electrons, most proposed schemes for laboratory relativistic-plasma studies focus on relativistic electrons (and possibly positrons). A promising near-term approach aims to achieve electron-only~RR, where ions serve as a stationary neutralizing background while electron flows can reach relativistic speeds. Modern and upcoming next-generation high-intensity lasers, including multi-petawatt lasers, offer the most promising path toward this goal. 
The most common approach involves focusing two laser pulses onto adjacent spots on a solid target, producing two expanding plasma plumes. Azimuthal magnetic fields, generated around each focal point by the Biermann battery, are naturally anti-parallel between the two plumes; hence, when the plumes expand and collide, a reconnecting current sheet forms at the interface. Many non-relativistic reconnection experiments have used this scheme at moderate laser intensities \citep[e.g.,][]{Stamper_PRL_1978,Nilson_PRL_2006, li07, willingale10, zhong10, fiksel14, zhong16, raymond_18, fox20, fiksel21, fox21, ping23}.
Pioneering studies of electron-only RR by \citet{raymond_18} employed a similar configuration but with high-intensity lasers. When an intense laser pulse hits a solid target, it heats a large number of electrons to relativistic energies in a small volume. These electrons fly outward, creating time-varying charge-separation sheath electric fields that expand along the target surface. The electron current generates a strong ($\sim$$10^{9}\,$G) azimuthal magnetic field \citep{Tatarakis_Nature_2002}---crucial for achieving high electron magnetization, a key requirement of~RR. Other proposed schemes involve splitting laser pulses with a thin foil \citep{Yi_NC_2018} or using near-critical density targets \citep{Gu_SR_2019}.

Ultimately, however, electron-only reconnection experiments are confined to relatively small (sub-ion) spatial and temporal scales, undermining their potential to probe RR in the large-scale, plasmoid-dominated regime and to study reconnection-driven nonthermal particle acceleration. Moreover, the common approach of using two colliding plasma plumes in High-Energy-Density (HED) experiments is naturally suited for high-beta reconnection, which is less favorable for RR-driven particle acceleration.

A different approach, aimed at studying low-beta reconnection in HED experiments, uses laser-powered capacitor coils \citep{pei16,yuan18,chien19,chien23,yuan23,zhang23}.
A recent experimental breakthrough employing this platform has demonstrated nonthermal electron acceleration by the reconnection electric field \citep{chien23}. 
Experimental schemes aimed at studying other acceleration mechanisms, e.g., first-order Fermi acceleration in the plasmoid-dominated regime, have been proposed using a multiple-coil geometry \citep{ji24}.
Looking ahead, upcoming higher-power, multi-petawatt lasers, in conjunction with optimized capacitor-coil setups, 
bring the relativistic-electron regime of magnetic reconnection within the reach of laboratory studies in the coming years \citep{law16,chien23}.

In summary, experimental investigations of magnetic reconnection in the relativistic regime are still in their infancy but are advancing rapidly. This progress is fueled by new and upcoming multi-petawatt laser (and other) facilities with enhanced diagnostics, state-of-the-art computational support, and strong motivations from fundamental physics and astrophysics. We are optimistic about substantial advancements in this exciting frontier.

\subsection{Future Prospects: Computational Tools}
PIC simulations have been the primary driver of recent advances in~RR. However, PIC models---particularly global PIC simulations of compact object magnetospheres---often suffer from a lack of scale separation between macroscopic scales and plasma scales.
Several prospects could enhance the capabilities of PIC simulations moving forward. The most important leap comes on the hardware side, with the advent of GPUs and reduced-instruction-set CPUs (such as ARM and RISC-V). As these technologies become more available, and integrate larger memory and higher communication bandwidth, the community is steadily transitioning to adapting existing PIC algorithms to this new hardware.

On the algorithmic side, curvilinear grids \citep[see, e.g.,][]{2015NewA...36...37B,Chen_14}, even in general-relativistic space-times \citep{parfrey_19}, are becoming more widely implemented. This approach allows global simulations of neutron-star and black-hole magnetospheres to extend their dynamic range, while still retaining sufficient spatial resolution. However, 3D spherical grids have a coordinate singularity at the polar axis, which makes results in this region inaccurate. The ultimate solution for simulating systems with spherical symmetry would be to implement in PIC a patched-spherical geometry, such as the cubed-sphere, which is already used in hydrodynamic \citep{2009ASPC..406..112R} and force-free \citep{2017PhRvD..96f3006C} codes. Additionally, there is a promising effort to implement adaptive mesh refinement (AMR) (e.g., \citealt{2004CoPhC.164..297V, 2008CoPhC.178..915F}) to enhance resolution conditionally in regions where the plasma scales become small (e.g., inside dense plasmoids) while dynamically reducing resolution in less critical areas. In a similar vein, hybrid approaches that evolve the upstream region with fluid-type methods (force-free or MHD) while describing the downstream with a kinetic framework may prove useful for local reconnection studies, as done in pulsar magnetospheres \citep{soudais_24}.

State-of-the-art PIC codes are now incorporating routines to reduce the number of simulation particles through resampling \citep{VRANIC201565} and merging \citep{mahlmann_22}. Recent global PIC simulations have also greatly benefited from hybrid integrators of the particle equations of motion. In the upstream, the motion of low-energy particles is reduced to that of their guiding center, alleviating the need to resolve their Larmor orbits \citep{2020ApJS..251...10B}; in contrast, the full equations of motion are used in the downstream. 

The PIC technique has an inherent disadvantage: it approximates the smooth plasma distribution function with a finite number of discrete macroparticles. This finite sampling leads to aliasing issues, which often result in numerical heating and non-physical instabilities. Furthermore, capturing subtle plasma processes becomes computationally challenging due to this numerical noise. Continuous kinetic algorithms---such as those solving the full Vlasov-Maxwell system on a phase-space grid (Vlasov codes)---are becoming more powerful and may prove to be an ideal tool for tackling, e.g., the reconnection onset problem.

As computational models steadily advance, they create new opportunities to deepen our understanding of the physics of RR. With cutting-edge advancements in hardware, innovative algorithms, and the integration of hybrid and adaptive techniques, the community is overcoming longstanding challenges in scale separation, bringing us closer to unveiling the role of RR in the most energetic astrophysical objects.
\\
\\
The progress made over the past decade in understanding the physics of RR has been truly transformative. Rapid advancements in three key areas---(\textit{i}) focused multi-wavelength and multi-messenger observations of relativistic astrophysical sources, (\textit{ii}) ground-breaking laboratory experiments, and (\textit{iii}) cutting-edge computational techniques—strongly suggest that further major breakthroughs are on the horizon. These efforts promise to unlock the secrets of the Universe's most extreme plasmas, shedding light on how high-energy astrophysical sources shine.

\section*{DISCLOSURE STATEMENT}
The authors are not aware of any affiliations, memberships, funding, or financial holdings that
might be perceived as affecting the objectivity of this review. 

\section*{ACKNOWLEDGMENTS}
We are extremely grateful to Hayk Hakobyan, Benoit Cerutti, Louise Willingale and Hantao Ji for contributing to this review: Hayk Hakobyan was instrumental in writing \sect{pulsars} and gave extensive feedback on \figg{radiative} and Table~\ref{tab1}; Benoit Cerutti led the drafting of Section \ref{sect:bhmag}; Louise Willingale and Hantao Ji provided important contributions to \sect{lab}.
We thank the editor, Eliot Quataert, for insightful comments and suggestions on our manuscript and Luca Comisso, Jens Mahlmann, Emanuele Sobacchi, Navin Sridhar and Hao Zhang for comments and help with figures. We are also grateful to several collaborators and colleagues for many inspiring discussions on RR over the years:  
Mitch Begelman, 
Andrei Beloborodov,
Amitava Bhattacharjee,
Roger Blandford, 
Benoit Cerutti,
Luca Comisso,
Jim Drake, 
Fan Guo, 
Hayk Hakobyan, 
Masahiro Hoshino,
Hantao Ji, 
Russell Kulsrud, 
Nuno Loureiro, 
Yuri Lyubarsky, 
Maxim Lyutikov, 
Mikhail Medvedev, 
Ramesh Narayan,
Maria Petropoulou,
Sasha Philippov,
Bart Ripperda,
Anatoly Spitkovsky, 
Hendrik Spruit,
Greg Werner, 
Masaaki Yamada, 
and
Ellen Zweibel.
LS acknowledges support by the 2019 Sloan Fellowship in Physics, by the 2020 Cottrell Scholar Award, by NSF through grants  AST-2307202, AST-2108201, PHY-1903412, ACI-1657507, by the Department of Energy through grants DE-SC0023015, DE-SC0021254, DE-SC0016542, and by NASA through ATP grants 80NSSC24K1238, 80NSSC24K1826, 80NSSC22K0667, 80NSSC20K1556, and NNX17AG21G. LS is also supported by a grant from the Simons Foundation (MP-SCMPS-00001470) and by NSF through the Multimessenger Plasma Physics Center (MPPC), grant NSF PHY-2206609. 
DAU gratefully acknowledges support by NASA through ATP Grants 80NSSC20K0545 and 80NSSC22K0828, and by NSF through Grants AST-1806084 and AST-1903335. DG acknowledges support by NSF through AST-2107802, AST-2107806 and AST-2308090 grants.

\bibliographystyle{ar-style2.bst}
\bibliography{araa.bib,references.bib,rec.bib}

\end{document}